\documentclass[%
superscriptaddress,
preprint,
preprintnumbers,
nofootinbib,
amsmath,amssymb,
aps,
]{revtex4-2}

\usepackage[english]{babel}
\usepackage[utf8]{inputenc}
\usepackage{hyperref}
\usepackage{slashed}
\usepackage{graphicx}
\usepackage{subfig}
\usepackage{multirow}
\usepackage{adjustbox}

\newcounter{notecount}

\usepackage{todonotes}

\newcommand{\smeft}{SMEFT}
\newcommand{\sm}{SM}
\newcommand{\WH}{W^+H}

\newcommand{\iu}{\mathrm{i}}

\newcommand{\bnq}{\begin{equation}}
\newcommand{\enq}{\end{equation}}

\newcommand{\imineqh}[2]{\vcenter{\hbox{\includegraphics[height=#2ex]{#1}}}}

\newcommand{\UbarSp}[1]{\bar{u}\left( #1 \right)}
\newcommand{\VSp}[1]{v\left( #1 \right)}

\newcommand{\Spab}[3]{\langle #1 \vert #2 \vert #3 ]}

\newcommand{\cuW}{C_{uW}}
\newcommand{\cdW}{C_{dW}}
\newcommand{\cHqthree}{C_{\phi q}^{(3)}}
\newcommand{\cHud}{C_{\phi ud}}
\newcommand{\sigsm}{\sigma_{\mathrm{\sm}}}
\newcommand{\sigI}{\sigma_{I}}
\newcommand{\siguW}{\sigma_{uW}}
\newcommand{\sigdW}{\sigma_{dW}}
\newcommand{\sigHqthree}{\sigma_{\phi q^{(3)}}}
\newcommand{\sigHud}{\sigma_{\phi ud}}

\usepackage{dowith}

\newcommand{\UpperCase}[1]{
  \expandafter\newcommand\csname bb#1\endcsname{{\mathbb{#1}}}
  \expandafter\newcommand\csname cal#1\endcsname{{\mathcal{#1}}}   
  \expandafter\newcommand\csname rm#1\endcsname{{\mathrm{#1}}}
  \expandafter\newcommand\csname bf#1\endcsname{{\mathbf{#1}}}
  \expandafter\newcommand\csname bold#1\endcsname{{\boldsymbol{#1}}}
  \expandafter\newcommand\csname hat#1\endcsname{\hat{#1}}
  \expandafter\newcommand\csname tilde#1\endcsname{\widetilde{#1}}
  \expandafter\newcommand\csname bar#1\endcsname{\overline{#1}}
  \expandafter\newcommand\csname frak#1\endcsname{\mathfrak{#1}}
  }
\DoWith\UpperCase ABCDEFGHILJKMNOPQRSTUVWXYZ\StopDoing

\newcommand{\LowerCase}[1]{
  \expandafter\newcommand\csname rm#1\endcsname{{\mathrm{#1}}} 
  \expandafter\newcommand\csname bf#1\endcsname{{\mathbf{#1}}} 
  \expandafter\newcommand\csname bold#1\endcsname{{\boldsymbol{#1}}}
  \expandafter\newcommand\csname hat#1\endcsname{\hat{#1}}
  \expandafter\newcommand\csname tilde#1\endcsname{\tilde{#1}}
  \expandafter\newcommand\csname bar#1\endcsname{\bar{#1}}
  \expandafter\newcommand\csname frak#1\endcsname{\mathfrak{#1}}
  }
\DoWith\LowerCase abcdefghijklmnoqrstuvwxyz\StopDoing

\newcommand{\LO}{\mathrm{LO}}
\newcommand{\NLO}{\mathrm{NLO}}
\newcommand{\NNLO}{\mathrm {NNLO}}

\newcommand{\order}[1]{\calO(#1)}

\newcommand{\as}{\alpha_{\rms}}
\newcommand{\GF}{G_\rmF}

\newcommand{\CA}{C_\rmA}
\newcommand{\CF}{C_\rmF}
\newcommand{\nf}{n_\rmf}
\newcommand{\TR}{T_\rmR}

\newcommand{\muF}{\mu_\rmF}
\newcommand{\muR}{\mu_\rmR}

\newcommand{\pTeplus}{p_{\rmT,e^+}}
\newcommand{\pTH}{p_{\rmT,H}}

\newcommand{\pTWplus}{p_{\rmT,W^+}}

\usepackage{enumitem}
\usepackage{array}          
\usepackage{amsmath}        
\usepackage{booktabs}       
\usepackage{setspace}
\usepackage{pifont} 

\usepackage{ragged2e}  
\makeatletter
\renewcommand{\@makecaption}[2]{%
  \vspace{3mm} 
  \justifying\small\linespread{1.1}\noindent\selectfont\textbf{#1:} #2
}
\makeatother

\makeatletter
\newcommand{\footnotestretch}[1]{%
  \begingroup
  \linespread{1.} 
  \let\footnote\@footnote 
  \endgroup
}
\makeatother

\allowdisplaybreaks 
\begin{document}


\preprint{KA-TP-02-2025}
\preprint{P3H-25-009}
\preprint{TTK-25-06}
\preprint{TTP25-003}
\preprint{TIF-UNIMI-2025-5}

\title{\texorpdfstring{\boldmath$WH$}{WH} production at the LHC within SMEFT
  at next-to-next-to-leading order QCD}

\def\ITP{
  Institute for Theoretical Physics, KIT, D-76128 Karlsruhe, Germany
}\def\TTP{
  Institute for Theoretical Particle Physics, KIT, D-76128 Karlsruhe, Germany
}
\def\IKP{
  Institute for Astroparticle Physics, KIT, D-76344 Eggenstein Leopoldshafen, Germany
}
\def\RWTH{
  Institute for Theoretical Particle Physics and Cosmology, RWTH~Aachen University, Sommerfeldstrasse 16, D-52056 Aachen, Germany
}
\def\TIF{
  Tif Lab, Dipartimento di Fisica, Università di Milano
}
\def\INFN{
  INFN, Sezione di Milano, Via Celoria 16, I-20133 Milano, Italy
}

\author{Marco Bonetti}
\email{marco.bonetti@kit.edu}
\affiliation{\IKP}
\affiliation{\ITP}
\affiliation{\RWTH}

\author{Robert V. Harlander}
\email{harlander@physik.rwth-aachen.de}
\affiliation{\RWTH}

\author{Dmitrii Korneev}
\email{kdmitry.de@gmail.com}
\affiliation{\RWTH}

\author{Ming-Ming Long}
\email{ming-ming.long@kit.edu}
\affiliation{\TTP}

\author{Kirill Melnikov}
\email{kirill.melnikov@kit.edu}
\affiliation{\TTP}

\author{Raoul Röntsch}
\email{raoul.rontsch@unimi.it}
\affiliation{\TIF}
\affiliation{\INFN}

\author{Davide Maria Tagliabue}
\email{davide.tagliabue@kit.edu}
\affiliation{\TTP}
\affiliation{\TIF}
\affiliation{\INFN}


\begin{abstract}
\noindent We study the impact of $q \bar q' W$ and $q \bar q' WH$ interactions,
originating from dimension-six operators of the Standard Model Effective Field
Theory (SMEFT), on the Higgs boson associated production $pp \to \WH$ at the LHC. We compute the next-to-next-to-leading order  QCD corrections to this process and compare these corrections in the Standard Model and SMEFT production mechanisms. 
\end{abstract}

\maketitle
\newpage 
\tableofcontents


\newpage
%

\section{Introduction}
\label{sec:intro}
Despite its enormous success in describing elementary particles and their
interactions, the Standard Model (\sm) of particle physics leaves many questions related to its structure unanswered, and some key phenomenological observations unexplained~\cite{ParticleDataGroup:2022pth}. 
A theory that addresses the structural shortcomings of the SM will likely be characterized 
by an energy scale which is so large that the direct production of new particles, related to such 
a theory, will not be possible at the Large Hadron Collider (LHC), and perhaps  even at  future 
particle colliders. For this reason, searches for physics beyond the \sm\ are currently focused on 
\emph{high-precision} studies of various  \sm\ processes at the LHC and elsewhere, with the hope of 
uncovering off-shell effects of as-of-yet unconfirmed quantum fields. 
 
The \sm\ Effective Field Theory (\smeft), an effective field
theory built from the \sm\ fields and gauge symmetries, is a helpful
theoretical tool to parametrize such effects in a largely model-independent
way. Assuming that the New Physics scale $\Lambda$ is much larger than the
electroweak scale, which is set by the Higgs vacuum expectation value
$v=246$~\,GeV, the \smeft\ Lagrangian describes effects from beyond the Standard Model (BSM)  physics in terms of local operators with mass dimensions $\kappa >4$.
These operators appear in the Lagrangian divided by 
$\Lambda^{\kappa-4}$ and multiplied with dimensionless Wilson coefficients which are assumed to be
of order one. At energies below $\Lambda$, the SMEFT contributions organize themselves into a hierarchical
structure, whose leading term is the \sm\ Lagrangian. The first sub-leading
term is of order $\Lambda^{-1}$ and arises from a single class of dimension-five
operators of the form $L_iL_jHH + \mathrm{h.c.}$, where $L_i$ denotes the
$i^\mathrm{th}$-generation lepton doublet, and $H$ is the Higgs
doublet~\cite{Weinberg:1979sa}. These dimension-five operators may play an important role in various aspects of lepton and quark flavor physics but they are not important for hadron collider phenomenology. 

At mass dimension six, the number of operators amounts to 3045 \cite{Buchmuller:1985jz, Grzadkowski:2010es, Alonso:2013hga, Henning:2015alf}.\footnote{For completeness, let us point out that the full set of operators is known through
mass dimension~12~\cite{Harlander:2023psl}.} Their structure is quite
diverse, ranging from a simple rescaling of \sm\ parameters to
operators that, if present, would lead to radically new phenomena such as
lepton- and baryon-number violation. However, at such low orders in the SMEFT expansion, the
set of operators relevant for a specific process is typically just a
relatively small subset of the full operator basis. This allows us to focus on individual SM production processes and evaluate the impact of the SMEFT operators on them.

The Higgs sector is a particularly promising avenue to search for New Physics since, on the
one hand, it is still comparatively poorly explored and, on the other hand,
its particular form in the \sm\ is largely motivated by minimalistic considerations. A
non-minimal character of this sector would imply non-vanishing Wilson
coefficients in the \smeft\ and could eventually manifest itself in the experimental data.

In this paper, we consider the process of associated Higgs production with a
charged electroweak gauge boson, $pp\to WH$, at the LHC.  Together with the
related process $pp\to ZH$, it can be used to study the structure of the Higgs
boson coupling to electroweak gauge bosons as well as the $b$-quark Yukawa
coupling by means of jet-substructure techniques~\cite{Butterworth:2008iy}.
Furthermore, the similarity of $ZH$ and $WH$ production provides additional opportunities
to search for New Physics effects by studying peculiar differences between the
two processes~\cite{Harlander:2018yns}.

The \sm\ predictions for $VH$ production ($V\in \{W,Z\}$) are under very good
control. The fully differential cross sections are available through NNLO
QCD~\cite{Ferrera:2011bk,Ferrera:2013yga,Ferrera:2014lca,Campbell:2016jau},
including the decay of the Higgs boson into bottom
quarks~\cite{Caola:2017xuq,Ferrera:2017zex,Gauld:2019yng}, and through NLO in the
electroweak coupling~\cite{Ciccolini:2003jy,Denner:2011id,Denner:2014cla}. For the total inclusive cross section even the N$^3$LO QCD corrections are known~\cite{Baglio:2022wzu}.

The effect of all dimension-six operators from \smeft\ which are relevant for
$VH$ production has been studied through NLO QCD in
Ref.~\cite{Alioli:2018ljm}.  At NNLO QCD, the effects of
higher-dimensional operators that lead to anomalous gauge-Higgs couplings were
considered in Ref.~\cite{Bizon:2021rww}.  Ref.~\cite{Haisch:2022nwz} evaluated
QCD-induced \smeft\ effects in $ZH$ production, focussing on their impact on
the subsequent decay of the Higgs boson into a $b\bar b$ pair, and the \smeft-induced re-scaling of the \sm\ couplings.

However, some operators that contribute to $VH$
production induce interactions that are qualitatively different from the
\sm\ ones.  Indeed, as we will show in the next section, some operators lead
to four-point vertices that involve two quarks, a Higgs boson and a vector
boson.  The kinematical structure implied by such interactions may allow one
to disentangle New Physics effects from the typically dominant
\sm\ contributions.  In particular, since the contributions of \smeft\ operators 
increase at higher invariant masses and final-state transverse momenta \cite{Franceschini:2017xkh},  
these phase-space regions constitute high-priority targets when searching for indirect New Physics effects.

One may distinguish CP-even and CP-odd dimension-six operators relevant 
for $VH$ production. The effect of the former type has been investigated, including
NNLO QCD corrections and the matching to a parton shower,
 for both $WH$ and $ZH$ production~\cite{Gauld:2023gtb}. In this paper,
on the other hand, we also take into account the CP-odd operators, focusing
on $\WH$ production, and including the NNLO QCD corrections at the partonic
level.

The remainder of the paper is organized as follows. In the next section, we
provide the relevant information about the partonic contribution to the
process $pp \to \WH$ in SMEFT including higher-dimensional operators and their
impact.  In Section~\ref{sec:ampli} we report the calculation of all 
scattering amplitudes required for
computing the NNLO QCD corrections in the presence of a non-standard SMEFT
interaction.  In Section~\ref{sec:pheno} we perform phenomenological studies 
discussing the interplay of higher-order QCD corrections and
higher-dimensional SMEFT operators in $\WH$ production at the LHC.
We conclude in Section~\ref{sec:concl}.
Additional information can be found in the Appendix and the ancillary file provided with this submission. 


\section{Process description}
\label{sec:SMEFT}
\begin{figure}[t]
    \centering
    \subfloat[\label{fig:SM_EFT_a}]{{\includegraphics[height=0.18\textheight]{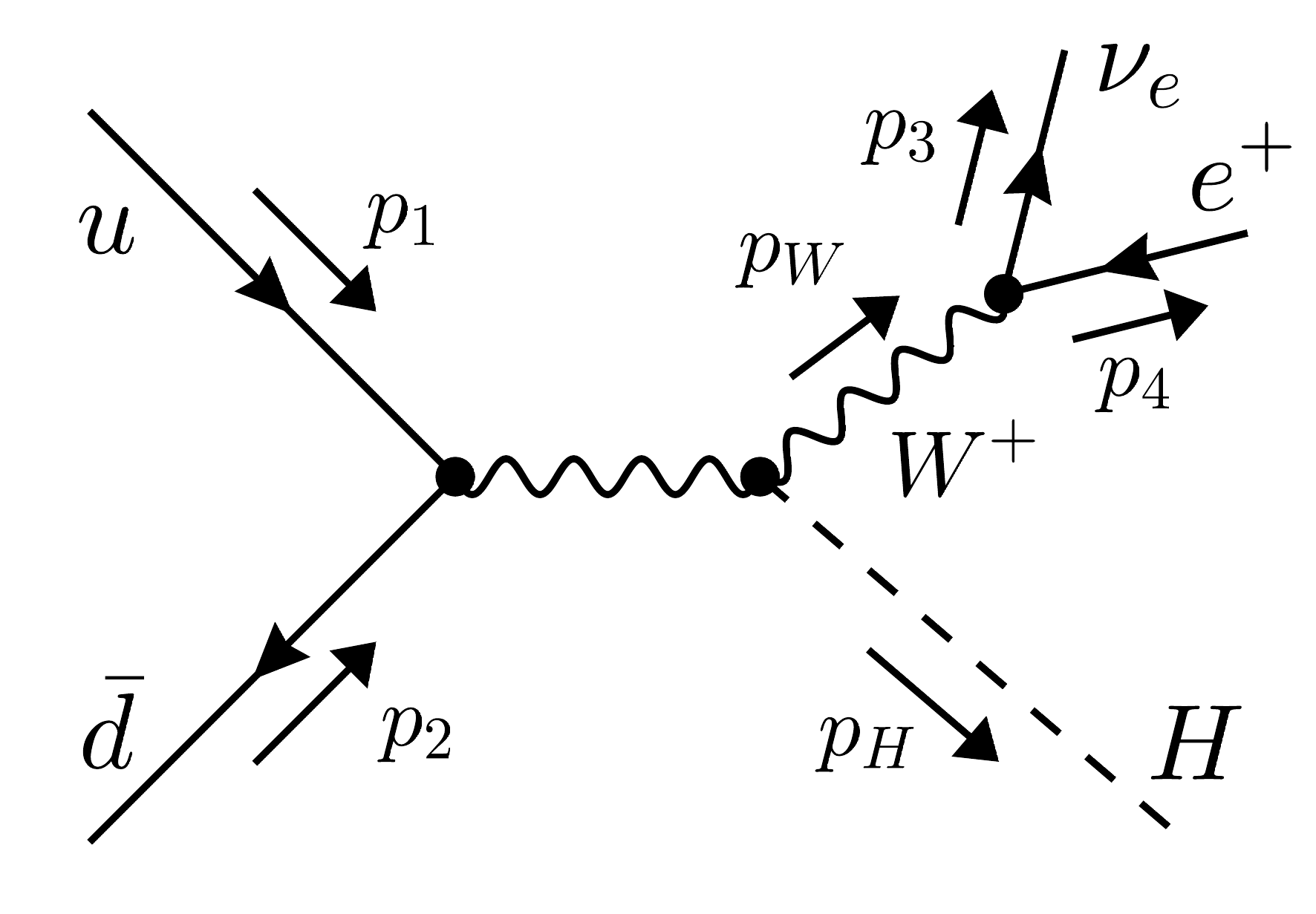}}}
    \hspace{0.05\textwidth}
    \subfloat[\label{fig:SM_EFT_b}]{{\includegraphics[height=0.18\textheight]{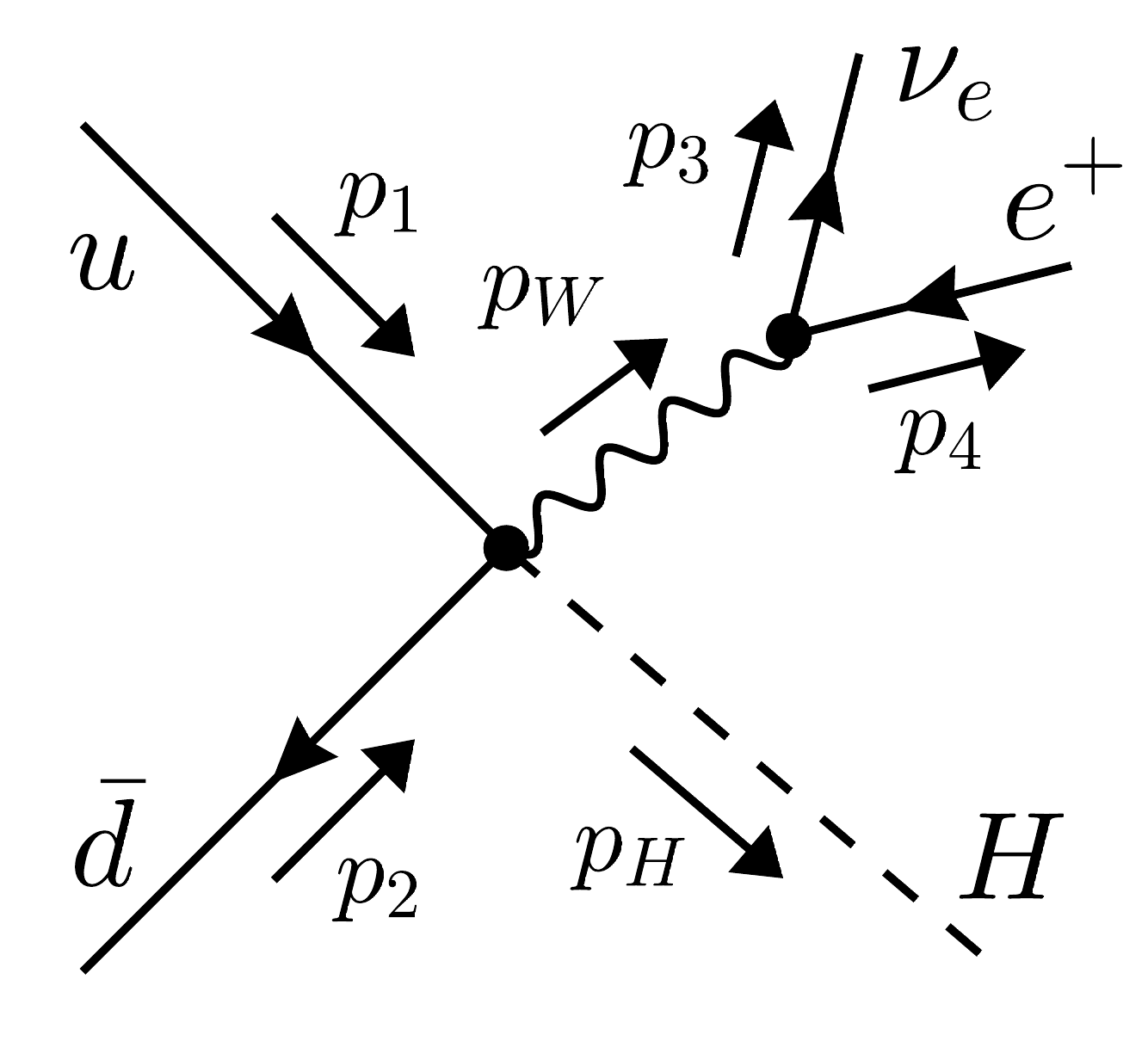}}}
    \caption{Examples of diagrams that contribute to $WH$ production at the LHC. (a)~contains both SM and SMEFT contributions, whilst (b) is present only in SMEFT. }
    \label{fig:SM_EFT}
\end{figure}

We consider the associated production of a $W^+$ boson with a Higgs boson in
the collision of a quark-antiquark pair, followed by the decay of the $W^+$
boson into an electron-neutrino and a positron,
\begin{equation}
    u(p_1 ) ~ \bar d (p_2 ) \to W^+\,(\to \nu_{e}(p_3) +  e^+(p_4)) + H \,.
    \label{eq1}
\end{equation}
In the SM, this process occurs through an intermediate off-shell
$W$ boson that is produced in $q \bar q$ collisions and then decays into a Higgs boson and an 
on-shell $W$ boson of momentum $p_W = p_3 + p_4$ .  
We do not consider decays of the  Higgs boson in this paper.
The inclusion of dimension-six SMEFT
operators leads to point-like interactions between the quark pair and the
Higgs and the $W$ bosons, c.f.\ Fig.~\ref{fig:SM_EFT_b}. This local interaction is induced by the following  SMEFT operators~\cite{Dedes:2017zog}:
\begin{align}
    \label{eq:operators}
    \begin{aligned}
        Q_{\phi q}^{(3)} & = \iu (\phi^\dagger\overleftrightarrow{D}_\mu^I\phi) (\overline{q}_\rmL \tau_I \gamma^\mu q_\rmL) \,,
        &
        Q_{\phi u d} & = \iu (\tilde{\phi}^\dagger D_\mu \phi) (\overline{u}_\rmR \gamma^\mu d_\rmR) \,, \\
        Q_{u W}          &= (\overline{q}_\rmL \sigma^{\mu\nu} u_\rmR) \tau_I \tilde{\phi} W^I_{\mu\nu}
        \,,\qquad &
        Q_{d W}          &= (\overline{q}_\rmL \sigma^{\mu\nu} d_\rmR) \tau_I \phi W^I_{\mu\nu}
        \,,
    \end{aligned}
\end{align}
where $q_\rmL=(u_\rmL,d_\rmL)$ denotes a left-handed quark doublet, and $u_\rmR$ and $d_\rmR$
are right-handed singlets. Here and in the following, by $u$ and $d$ we denote
the up- and down-type quarks, respectively.  The
Higgs doublet is denoted by $\phi$, its dual by $\tilde{\phi} = \iu\tau_2
\phi^*$, $W^I_{\mu\nu}$ is the field-strength tensor of the $\mathrm{SU}(2)_\rmL$
gauge group, $\tau^I$ are the Pauli matrices, $\sigma^{\mu\nu} =
\iu~[\gamma^\mu,\gamma^\nu]/2$, and
\begin{align}
  (\phi^\dagger\overleftrightarrow{D}_\mu^I\phi) =
  \phi^\dagger \tau^I (D_\mu \phi) - (D_\mu \phi^\dagger) \tau^I \phi
    \,.
\end{align}
The addition to the 
SM Lagrangian relevant 
for this paper reads 
\begin{equation}
\delta {\cal L} =
\frac{C_{\phi q}^{(3)} }{\Lambda^2} Q^{(3)}_{\phi q}
+\frac{C_{\phi ud}}{\Lambda^2} Q_{\phi ud}
+\frac{C_{uW}}{\Lambda^2}Q_{uW}
+ \frac{C_{dW}}{\Lambda^2} Q_{dW} \,.
\label{eq4}
\end{equation}
The four quantities 
$C_{\phi q}^{(3)}, C_{\phi ud}, 
C_{uW}, C_{dW}$ are the Wilson coefficients. 
The Feynman rules relevant for our calculation induced by the operators of
Eq.~\eqref{eq:operators} are given by
\begin{align}
\label{eq:qqwh_vertex_1}
    \imineqh{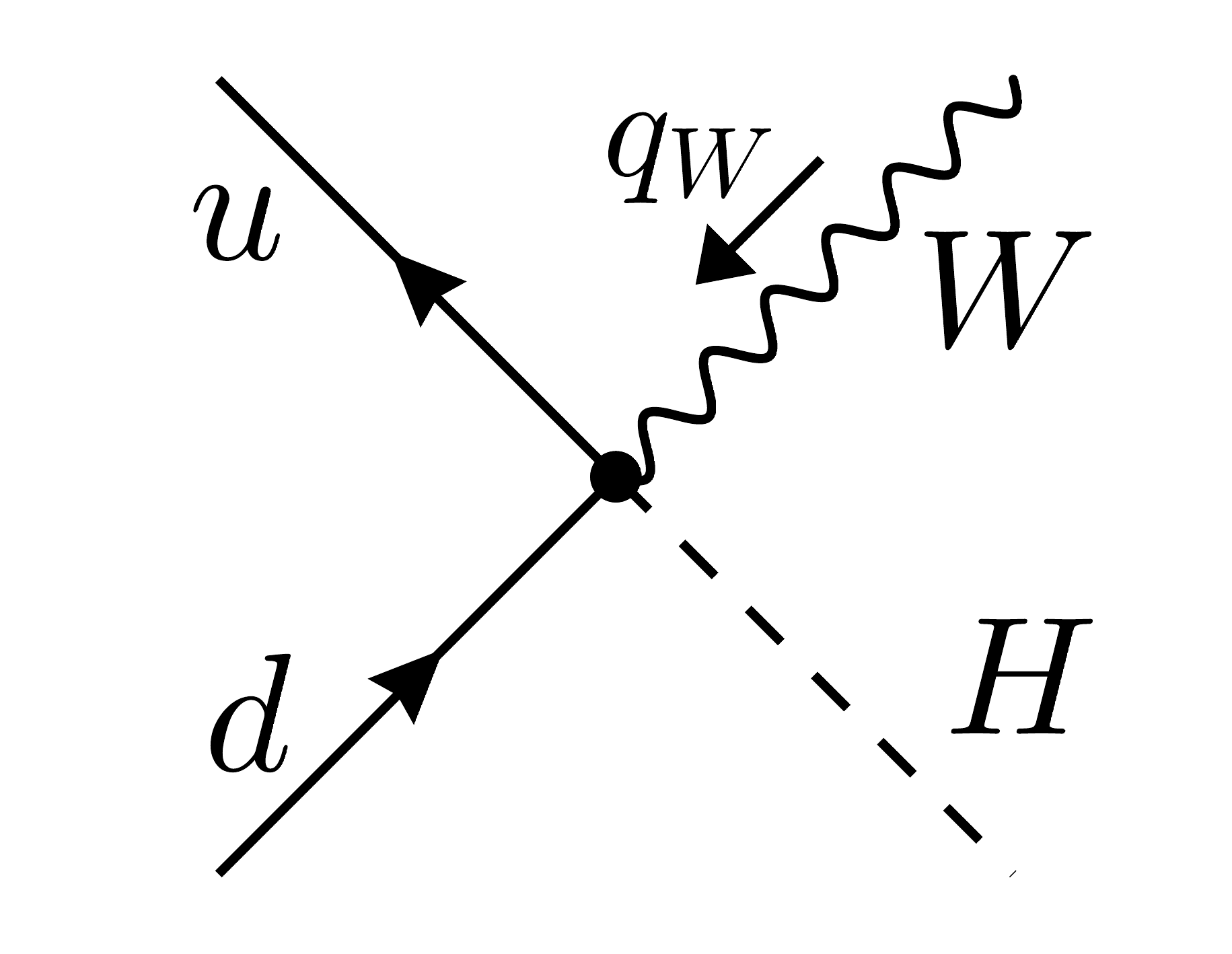}{14} \!\!\!&=
    C_1 \gamma_\mu \mathbb{P}_\rmL +
    C_2 \gamma_\mu \mathbb{P}_\rmR +
    C_3 \sigma_{\mu\nu} q_W^\nu \mathbb{P}_\rmL +
    C_4 \sigma_{\mu\nu} q_W^\nu \mathbb{P}_\rmR
    \,,
    \\
\label{eq:qqwh_vertex_2}
    \imineqh{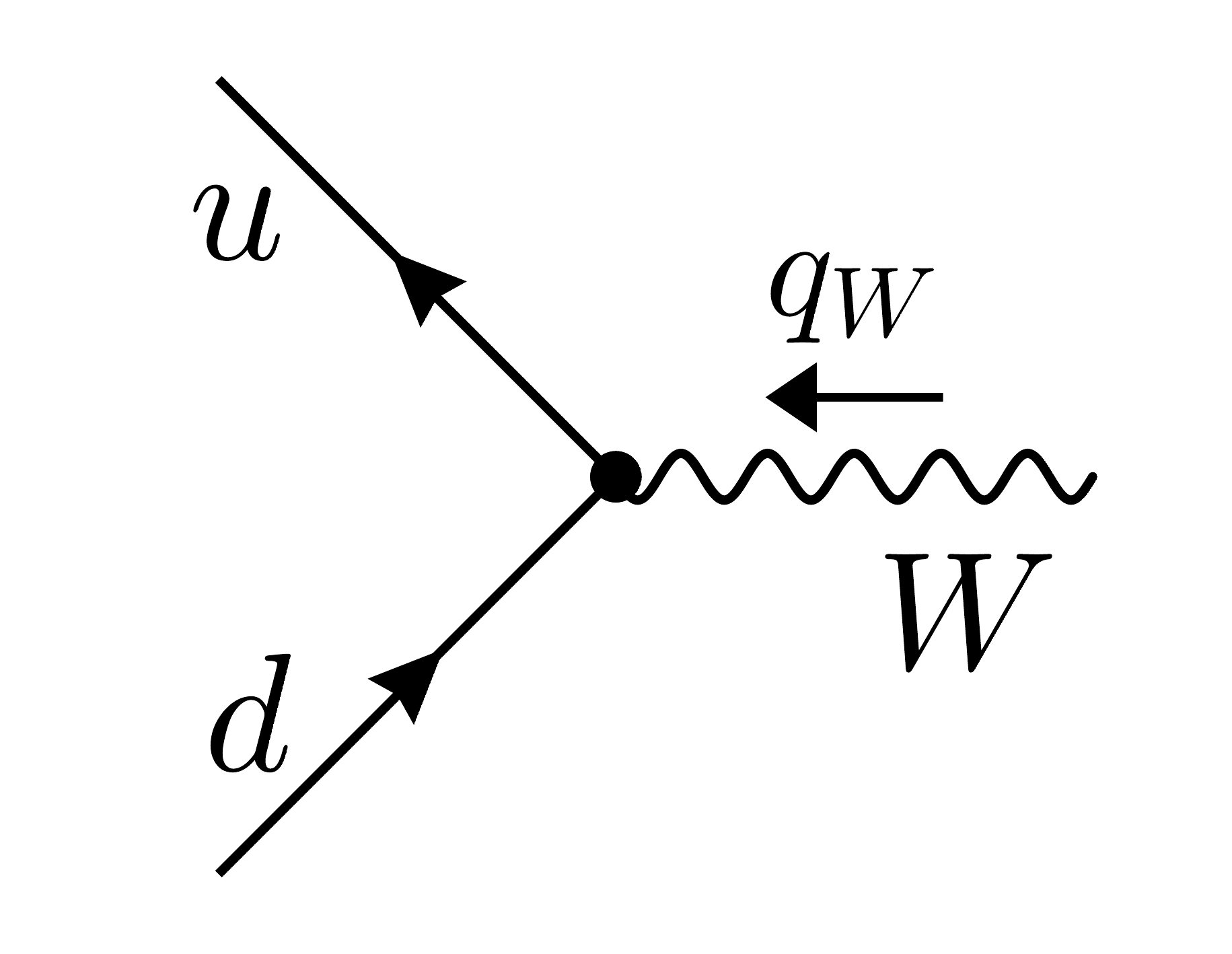}{14} \!\!\!&=
    \left(S_1 + \frac{v}{2} C_1\right) \gamma_\mu \mathbb{P}_\rmL +
    \frac{v}{2} C_2 \gamma_\mu \mathbb{P}_\rmR +
    v C_3 \sigma_{\mu\nu} q_W^\nu \mathbb{P}_\rmL +
    v C_4 \sigma_{\mu\nu} q_W^\nu \mathbb{P}_\rmR
    \,,
    \\
\label{eq:qqwh_vertex_3}
    \imineqh{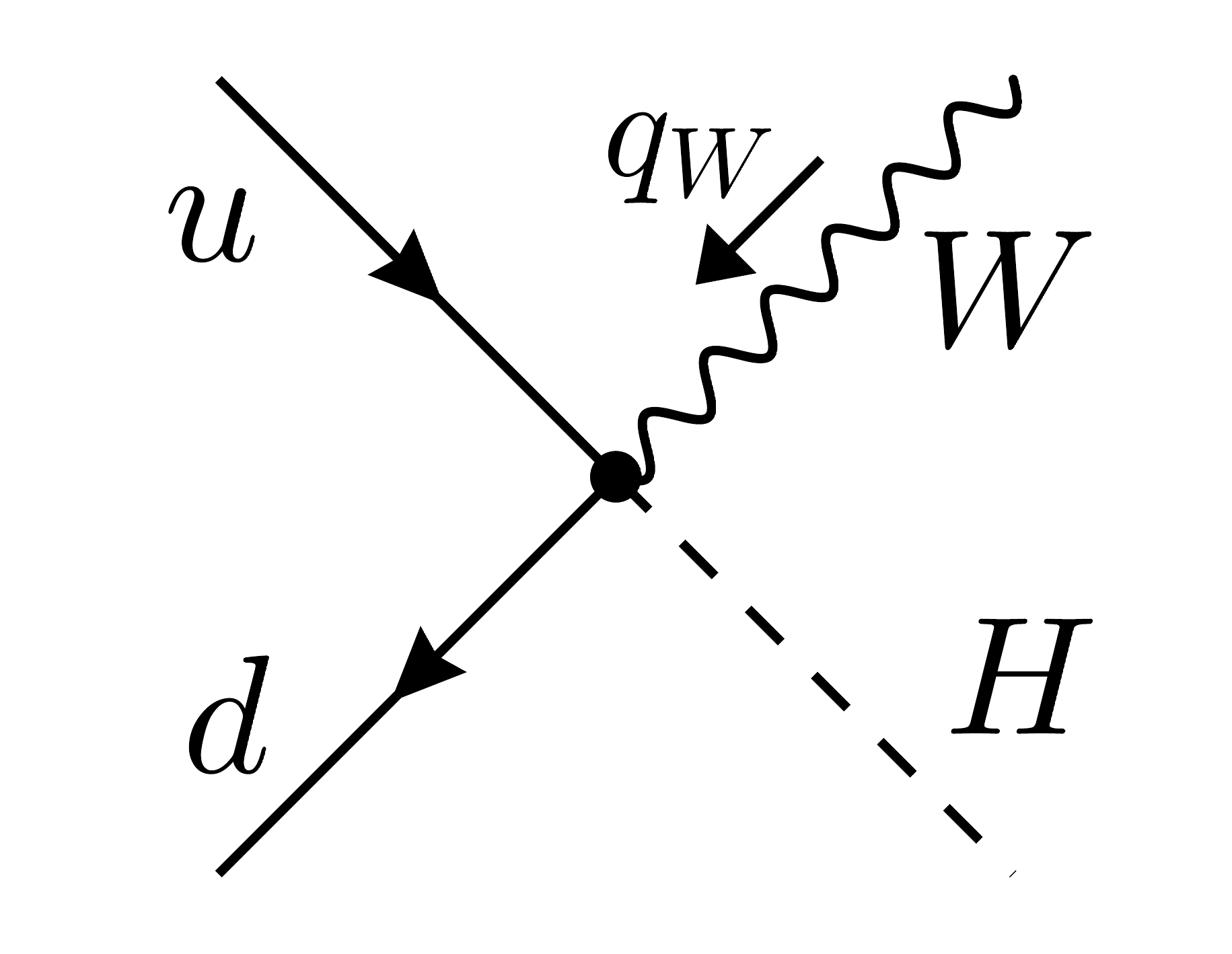}{14} \!\!\!&=
    -C_1^* \gamma_\mu \mathbb{P}_\rmL -
    C_2^* \gamma_\mu \mathbb{P}_\rmR +
    C_3^* \sigma_{\mu\nu} q_W^\nu \mathbb{P}_\rmR +
    C_4^* \sigma_{\mu\nu} q_W^\nu \mathbb{P}_\rmL
    \,,
    \\
\label{eq:qqwh_vertex_4}
    \imineqh{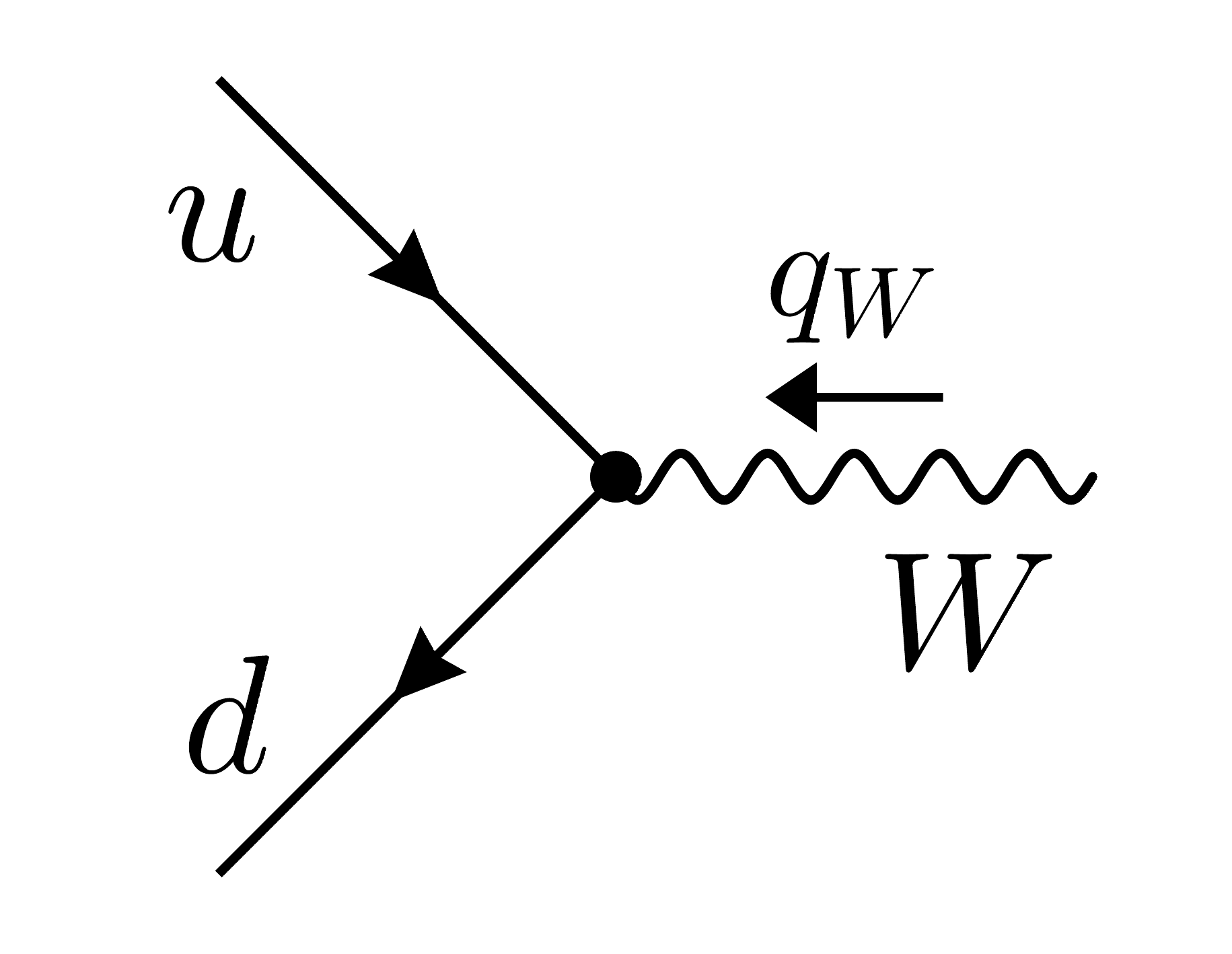}{14} \!\!\!&=
    \left(S_1 - \frac{v}{2} C_1^*\right) \gamma_\mu \mathbb{P}_\rmL -
    \frac{v}{2} C_2^* \gamma_\mu \mathbb{P}_\rmR +
    v C_3^* \sigma_{\mu\nu} q_W^\nu \mathbb{P}_\rmR +
    v C_4^* \sigma_{\mu\nu} q_W^\nu \mathbb{P}_\rmL
    \,,
\end{align}
where $q_W$ is the incoming momentum of the $W$ boson, $\mathbb{P}_{\rmL,\rmR} = (1
\pm \gamma_5)/2$ are the projection operators onto left- and right-handed helicity
states,  $S_1= -\iu g/\sqrt{2}$ is the SM $Wq\bar q'$  coupling, and $v$ is the Higgs field vacuum expectation value in the \sm. The four constants 
$C_{1,2,3,4}$ are related to the Wilson coefficients of the four
operators, c.f.\ Eq.~(\ref{eq4}).
They read
\begin{align}
    \begin{aligned}\label{eq:Wilson}
        C_1 &= -\iu \sqrt{2} g v \frac{C_{\phi q}^{(3)}}{\Lambda^2}
        \,,\qquad&
        C_2 &= -\frac{\iu g v}{\sqrt{2}} \frac{\cHud}{\Lambda^2}
        \,,\\
        C_3 &= -2 \frac{(\cuW)^*}{\Lambda^2}
        \,,\qquad&
        C_4 &= -2 \frac{\cdW}{{\Lambda^2}}
        \,.
    \end{aligned}
\end{align}
Furthermore, we use  $g = 2 m_W/v = e /\sin\theta_\rmW$, where $m_W$ is the mass of the $W$ boson,  $e>0$ is the positron charge and $\theta_\rmW$ is the weak mixing angle.\footnote{All the
parameters listed here are SM-like, thanks to the fact that we do not consider
SMEFT corrections involving $Q_{\phi W B}$ or $Q_{\phi D}$ operators, see Ref.~\cite{Dedes:2017zog}.}  The operators $Q_{uW}$ and $Q_{dW}$ also induce interactions with two gauge
bosons which contribute at higher orders in the electroweak coupling and are
thus irrelevant within the scope of this paper. The same comment applies to Feynman
rules involving two Higgs bosons arising from $Q_{\phi q}^{(3)}$ and $Q_{\phi ud}$.

In principle, the Wilson coefficients in Eq.~\eqref{eq:Wilson} depend on the
quark generation. For simplicity, we will neglect this dependence.  Furthermore, for our
purpose of investigating the effects of non-SM interactions and their
interplay with QCD corrections, we also replace the CKM matrix with the identity.\footnote{\label{fn:ckm}For the SM $WH$ 
associated production, this approximation is accurate to about one percent.}
Finally, we do not consider possible  SMEFT corrections to the decay of the
$W$ boson into a positron and an electron-neutrino.


\section{Calculation of the amplitudes}
\label{sec:ampli}

\begin{figure}
\centering
	\subfloat[]{{\includegraphics[height=0.15\textheight]{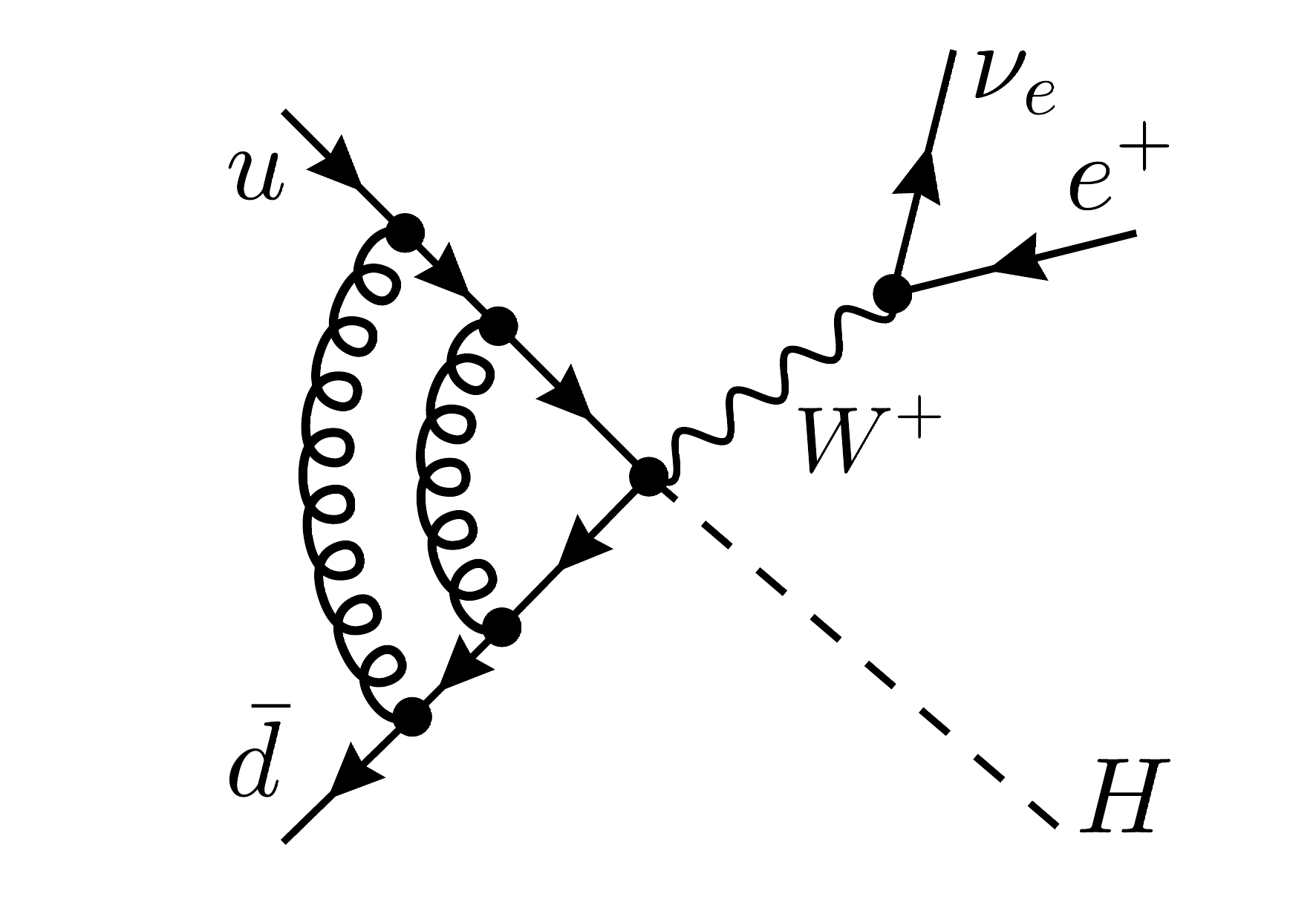}}}
    \quad
    \subfloat[]{{\includegraphics[height=0.15\textheight]{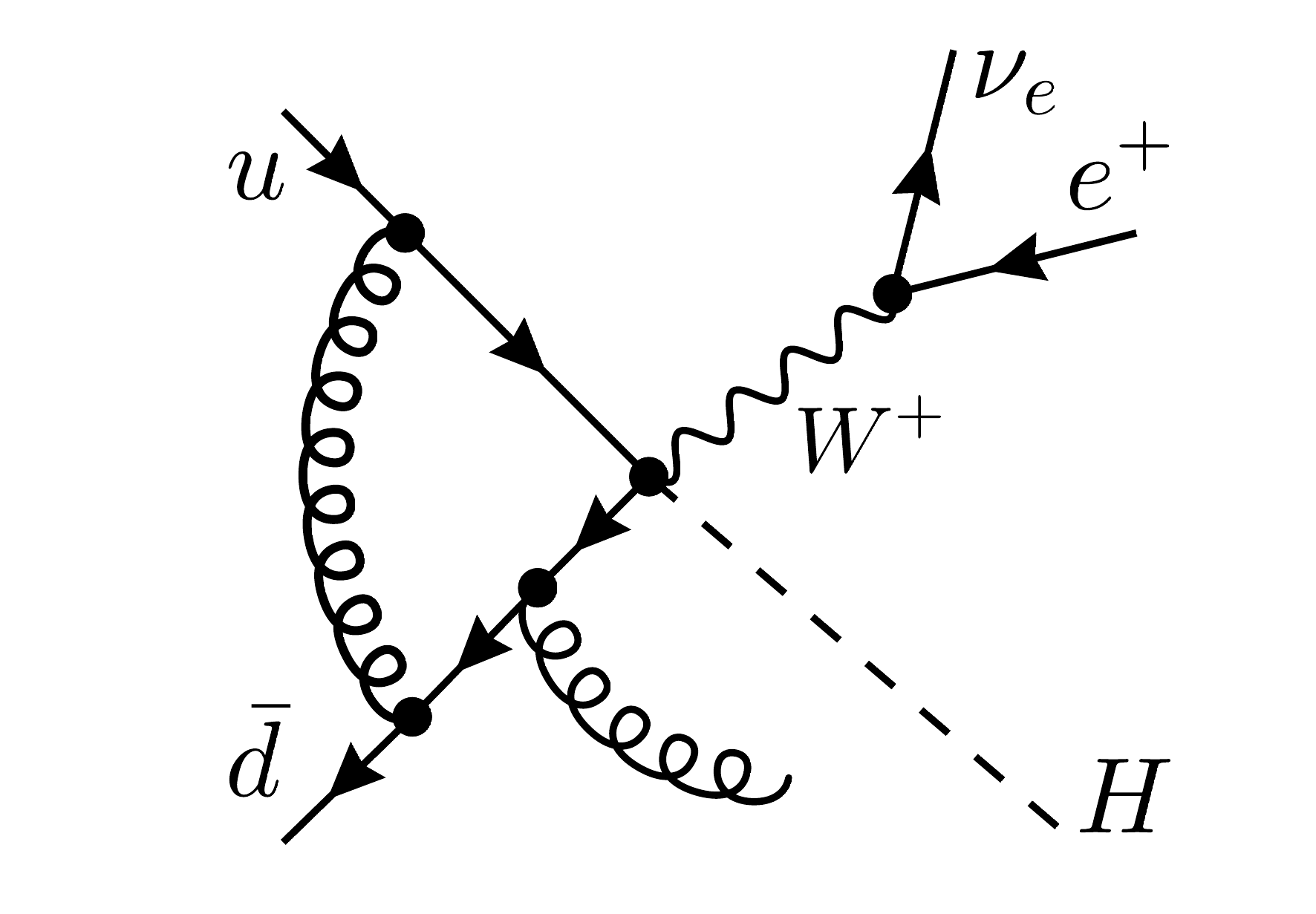}}}
    \quad
    \subfloat[]{{\includegraphics[height=0.15\textheight]{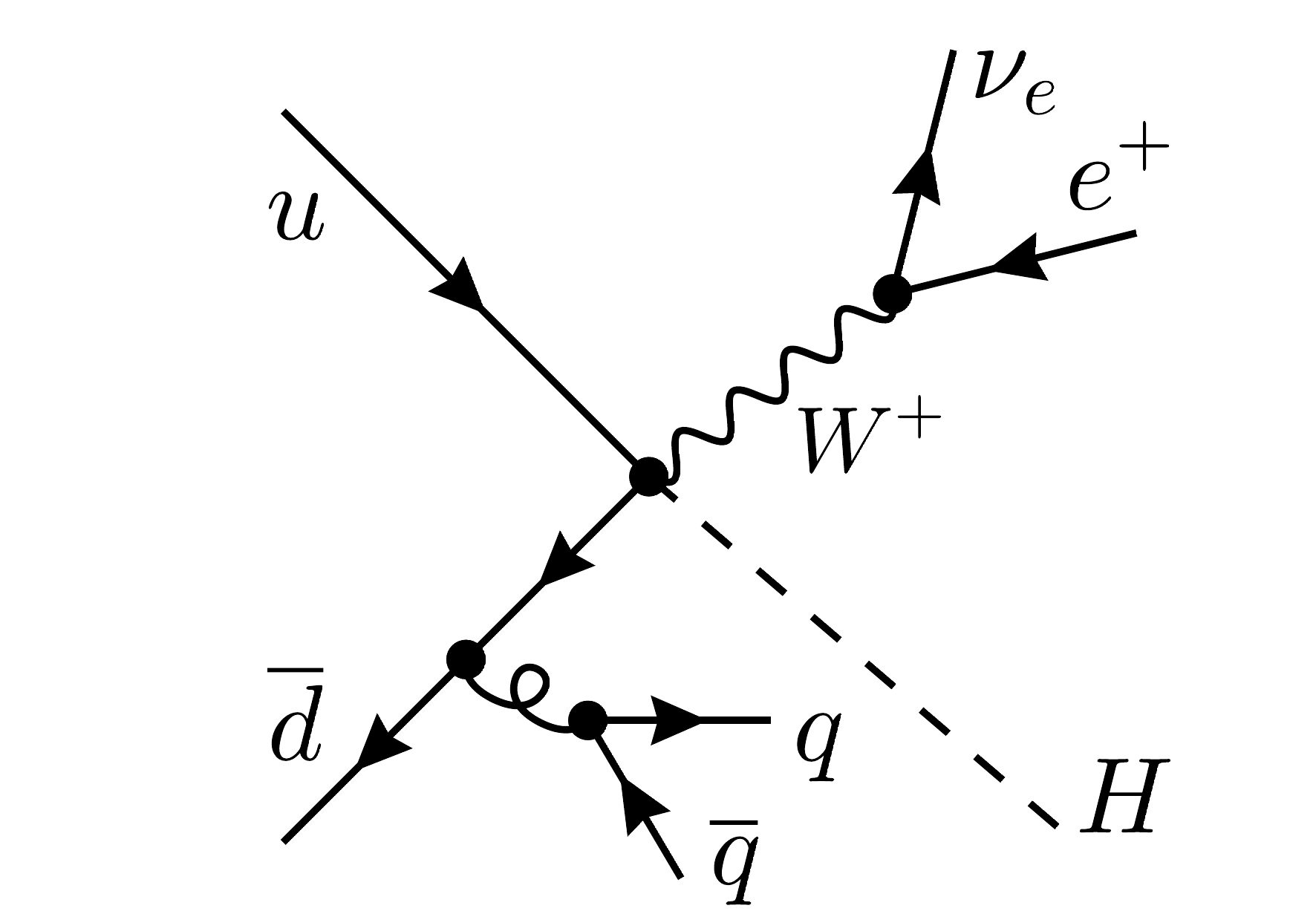}}}
    \caption[]{\centering Representative NNLO contributions to $pp \to W^+H $.}
    \label{fig:SM_EFT:loop}
\end{figure}

To compute NNLO QCD corrections to the process in Eq.~(\ref{eq1}), we need to
account for double-virtual, real-virtual and double-real contributions.
Hence, we require the scattering amplitudes for the following processes:
\begin{itemize}
\item
$0 \to \overline{u}d\nu_{e}e^+H$
through  two loops;
\item
$0 \to \overline{u}d\nu_{e}e^+H+g$
through  one loop; 
\item 
$0 \to \overline{u}d\nu_{e}e^+H+gg$ 
and $0 \to \overline{u}d\nu_{e}e^+H+ q \overline q$ at tree-level,
\end{itemize}
where $q \in \{u,d,c,s,b\}$.\footnote{Throughout the paper we consider $\nf=5$ massless quarks.}  Representative diagrams for each of these
three contributions are shown in Fig.\,\ref{fig:SM_EFT:loop} for the pointlike $q\bar{q}'WH$ interactions.

\begin{figure}
\centering
	\subfloat[]{{\includegraphics[height=0.15\textheight]{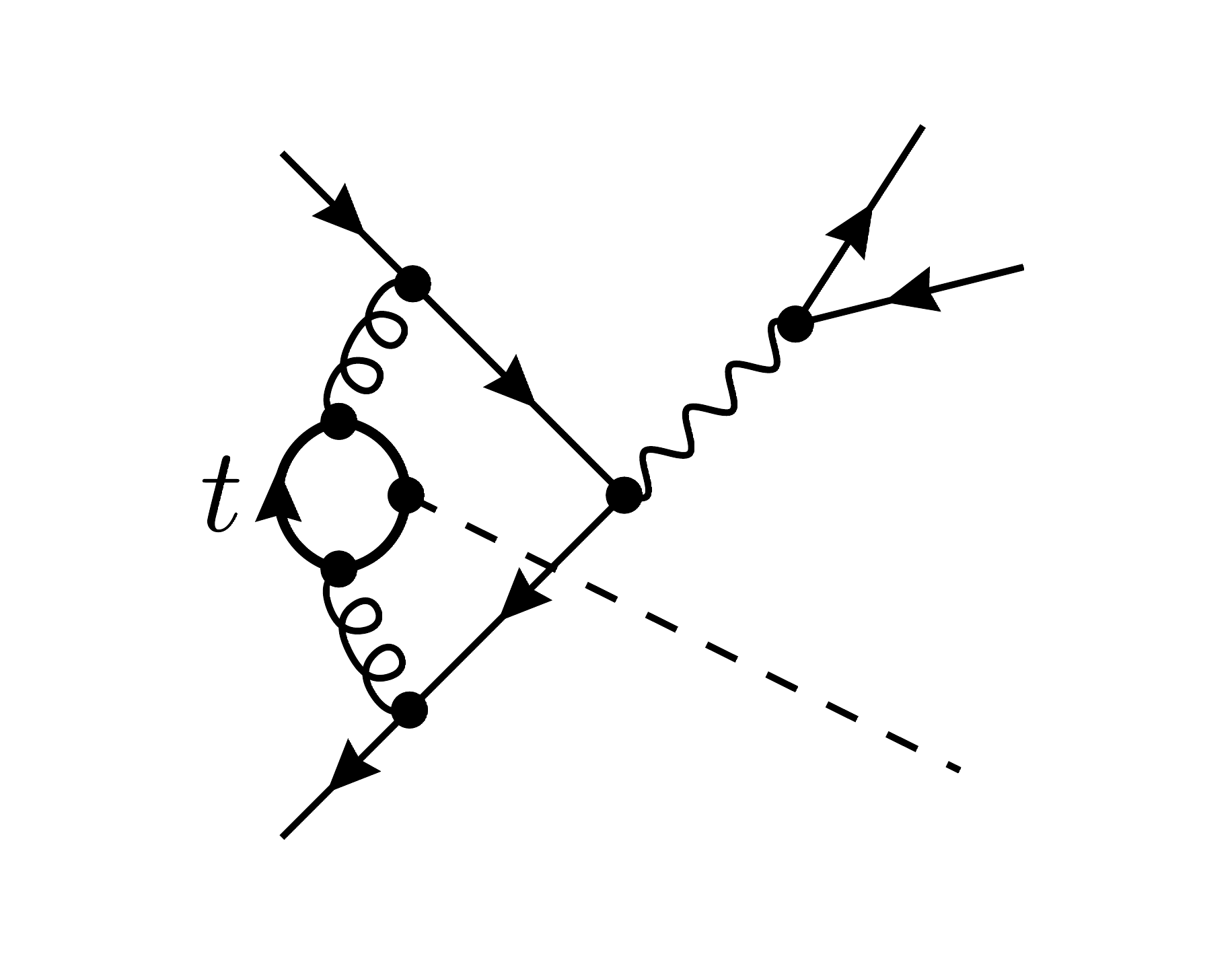}}\label{fig:top-loop:yt}}
    \qquad
    \subfloat[]{{\includegraphics[height=0.15\textheight]{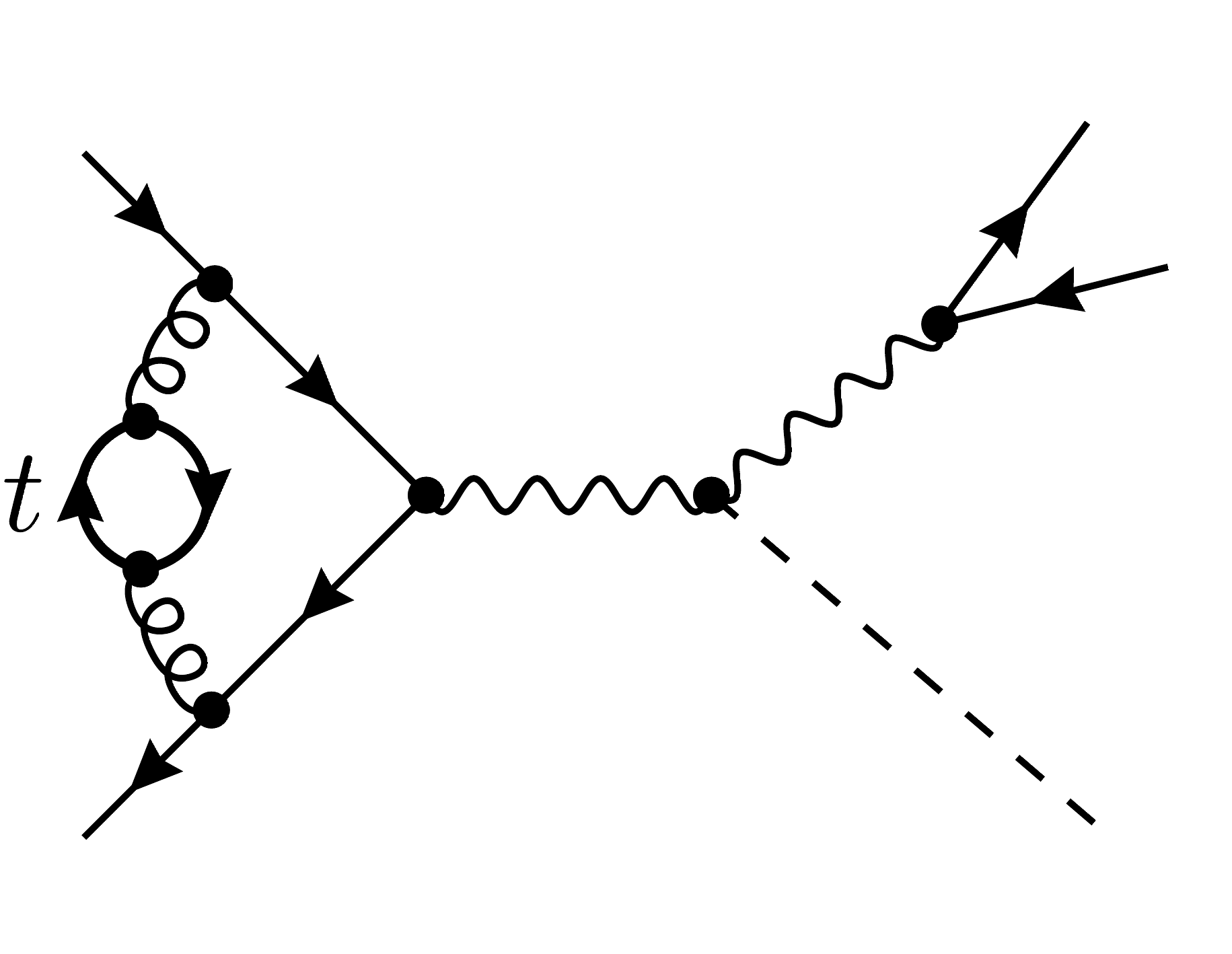}}\label{fig:top-loop:loop}}
    \caption[]{Contributions involving closed top-quark loops. They are
      neglected in this paper for reasons discussed in the main text.}
    \label{fig:top-loop}
\end{figure}

For completeness, let us remark that there are contributions involving a
closed top-quark loop from which the Higgs boson is radiated, see, e.g.,
Fig.\,\ref{fig:top-loop:yt}. In the \sm, such
terms were first considered in Ref.\,\cite{Brein:2011vx}. They are proportional to the second power of the strong coupling constant $\as$, and are finite and gauge invariant. They are sensitive to
BSM modifications of the top Yukawa coupling, which is not the topic of our
current paper. We therefore defer the study of these terms to future
work. Other contributions involving closed top-quark loops factorize into
Drell-Yan-like $W$ production and subsequent decay into $WH$, see Fig.~\ref{fig:top-loop:loop}, for example. In the \sm, they are suppressed by two to
three orders of magnitude w.r.t.\ the remaining NNLO
contributions~\cite{Kniehl:1990iva,Brein:2003wg,Brein:2011vx}. We also expect a
similar degree of suppression if the amplitude is mediated by the SMEFT operators, and neglect such contributions in our calculation. 

Since we consider the production 
of the $W$ boson in the narrow width approximation, any amplitude we need to compute can be written 
as a product of hadronic and leptonic currents, divided by the mass and the
width of the $W$ boson,
\bnq 
\mathcal{A} =
-\frac{\mathcal{H}_{\mu}\mathcal{L}^{\mu} }{\Gamma_W m_W} \,.
\label{eq6}
\enq
The denominator  arises  from the on-shell
($p_W^2 \to m_W^2$)
limit of the $W$ boson propagator 
\bnq
\lim_{p_W^2 \to m_W^2} \frac{-\iu g^{\mu\nu}}{p_W^2-m_W^2+\iu\Gamma_W m_W} 
= \frac{- g^{\mu \nu}}{\Gamma_W m_W}
\, .
\enq
  The leptonic current is always
left-handed; in the spinor-helicity formalism (for a review, see
Ref.\,\cite{Elvang:2013cua}), we can write it as 
\bnq
\mathcal{L}^{\mu}(p_3,p_4) = \Spab{3}{\gamma^{\mu}}{4} \,.  
\enq 

The effective vertices shown in Eqs.~\eqref{eq:qqwh_vertex_1} to
\eqref{eq:qqwh_vertex_4} contain terms either with one Dirac matrix
$\gamma^\mu$ or with $\sigma^{\mu \nu}$. It is therefore convenient to split the hadronic current accordingly. We write
\bnq
    \mathcal{H}^{\mu} = \mathcal{H}_1^{\mu} + \mathcal{H}_2^{\mu} \,,
\enq
where $\mathcal{H}_1^{\mu}$ corresponds to terms containing $\gamma^\mu$ and $\mathcal{H}_2^{\mu}$ to the terms containing $\sigma^{\mu \nu}$. Similarly, we express the amplitude as a sum of two terms
\bnq
    \mathcal{A} =
    \mathcal{A}_1 + \mathcal{A}_2 \,,
\enq
where 
\bnq
    \mathcal{A}_i = -\frac{\mathcal{H}_{i,\mu}\mathcal{L}^{\mu} }{\Gamma_W m_W} \,,
    \qquad\
    i = 1,2 \,.
\enq
The two amplitudes ${\cal A}_{1,2}$ can be computed separately. Furthermore, each of the amplitudes ${\cal A}_{1,2}$ contains left- and right-handed projectors. These projectors select specific helicities of external quarks; for this reason, we discard these projections when computing  helicity amplitudes and account for them later by choosing  appropriate helicities of the  quarks
that contribute to matrix elements of particular operators. 
Furthermore, since the $\gamma^\mu$ vertex conserves helicity, while the $\sigma^{\mu \nu}$ vertex flips it, the two amplitudes ${\cal A}_1$ and ${\cal A}_2$ do not interfere and can be studied independently of each other. This implies that the $\sigma^{\mu\nu}$ part of the amplitude does not interfere with the \sm\ amplitude and that its contribution to the $WH$ production cross section is suppressed by $1/\Lambda^4$, in spite of the fact that 
it originates from a dimension-six operator. We will comment on this further in the following section.

We employ standard techniques for higher-order calculations to compute the amplitudes ${\cal A}_{1,2}$. 
To this end, we generate Feynman diagrams  using \texttt{qgraf}~\cite{Nogueira:1991ex} and 
process them using \texttt{FORM}~\cite{Ruijl:2017dtg} and \texttt{Mathematica}~\cite{Mathematica}.
We describe particularities related to the calculation of one- and two-loop amplitudes below. 

\subsection{Virtual corrections to $0 \to \overline{u}d\nu_{e}e^+H$}

In this section, we discuss the calculation of one- and two-loop virtual
corrections to the process $0 \to \overline{u}d\nu_{e}e^+H$. The vector 
($\gamma^\mu$) part of the hadronic amplitude is described by the familiar quark form factor (see Refs.~\cite{Garland:2002ak, Gehrmann:2010ue} for the more  recent results).  We therefore write 
\bnq
    \label{PV:H1}
    \mathcal{H}_1^{\mu} = \mathcal{F}_{1,1}(s_{12}) \overline u(p_2)\gamma^{\mu }v(p_1) \,,
\enq
where $u$ and $v$ are Dirac spinors, and  $\mathcal{F}_{1,1}$ is the two-loop form factor which depends on $s_{12} = (p_1 + p_2)^2$. Throughout this section, we will use the convention that the four-momenta of all particles are outgoing; it is straightforward to transform the results to physical kinematics by analytic continuation.

To the best of our knowledge, virtual corrections to amplitudes with the tensor ($\sigma^{\mu \nu}$) currents  are unknown.
Thanks to the fact that such a current is helicity-flipping, the number of independent Lorentz structures that may appear is restricted.
We specify the hadronic current $\mathcal{H}_2$ for three- and four-point interaction vertices; as we will see, they are different because of the momentum flow in the tensor current. 

For the four-point case, the most general form  of the amplitude $\mathcal{H}_{2}^{\mu}$ reads
\bnq
    \mathcal{H}_{2, \text{4pt}}^{\mu} = \mathcal{F}_{1} \UbarSp{p_2}[\gamma ^{\mu
  },\slashed{p}_{W}]\VSp{p_1} + \mathcal{F}_{2}
\UbarSp{p_2}\VSp{p_1} p_1^{\mu } + \mathcal{F}_{3} \UbarSp{p_2}\VSp{p_1}
p_2^{\mu } \,.
    \label{PV:H2_4pt}
\enq
We note that in this case  $p_W = -p_H - p_1 - p_2$. However, in the case of the three-point interactions, $p_W$ is replaced by the momentum of the intermediate $W$ boson $p_{W^*} = -p_1 - p_2$ and the above expression simplifies. Hence, for the three-point interaction case, we write
\bnq
\mathcal{H}_{2, \text{3pt}}^{\mu} = 
  \mathcal{F}_{4} \;\UbarSp{p_2}\VSp{p_1} p_1^{\mu } 
+ \mathcal{F}_{5} \;\UbarSp{p_2}\VSp{p_1} p_2^{\mu } \,.
    \label{PV:H2_3pt}
\enq

Thanks to the structure of the interaction vertex that involves $\sigma^{\mu \nu} p_{W,\nu}$,  the amplitude $\mathcal{H}_2$ vanishes  when contracted with the $W$ boson momentum, resulting in
\bnq
    \mathcal{H}_{2, \text{4pt}} \cdot p_{W} = 0 \,,
    \qquad\qquad
    \mathcal{H}_{2, \text{3pt}} \cdot p_{W^*} = -\mathcal{H}_{2, \text{3pt}} \cdot (p_1 + p_2)  = 0
        \,.
\label{eq13}
\enq
We  use Eq.~(\ref{eq13}) to  rewrite $\mathcal{H}_{2, \text{4pt}}$ in terms of two independent form factors
\bnq
    \mathcal{H}_{2, \text{4pt}}^{\mu} = 
      \mathcal{F}_{2,1} \mathcal{T}_1^{\mu}  
    + \mathcal{F}_{2,2} \mathcal{T}_2^{\mu}
    \,,
    \label{eq19}
\enq
and $\mathcal{H}_{2, \text{3pt}}$ in terms of a single form factor
\bnq
    \mathcal{H}_{2, \text{3pt}}^{\mu} = 
      \mathcal{F}_{2,3} \UbarSp{p_2}\VSp{p_1}\left( p_1^{\mu } - p_2^{\mu } \right)
    \,.
    \label{eq20}
\enq
 In Eq.~(\ref{eq19})
we have used the following quantities  
\bnq
\begin{split}
    \mathcal{T}_1^{\mu} &= \UbarSp{p_2}[\gamma ^{\mu },\slashed{p}_W]\VSp{p_1}, \;\;\;
    \mathcal{T}_2^{\mu} = \UbarSp{p_2}\VSp{p_1} \left( s_{2W} p_1^{\mu }-s_{1W} p_2^{\mu } \right)
    ,
\end{split}
\enq
with $s_{iW} = 2 p_i \cdot p_{W}$, $i = 1,2$, to write the amplitude.

The form factors can be extracted by applying the projection operators to the matrix element of the current $\mathcal{H}_2^\mu$. We do this separately for contributions that originate from three- and four-point interaction vertices.  In the former case, projecting on $\mathcal{F}_{2,3}$  is straightforward. In the latter case, we write
\bnq
\label{eq:projecting_ff_PV}
    \mathcal{F}_{2,j}
    =
    \sum_{\mathrm{pol}} \mathcal{P}_{j} \; \epsilon_W^{*} \cdot \mathcal{H}_{2, \text{4pt}}
    \,,
\enq
where $\epsilon_W^{*}$ is the polarization vector of the $W$ boson and we sum over polarizations of all external particles including quarks and the $W$ boson. The projection operators ${\cal P}_{j=1,2}$ can be written as linear combinations of the tensors $\mathcal{T}_1^{\dagger}$ and $\mathcal{T}_2^{\dagger}$ 
\bnq
    \label{pv:proj}
    \mathcal{P}_j = \mathbf{R}_{jk} \, \mathcal{T}_{k,\mu}^{\dagger} \, \epsilon_W^\mu
    \,,
\enq
where the matrix $\mathbf{R}$ reads
\bnq
    \label{eq:matR_PV}
    \mathbf{R} =
    \frac{1}{\mathcal{N}}
    \begin{pmatrix}
        s_{12}^2 s_{1W} s_{2W}~~~~& -2 s_{12} s_{1W} s_{2W} \\
        -2 s_{12} s_{1W} s_{2W}~~~~& 4 (d-2) s_{1W} s_{2W}-4 (d-3) s_{12} p_W^2
    \end{pmatrix}
    \, .
\enq
The quantity ${\cal N}$ is given by  
\bnq
    \mathcal{N} = 8 (3-d) s_{12}^2 s_{1W} s_{2W} \left(s_{12} p_W^2-s_{1W} s_{2W}\right),
\enq 
where $d = 4-2 \epsilon$ is the space-time dimension. The sum over $W$ polarizations is given by the standard  expression  
\begin{equation}
    \sum \epsilon^{*,\mu}_W
    \epsilon^{\nu}_W  
    = -g^{\mu \nu} + \frac{ p_W^\mu p_W^\nu}{m_W^2} \,,
\end{equation}
where the second term can be dropped because of Eq.~\eqref{eq13}. 

Once the current $\mathcal{H}_2^\mu$ is projected onto individual form
factors, we make use of integration-by-parts
identities~\cite{Chetyrkin:1981qh,Laporta:2000dsw} to reduce all one- and
two-loop integrals to master integrals. For this purpose, we have used
\texttt{Reduze}~\cite{vonManteuffel:2012np} and
\texttt{Kira}~\cite{Maierhofer:2017gsa,Klappert:2020nbg,Lewis} interfaced with
\texttt{FireFly}~\cite{Klappert:2019emp,Klappert:2020aqs}.
The master integrals required to compute $\mathcal{H}_1^\mu$ and
$\mathcal{H}_2^\mu$ are identical; they can be borrowed from Refs.~\cite{Gehrmann:1999as, Gehrmann:2010ue}.
Although, as mentioned earlier, we can take the vector form factor
${\cal F}_{1,1}$ from the literature, we also recomputed it with the same
setup we used to calculate the current ${\cal H}_2^\mu$, and found agreement
with the known results in Ref.~\cite{Gehrmann:2010ue}.

The calculation of the helicity amplitudes is straightforward. Starting from
Eqs.~\eqref{PV:H1}, \eqref{PV:H2_4pt}, and \eqref{PV:H2_3pt}, using Eq.~\eqref{eq6}, and fixing the 
helicities of  external partons, we obtain
\bnq
\begin{split}
    \mathcal{A}_1(p_1^{+},p_2^{-},p_3^{-},p_4^{+}) &= \frac{2 \mathcal{F}_{1,1}}{m_W\Gamma_W}   \; \langle 23 \rangle [14] \,, \\
    \mathcal{A}_1(p_1^{-},p_2^{+},p_3^{-},p_4^{+}) &=  \frac{2 \mathcal{F}_{1,1}}{m_W\Gamma_W} \; \langle 13 \rangle [24] \,, \\
    \mathcal{A}_2(p_1^{-},p_2^{-},p_3^{-},p_4^{+}) &= -\frac{2\mathcal{F}_{2,1} }{m_W \Gamma_W}  \langle 13 \rangle \langle 23 \rangle [34] - \frac{\langle 12 \rangle}{2 m_W \Gamma_W}  \; K \,, \\
    \mathcal{A}_2(p_1^{+},p_2^{+},p_3^{-},p_4^{+}) &= \frac{ 2\mathcal{F}_{2,1}}{m_W \Gamma_W}  [14] [24] \langle 34 \rangle -\frac{[12]}{2 m_W \Gamma_W}  \; K \,,
\end{split}
\enq
where $K$ is given by
\begin{equation}
K = \langle 13 \rangle [14]\left(s_{2W}\mathcal{F}_{2,2}+2m_W^2\Pi_{W^*}\mathcal{F}_{2,3}\right)-\langle 23 \rangle [24]\left(s_{1W}\mathcal{F}_{2,2}+2m_W^2\Pi_{W^*}\mathcal{F}_{2,3}\right) \,,
\end{equation}
with\footnote{We note that it is customary to include the width term $\rmi\, m_W \Gamma_W$ in the propagator of the intermediate $W$ boson, although it cannot go on the mass shell since $s_{12} > (m_W + m_H)^2$.}
\bnq
\Pi_{W^*} = \frac{1}{s_{12}-m_W^2 + \rmi \, m_W\Gamma_W} \,.
\enq
All other helicity amplitudes vanish thanks to the left-handed coupling between the $W$ boson and the leptonic current.

Physical helicity amplitudes are constructed from the $\mathcal{A}_i$ by restoring the electroweak couplings:
\bnq
\begin{split}
    \mathcal{M}(p_1^{+},p_2^{-},p_3^{-},p_4^{+}) &= \delta_{i_1 i_2} C_W \left[C_1\left(1+m_W^2\Pi_{W^*}\right)+\frac{2m_W^2\Pi_{W^*}}{v}S_1\right] \mathcal{A}_1(p_1^{+},p_2^{-},p_3^{-},p_4^{+}) \,, \\
    \mathcal{M}(p_1^{-},p_2^{+},p_3^{-},p_4^{+}) &= \delta_{i_1 i_2} C_W C_2\left(1+m_W^2\Pi_{W^*}\right) \mathcal{A}_1(p_1^{-},p_2^{+},p_3^{-},p_4^{+}) \,, \\
    \mathcal{M}(p_1^{-},p_2^{-},p_3^{-},p_4^{+}) &= \delta_{i_1 i_2} C_W C_3 \mathcal{A}_2(p_1^{-},p_2^{-},p_3^{-},p_4^{+}) \,, \\
    \mathcal{M}(p_1^{+},p_2^{+},p_3^{-},p_4^{+}) &= \delta_{i_1 i_2} C_W C_4 \mathcal{A}_2(p_1^{+},p_2^{+},p_3^{-},p_4^{+}) \,,
\end{split}
\label{eq:bare_hel_amp}
\enq
where $C_W = -\rmi e/(\sqrt{2}\sin{\theta_\rmW})$ is the strength of the $W$ boson coupling to the $e^+\nu_e$ pair, $S_1$ and $C_i$ have been defined in Sec.~\ref{sec:SMEFT}, and $i_{1,2}$ are color indices of the anti-quark and quark.

The form factors that we obtained contain both ultraviolet (UV) and infrared (IR) divergences, appearing as poles in the dimensional regularization parameter $\epsilon$. 
For the vector current form factor, only the strong coupling constant needs to be renormalized.  The relation between bare ($\alpha_{\rms,0}$) and renormalized  ($\as$) strong coupling constants  reads (see e.g.~\cite{Catani:1998bh})
\bnq \alpha_{\rms,0} \mu_0^{2\epsilon}
    S_{\epsilon} = \as \mu^{2\epsilon} \left[ 1
    -\frac{\beta_0}{\epsilon}\frac{\as}{2\pi} +\mathcal{O}(\as^2)\right], 
\enq 
where $S_{\epsilon} = (4\pi)^{\epsilon} e^{-\epsilon
  \gamma_{\text{E}}}$,  
\begin{align}
    \beta_0 &= \frac{11}{6}\CA - \frac{2}{3}\nf \TR \,,
\end{align}
with $\CA=3$, $\TR = 1/2$, and $\nf$ denoting the number of light quark flavors.

In contrast to the case of the vector current, the tensor current $\sigma_{\mu \nu} p_W^\nu$ is not conserved and, therefore,  requires UV renormalization. We perform the renormalization in the  $\overline {\rm MS}$-scheme. The corresponding renormalization constant can be extracted from Ref.~\cite{Alioli:2018ljm} and reads
\bnq
\begin{split}
    Z_{\mathcal{H}_2} = &\; 
    1
    + 
    \frac{\CF}{2 \epsilon}\frac{\as}{2\pi}
    +  \left(\frac{\as}{2\pi}\right)^2 \CF
    \bigg [
    \CA \left(-\frac{11}{24 \epsilon^2}+\frac{257}{144 \epsilon}\right)
    \\
    &   + \CF \left(\frac{1}{8 \epsilon^2}-\frac{19}{16 \epsilon}\right)
        + \nf \TR \left(\frac{1}{6 \epsilon^2}-\frac{13}{36 \epsilon}\right)
    \bigg]
    +
    \mathcal{O}(\as^3) \,,
    \, 
\end{split}
\enq 
with $\CF=4/3$.  Once the UV renormalization is performed, we find that
the remaining infrared pole structure of all amplitudes computed in this paper
agrees with Catani's universal formula~\cite{Catani:1998bh}.  Additional
details can be found in Appendix~\ref{sec:app2}.

As a check, we also evaluated the one- and two-loop amplitudes
using an approach which generalizes the Passarino-Veltman reduction~\cite{Passarino:1978jh}. At two-loops, we used a combination of the 
Passarino-Veltman tensorial reduction and integration-by-parts identities to
express tensor integrals through master integrals.  In this way, the
decomposition of  amplitudes into invariant form factors shown in
Eqs.~\eqref{PV:H1}, \eqref{eq19}, and \eqref{eq20} appears as a natural result of the calculation.

A further consistency check can be performed by realizing that the Feynman rules for the three-point and the four-point SMEFT vertices become identical in the (formal) soft-Higgs limit,  $p_H\to0$ (up to the coupling constants). We have explicitly checked this limit of the computed amplitudes, finding agreement among the different expressions. 

Finally, we have checked the Born and the renormalized one-loop amplitudes numerically for multiple phase-space points against 
\texttt{OpenLoops}~\cite{Buccioni:2019sur},
\texttt{MadGraph}~\cite{Alwall:2014hca}, and
\texttt{GoSam}~\cite{Cullen:2011ac,GoSam:2014iqq}, finding perfect agreement. Results for renormalized helicity amplitudes computed in this paper can be found in the ancillary files provided with this submission.

\subsection{One-loop amplitudes for   $0 \to \overline{u}d\nu_{e}e^+H+g$}

The second ingredient required for the NNLO calculation is the one-loop amplitude for
the process $0 \to \overline{u}d\nu_{e}e^+H+g(p_5)$.  Similar to the
two-loop case, we need to consider two currents ${\cal H}_1^{\mu}$ and
${\cal H}_2^{\mu}$ and select the appropriate helicity configurations to
discuss their left- and right-handed components.  The amplitudes $0 \to \overline{u}d\nu_{e}e^+H+g$ for the vector current can be extracted from
Ref.~\cite{Garland:2002ak}, whereas the matrix element of the current with
$\sigma^{\mu \nu} p_{W,\nu}$ is unknown.

As discussed in Ref.\,\cite{Garland:2002ak}, the matrix element of the current
$\mathcal{H}_1^{\mu}$ can be decomposed into seven independent Lorentz
structures. The calculation of the matrix element of
$\mathcal{H}_2^{\mu}$ can be performed following similar steps, but it is much
more complicated. In fact, after imposing two sets of conditions
\bnq 
\begin{split}
\left.\mathcal{H}_{2, \text{4pt}}^{\mu}\right|_{\epsilon_5^* \rightarrow p_5} &= 0 \,,
\qquad \mathcal{H}_{2, \text{4pt}} \cdot p_{W} = 0 \,, \\
\left.\mathcal{H}_{2, \text{3pt}}^{\mu}\right|_{\epsilon_5^* \rightarrow p_5} &= 0 \,,
\qquad \mathcal{H}_{2, \text{3pt}} \cdot \left(p_{1}+p_{2}+p_{5}\right) = 0 \,,
\end{split}
\label{eq29}
\enq 
where $\epsilon_5$ is the polarization vector of the radiated gluon, the matrix element of $\mathcal{H}_2^{\mu}$ can be written as a linear
combination of tensor structures
\begin{equation}
    \calH_2^\mu 
    = 
    \sum \limits_{i=1}^{i_{\rm max}} \mathcal{F}^{(g)}_{2,i} \, {\cal T}_i^\mu \,,
\end{equation}
where $i_{\rm max}=7 \, (16)$ for the three-point (four-point) SMEFT vertex.
For the three-point vertex, they read
\begin{align}
\mathcal{T}_{1}^{\mu} & = 
\left[\left(s_{12}+s_{15}\right) p_5^{\mu }-\left(s_{15}+s_{25}\right) p_1^{\mu }\right] \UbarSp{p_2}\slashed{p}_5\slashed{\epsilon}^*_5\VSp{p_1} \,, \nonumber\\
\mathcal{T}_{2}^{\mu} & = 
\left[\left(s_{12}+s_{25}\right) p_5^{\mu }-\left(s_{15}+s_{25}\right) p_2^{\mu }\right] \UbarSp{p_2}\slashed{p}_5\slashed{\epsilon}^*_5\VSp{p_1} \,, \nonumber\\
\mathcal{T}_{3}^{\mu} & = 
\frac{2 \left(p_1\cdot \epsilon^*_5\right) \UbarSp{p_2}\gamma ^{\mu }\slashed{p}_5\VSp{p_1}}{s_{15}}+\frac{2 p_5^{\mu } \UbarSp{p_2}\slashed{p}_5\slashed{\epsilon}^*_5\VSp{p_1}}{s_{15}+s_{25}}-\UbarSp{p_2}\gamma ^{\mu }\slashed{\epsilon}^*_5\VSp{p_1} \,, \nonumber\\
\mathcal{T}_{4}^{\mu} & = 
\frac{\left(s_{12}+s_{15}\right) \left(s_{25} \left(p_1\cdot \epsilon^*_5\right)-s_{15} \left(p_2\cdot \epsilon^*_5\right)\right) \UbarSp{p_2}\gamma ^{\mu }\slashed{p}_5\VSp{p_1}}{2 s_{15}} \nonumber\\ 
  &\qquad \qquad -p_1^{\mu } \UbarSp{p_2}\VSp{p_1} \left(s_{25} \left(p_1\cdot \epsilon^*_5\right)-s_{15} \left(p_2\cdot \epsilon^*_5\right)\right) \,, \nonumber\\
\mathcal{T}_{5}^{\mu} & = 
\frac{\left(s_{12}+s_{25}\right) \left(s_{25} \left(p_1\cdot \epsilon^*_5\right)-s_{15} \left(p_2\cdot \epsilon^*_5\right)\right) \UbarSp{p_2}\gamma ^{\mu }\slashed{p}_5\VSp{p_1}}{2 s_{15}} \label{eq:tensor_VR_3p}\\ 
   &\qquad \qquad -p_2^{\mu } \UbarSp{p_2}\VSp{p_1} \left(s_{25} \left(p_1\cdot \epsilon^*_5\right)-s_{15} \left(p_2\cdot \epsilon^*_5\right)\right) \,, \nonumber\\
\mathcal{T}_{6}^{\mu} & = 
\frac{\left(s_{15}+s_{25}\right) \left(s_{15} \left(p_2\cdot \epsilon^*_5\right)-s_{25} \left(p_1\cdot \epsilon^*_5\right)\right) \UbarSp{p_2}\gamma ^{\mu }\slashed{p}_5\VSp{p_1}}{2 s_{15}} \nonumber\\
    &\qquad \qquad +p_5^{\mu } \UbarSp{p_2}\VSp{p_1} \left(s_{25} \left(p_1\cdot \epsilon^*_5\right)-s_{15} \left(p_2\cdot \epsilon^*_5\right)\right) \,, \nonumber\\
\mathcal{T}_{7}^{\mu} & = 
\frac{2 \left(p_2\cdot \epsilon^*_5\right) \UbarSp{p_2}\gamma ^{\mu }\slashed{p}_5\VSp{p_1}}{s_{25}}+\frac{2 p_5^{\mu } \UbarSp{p_2}\slashed{p}_5\slashed{\epsilon}^*_5\VSp{p_1}}{s_{15}+s_{25}} \nonumber\\
   &\qquad \qquad -\frac{4 p_5^{\mu } \UbarSp{p_2}\VSp{p_1} \left(p_2\cdot \epsilon^*_5\right)}{s_{25}}+\UbarSp{p_2}\slashed{\epsilon}^*_5\gamma ^{\mu }\VSp{p_1}\,,  \nonumber
\end{align}
whereas for the four-point vertex we find
\begin{align}
\mathcal{T}_{1}^{\mu} & = \UbarSp{p_2}[\gamma^{\mu}, \slashed{p}_W]\slashed{p}_5\slashed{\epsilon}^*_5\VSp{p_1} ,\nonumber\\ 
\mathcal{T}_{2}^{\mu} & = \UbarSp{p_2}\slashed{\epsilon}^*_5\slashed{p}_5[\gamma^{\mu}, \slashed{p}_W]\VSp{p_1} ,\nonumber\\
\mathcal{T}_{3}^{\mu} & = \left(s_{1W} p_5^{\mu }-s_{5W} p_1^{\mu }\right) \UbarSp{p_2}\slashed{p}_5\slashed{\epsilon}^*_5\VSp{p_1} ,\nonumber\\
\mathcal{T}_{4}^{\mu} & = \left(s_{2W} p_5^{\mu }-s_{5W} p_2^{\mu }\right) \UbarSp{p_2}\slashed{p}_5\slashed{\epsilon}^*_5\VSp{p_1} ,\nonumber\\
\mathcal{T}_{5}^{\mu} & = \UbarSp{p_2}[\gamma^{\mu}, \slashed{p}_W]\VSp{p_1} \left[s_{25} \left(\epsilon_5^*\cdot  p_1\right)-s_{15} \left(\epsilon_5^*\cdot  p_2\right)\right] ,\nonumber\\
\mathcal{T}_{6}^{\mu} & = 
\Big[p_1^{\mu } \UbarSp{p_2}\slashed{p}_W\slashed{p}_5\VSp{p_1} -\frac{1}{2} s_{1W} \UbarSp{p_2}\gamma ^{\mu }\slashed{p}_5\VSp{p_1}\Big] 
\left[s_{25} \left(\epsilon_5^*\cdot  p_1\right)-s_{15} \left(\epsilon_5^*\cdot  p_2\right)\right] ,\nonumber\\
\mathcal{T}_{7}^{\mu} & = 
\Big[p_2^{\mu } \UbarSp{p_2}\slashed{p}_W\slashed{p}_5\VSp{p_1} -\frac{1}{2} s_{2W} \UbarSp{p_2}\gamma ^{\mu }\slashed{p}_5\VSp{p_1}\Big] 
\left[s_{25} \left(\epsilon_5^*\cdot  p_1\right)-s_{15} \left(\epsilon_5^*\cdot  p_2\right)\right] ,\nonumber\\
\mathcal{T}_{8}^{\mu} & = 
\Big[p_5^{\mu } \UbarSp{p_2}\slashed{p}_W\slashed{p}_5\VSp{p_1} -\frac{1}{2} s_{5W} \UbarSp{p_2}\gamma ^{\mu }\slashed{p}_5\VSp{p_1}\Big] 
\left[s_{25} \left(\epsilon_5^*\cdot  p_1\right)-s_{15} \left(\epsilon_5^*\cdot  p_2\right)\right] ,\nonumber\\
\mathcal{T}_{9}^{\mu} & = 
\frac{1}{2} s_{25} p_1^{\mu } \UbarSp{p_2}\slashed{p}_W\slashed{\epsilon}^*_5\VSp{p_1}+\frac{1}{2} s_{1W} \left(\epsilon_5^*\cdot  p_2\right) \UbarSp{p_2}\gamma ^{\mu }\slashed{p}_5\VSp{p_1} \nonumber\\
& \qquad \qquad -p_1^{\mu } \left(\epsilon_5^*\cdot  p_2\right) \UbarSp{p_2}\slashed{p}_W\slashed{p}_5\VSp{p_1}-\frac{1}{4} s_{25} s_{1W} \UbarSp{p_2}\gamma ^{\mu }\slashed{\epsilon}^*_5\VSp{p_1} ,\nonumber\\
\mathcal{T}_{10}^{\mu} & = 
\frac{1}{2} s_{25} p_2^{\mu } \UbarSp{p_2}\slashed{p}_W\slashed{\epsilon}^*_5\VSp{p_1}+\frac{1}{2} s_{2W} \left(\epsilon_5^*\cdot  p_2\right) \UbarSp{p_2}\gamma ^{\mu }\slashed{p}_5\VSp{p_1} \nonumber\\
& \qquad \qquad -p_2^{\mu } \left(\epsilon_5^*\cdot  p_2\right) \UbarSp{p_2}\slashed{p}_W\slashed{p}_5\VSp{p_1}-\frac{1}{4} s_{25} s_{2W} \UbarSp{p_2}\gamma ^{\mu }\slashed{\epsilon}^*_5\VSp{p_1} , \label{eq:tensor_VR_4p}\\
\mathcal{T}_{11}^{\mu} & = 
\frac{1}{2} s_{25} p_5^{\mu } \UbarSp{p_2}\slashed{p}_W\slashed{\epsilon}^*_5\VSp{p_1}+\frac{1}{2} s_{5W} \left(\epsilon_5^*\cdot  p_2\right) \UbarSp{p_2}\gamma ^{\mu }\slashed{p}_5\VSp{p_1} \nonumber\\
& \qquad \qquad -p_5^{\mu } \left(\epsilon_5^*\cdot  p_2\right) \UbarSp{p_2}\slashed{p}_W\slashed{p}_5\VSp{p_1}-\frac{1}{4} s_{25} s_{5W} \UbarSp{p_2}\gamma ^{\mu }\slashed{\epsilon}^*_5\VSp{p_1} ,\nonumber\\
\mathcal{T}_{12}^{\mu} & = 
p_5^{\mu } \UbarSp{p_2}\VSp{p_1} \left(\epsilon_5^*\cdot  p_W\right)-\frac{1}{4} s_{5W} \UbarSp{p_2}\gamma ^{\mu }\slashed{\epsilon}^*_5\VSp{p_1}-\frac{1}{4} s_{5W} \UbarSp{p_2}\slashed{\epsilon}^*_5\gamma ^{\mu }\VSp{p_1} ,\nonumber\\
\mathcal{T}_{13}^{\mu} & = 
\UbarSp{p_2}\VSp{p_1} \left(p_1^{\mu } \left(s_{15} \left(\epsilon_5^*\cdot  p_W\right)-s_{5W} \left(\epsilon_5^*\cdot  p_1\right)\right)+s_{1W} p_5^{\mu } \left(\epsilon_5^*\cdot  p_1\right)\right) \nonumber\\
& \qquad \qquad -\frac{1}{4} s_{15} s_{1W} \UbarSp{p_2}\gamma ^{\mu }\slashed{\epsilon}^*_5\VSp{p_1}-\frac{1}{4} s_{15} s_{1W} \UbarSp{p_2}\slashed{\epsilon}^*_5\gamma ^{\mu }\VSp{p_1} ,\nonumber\\
\mathcal{T}_{14}^{\mu} & = 
\UbarSp{p_2}\VSp{p_1} \left(p_2^{\mu } \left(s_{15} \left(\epsilon_5^*\cdot  p_W\right)-s_{5W} \left(\epsilon_5^*\cdot  p_1\right)\right)+s_{2W} p_5^{\mu } \left(\epsilon_5^*\cdot  p_1\right)\right) \nonumber\\ 
& \qquad \qquad  -\frac{1}{4} s_{15} s_{2W} \UbarSp{p_2}\gamma ^{\mu }\slashed{\epsilon}^*_5\VSp{p_1}-\frac{1}{4} s_{15} s_{2W} \UbarSp{p_2}\slashed{\epsilon}^*_5\gamma ^{\mu }\VSp{p_1} ,\nonumber\\
\mathcal{T}_{15}^{\mu} & = 
\UbarSp{p_2}\VSp{p_1} \left(p_1^{\mu } \left(s_{25} \left(\epsilon_5^*\cdot  p_W\right)-s_{5W} \left(\epsilon_5^*\cdot  p_2\right)\right)+s_{1W} p_5^{\mu } \left(\epsilon_5^*\cdot  p_2\right)\right) \nonumber\\ 
& \qquad \qquad  -\frac{1}{4} s_{25} s_{1W} \UbarSp{p_2}\gamma ^{\mu }\slashed{\epsilon}^*_5\VSp{p_1}-\frac{1}{4} s_{25} s_{1W} \UbarSp{p_2}\slashed{\epsilon}^*_5\gamma ^{\mu }\VSp{p_1} ,\nonumber\\
\mathcal{T}_{16}^{\mu} & = 
\UbarSp{p_2}\VSp{p_1} \left(p_2^{\mu } \left(s_{25} \left(\epsilon_5^*\cdot  p_W\right)-s_{5W} \left(\epsilon_5^*\cdot  p_2\right)\right)+s_{2W} p_5^{\mu } \left(\epsilon_5^*\cdot  p_2\right)\right) \nonumber\\ 
& \qquad \qquad  -\frac{1}{4} s_{25} s_{2W} \UbarSp{p_2}\gamma ^{\mu }\slashed{\epsilon}^*_5\VSp{p_1}-\frac{1}{4} s_{25} s_{2W} \UbarSp{p_2}\slashed{\epsilon}^*_5\gamma ^{\mu }\VSp{p_1}
\,.
\nonumber
\end{align}
In the above, we have defined $s_{ij} = 2 p_i\cdot p_j$. 
We note that the reference vector for the gluon polarization 
 has not been fixed in the above expressions. 

We construct projection operators using the tensors in Eqs.~\eqref{eq:tensor_VR_3p}, \eqref{eq:tensor_VR_4p} to compute the form factors $\mathcal{F}_{2,i}^{(g)}$.  Explicit expressions for the projection operators can be found in the ancillary files provided with this submission.\footnote{
We note that their construction requires the inversion of a very complicated and
dense matrix, which turns out to be computationally expensive using standard
\texttt{Mathematica} routines such as \texttt{Inverse}. We rely instead on the
\texttt{FFInverse} function provided by the package
\texttt{FiniteFlow}~\cite{Peraro:2019svx}.} 
Once the amplitudes are projected onto the individual form factors, we use
integration-by-parts identities to reduce them to master integrals.  We employ the analytic
expressions for master integrals computed in Refs.~\cite{Gehrmann:2000zt,Gehrmann:2001ck,Garland:2002ak}. 
Infrared divergencies in the amplitudes are described by Catani's formula~\cite{Catani:1998bh} which also serves as a useful check of the calculation. 

Finally, we note that we computed the bare amplitudes using the 
Passarino-Veltman reduction as a cross-check, finding agreement with the result obtained using
projection operators.  We also compared  the final result for the amplitude with 
\texttt{OpenLoops}~\cite{Buccioni:2019sur},
\texttt{MadGraph}~\cite{Alwall:2014hca}, and
\texttt{GoSam}~\cite{Cullen:2011ac, GoSam:2014iqq} and found perfect agreement. 
The renormalized helicity amplitudes for the process $0 \to \overline{u}d\nu_{e}e^+H+g$ are provided in the ancillary files.

\subsection{Double-real emissions}

The last ingredient for the NNLO computation comprises amplitudes for two real-emission processes 
$0 \to \overline{u}d\nu_{e}e^+H+gg$ 
and $0 \to \overline{u}d\nu_{e}e^+H+ q \overline q$.  The calculation relies
on the spinor-helicity methods and is completely straightforward.  Amplitudes for
the vector current $\mathcal{H}_1^{\mu}$ can be extracted from
Refs.~\cite{Brein:2012ne, Boughezal:2016wmq}; amplitudes for the $\sigma^{\mu \nu}
p_{W,\nu}$ current are new and are provided in the ancillary files.

\subsection{Combining amplitudes}
Ultraviolet-renormalized loop amplitudes contain $1/\epsilon$ poles
arising from soft and collinear virtual partons. Similarly, the real-emission amplitudes develop singularities when the emitted partons become soft or collinear to other partons. This implies that the real-emission amplitudes cannot be integrated over the full radiative phase space in four space-time dimensions. Removing the singularities arising from the real-emission corrections and combining them with those from the virtual corrections to arrive at an infrared-finite final result requires a subtraction scheme. For the calculations described in this paper, we employ the nested soft-collinear subtraction scheme~\cite{Caola:2017dug, Caola:2019nzf}, which has previously been used in several studies of $VH$ phenomenology~\cite{Caola:2017xuq, Behring:2020uzq, Bizon:2021rww}. 
We note that since the infrared and collinear limits of the real-emission amplitudes are universal, upgrading the calculations reported in Refs.~\cite{Caola:2017xuq, Behring:2020uzq, Bizon:2021rww} to include the EFT contributions is straightforward since only hard, four-dimensional amplitudes have to be provided anew. Furthermore, the numerical code used in Refs.~\cite{Caola:2017xuq, Behring:2020uzq, Bizon:2021rww} and in other similar computations~\cite{Brein:2012ne, Boughezal:2016wmq} is modular, allowing the implementation of the amplitudes discussed above in a straightforward way. To check the implementation, we confirmed that the subtraction terms have the same behavior as the amplitudes in the various unresolved limits and also that the SM results for $W^+H$ production are reproduced~\cite{Behring:2020uzq,Bizon:2021rww}.

\section{Phenomenological results}
\label{sec:pheno}

In this section, we present phenomenological results through NNLO in perturbative QCD. 
We consider the process $pp \to W^+(\to e^+ \nu_e)\; H$ and remind the reader that, for the sake of simplicity,  the  Higgs is treated as a stable particle. 
We begin by listing the physical parameters and selection criteria used in the calculation.

We set the Higgs boson mass to $m_H = 125$~GeV and the $W$ boson mass to $m_W = 80.399$~GeV.  
We use the Fermi constant $\GF = 1.16639 \times  10^{-5} ~ {\rm GeV}^{-2}$, and the sine squared of the weak mixing angle
$\sin^2 \theta_\rmW = 1-m_W^2/m_Z^2$ with $m_Z=91.1876$~GeV. The width of the $W$ boson is taken to be $\Gamma_W = 2.1054$~GeV. 
Furthermore,  we take the SMEFT scale $\Lambda$ to be ${\rm 1}~{\rm TeV}$. 
Finally, we treat the decay process $W^+ \to \nu e^+$ 
in the narrow-width approximation. Using the above parameters, the branching ratio evaluates to ${\rm Br}(W \to \nu_e e^+) = 0.10801$. 

To present numerical results we employ the  parton distribution function (PDF) set
{\tt NNPDF31${}_{-}$nnlo${}_{-}$as${}_{-}$0118${}$}.   We use NNLO PDFs to compute LO, NLO and NNLO cross sections in what follows. The values of the strong coupling 
constant $\as$ are obtained from the NNPDF routines. 
We choose $\mu = \muR = \muF = \sqrt{(p_{W^+} +p_H)^2}/2$ as central renormalization/factorization scale for the production process, and estimate uncertainties related to higher-order QCD
effects by varying the scale by a factor of two in either direction. 

In order to keep our analysis as broad as possible, we apply very loose kinematic cuts to define the fiducial phase space. 
We require that the transverse momentum of the positron originating from the $W^+$ decay is $\pTeplus > 25~\mathrm{GeV}$, and its rapidity satisfies $|\eta_{e^+}| < 2.5$. In order to ensure the  kinematic regime in which SMEFT is applicable, we further require that the $W^+$-boson transverse momentum satisfies $\pTWplus^{\rm max} < 250~\mathrm{GeV}$.
This cut does not completely exclude  events with high invariant mass $m_{\WH}$. 
However, at leading order for fixed high $m_{\WH}$, the transverse momentum of the $W$ boson is constrained to be $\pTWplus \lesssim  m_{\WH}/2$. 
Hence, for $m_{\WH} > 542~{\rm GeV}$,  the fraction of $W$-bosons that pass the cut decreases with the increase of $m_{\WH}$.
Together with the decreasing parton flux at high Bjorken $x$, this leads to a  significant suppression of high invariant mass contributions and ensures the applicability of SMEFT for our analyses. 
Furthermore, in $WH$ studies, an additional cut $\pTWplus > 150~\mathrm{GeV}$ is sometimes imposed to emphasize the boosted Higgs contributions. In what follows we present results for fiducial cross sections with and without this cut.

The modified \sm\ Lagrangian considered in this paper contains four
dimension-six operators, see Eqs.\,(\ref{eq:operators}), (\ref{eq4}).
As discussed in the previous section, these operators lead to scattering amplitudes characterized by different helicity selection rules for the external quarks. 
Only one of these operators, $Q_{\phi q}^{(3)}$, leads to an amplitude that interferes with the \sm\ one, whereas  the contributions of all other operators  interfere neither with the SM, nor with each other. 
This feature remains true through NNLO in QCD. Consequently, in the context of the SM extension  discussed in this paper, one can write the most general differential cross section as the sum of six terms weighted with the corresponding coefficients:
\begin{equation}  
  \rmd \sigma = \rmd \sigsm + \cuW^2\,  \rmd \siguW + \cdW^2 \,  \rmd \sigdW
  + \cHqthree \,  \rmd \sigI + \big(\cHqthree\big)^2  \, \rmd \sigHqthree
  +\cHud^2 \, \rmd \sigHud \,.
\label{eq:xsecs_coeffs}
\end{equation}
We obtain the cross sections\footnote{It is not quite proper to call these six contributions ``cross sections'' since one of them is the interference and, therefore, does not have to be  positive-definite, but we will do this nonetheless.
} $\rmd \siguW$, $\rmd \sigdW$, $\rmd \sigHqthree$ and $\rmd \sigHud$ by setting the associated coupling coefficient to one and all other contributions, including that of the SM, to zero. To obtain the interference term $\rmd \sigma_I$, we take the difference between the calculation that yields $\rmd \sigHqthree$ and a similar calculation in which the SM coupling is not removed.

\begin{table}[!t] 
    \centering
    \caption{Fiducial cross sections (in fb) at LO, NLO and NNLO for $pp \to  W^+(\to \nu_e e^+)H$ production at the 13 TeV LHC. The subscripts indicate the maximum and minimum values obtained by varying the factorization/renormalization by a factor of two in either direction. The fifth and sixth columns show the NLO and NNLO $k$-factors. Statistical errors are below  the displayed  digit, and are not shown. 
    The table below is derived with an additional cut $p_{\rmT,W^+} > 150~\mathrm{GeV}$. See text for details.}
    \vspace{3mm}
    \renewcommand{\arraystretch}{1.7} 
    
    \begin{tabular}{
    >{\centering\arraybackslash}m{20mm} | 
    >{\centering\arraybackslash}m{23mm}
    >{\centering\arraybackslash}m{23mm} 
    >{\centering\arraybackslash}m{23mm} |
    >{\centering\arraybackslash}m{23mm}
    >{\centering\arraybackslash}m{23mm}
    }
        \hline
        \multicolumn{6}{c}{\textbf{Fiducial cross sections at LO, NLO and NNLO}} \\
        \hline\hline
            \textit{Case}    & $\sigma_\LO$    & $\sigma_\NLO$     & $\sigma_\NNLO$    & $k_\NLO$     & $k_\NNLO$ \\
            \hline 
            SM   & $ 47.84^{ +1.65 }_{ -2.10 } $& $ 58.12^{ +0.59 }_{ -0.26 }  $ & $ 58.61^{ +0.08 }_{ -0.06 }  $ & 1.215 & 1.008  \\
            $uW$ & $ 59.35^{ +0.56 }_{ -0.93 }  $  & $ 79.98^{ +4.14 }_{ -3.31 }  $ & $ 83.14^{ +3.85 }_{ -3.22 }  $ & 1.347 & 1.040 \\
            $dW$ & $ 38.98^{ +0.36 }_{ -0.61 }  $ & $ 52.96^{ +2.78 }_{ -2.23 } $  & $ 55.17^{ +2.58 }_{ -2.17 }  $ & 1.358 & 1.042  \\
            $\phi q^{(3)}$ & $ 58.65^{ +0.59 }_{ -0.97 }  $ & $ 69.74^{ +0.96 }_{ -0.61 }  $ & $ 70.12^{ +0.08 }_{ -0.03 }  $ & 1.189 & 1.006  \\
            $I$ & $ 88.47^{ +2.10 }_{ -2.80 }  $ &  $ 106.1^{ +1.21 }_{ -0.64 }  $ & $ 106.8^{ +0.25 }_{ -0.11 }  $ & 1.200 & 1.007  \\
            $\phi ud$ & $ 16.13^{ +0.16 }_{ -0.27 }  $ &  $ 19.11^{ +0.24 }_{ -0.15 } $&  $ 19.20^{ +0.01 }_{ -0.01 }  $ & 1.185 & 1.005  \\
            \hline
        \multicolumn{6}{c}{\vspace{3mm}} \\ 
        \hline
        \multicolumn{6}{c}{\textbf{Fiducial cross sections at LO, NLO and NNLO ($\boldsymbol{ p_{\rmT,W^+} > 150~\mathrm{GeV}}$)}} \\
        \hline\hline
            \textit{Case}    & $\sigma_\LO$    & $\sigma_\NLO$     & $\sigma_\NNLO$    & $k_\NLO$     & $k_\NNLO$ \\
            \hline 
            SM & $ 6.357^{ +0.021 }_{ -0.057 }  $  & $ 8.121^{ +0.191 }_{ -0.139 }  $  & $ 8.314^{ +0.040 }_{ -0.050 }  $ & 1.278 & 1.024  \\
            $uW$ & $ 30.46^{ +0.02 }_{ -0.15 } $ & $ 40.82^{ +2.26 }_{ -1.81}  $ & $ 42.39^{ +1.92 }_{ -1.62 }  $ & 1.340 & 1.039  \\
            $dW$ & $ 20.65^{ +0.02}_{ -0.11 }  $ & $ 27.83^{ +1.57 }_{ -1.25 }  $ & $ 28.94^{ +1.33 }_{ -1.12}  $ & 1.348 & 1.040  \\
            $\phi q^{(3)}$ & $ 30.54^{ +0.00 }_{ -0.13 }  $ & $ 36.03^{ +0.58 }_{ -0.41 }  $ & $ 36.23^{ +0.04 }_{ -0.05 }  $ & 1.180 & 1.006  \\
            $I$ & $ 26.22^{ +0.04 }_{ -0.14 }  $ & $ 31.61^{ +0.54 }_{ -0.39 }  $ & $ 31.89^{ +0.10 }_{ -0.14 }  $ & 1.204 & 1.010 \\
            $\phi ud$ & $ 8.065^{ +0.004 }_{ -0.040 }  $ &  $ 9.501^{ +0.147 }_{ -0.106 }  $ & $ 9.552^{ +0.003 }_{ -0.014 }  $ & 1.178 & 1.005 \\
            \hline
    \end{tabular}
    \label{tab:xsecs2}
\end{table}

We begin by discussing 
the different contributions to Eq.~\eqref{eq:xsecs_coeffs}.
In Table~\ref{tab:xsecs2}, we present the fiducial cross sections for the quantities defined there 
at LO, NLO and NNLO, without and with the cut $\pTWplus > 150~\mathrm{GeV}$.
A striking feature of these results is that the cross sections 
associated with SMEFT operators are large: they are comparable to the SM cross sections
for basic fiducial cuts, but start exceeding it if,  in addition, the $\pTWplus > 150~{\rm GeV}$ cut is applied.  We note that this cut effectively increases the invariant mass of the 
$\WH$ final states that contribute to the corresponding cross sections. 

\begin{figure}[pt]
    \centering
	\subfloat[]{{\includegraphics[width=0.8\linewidth]{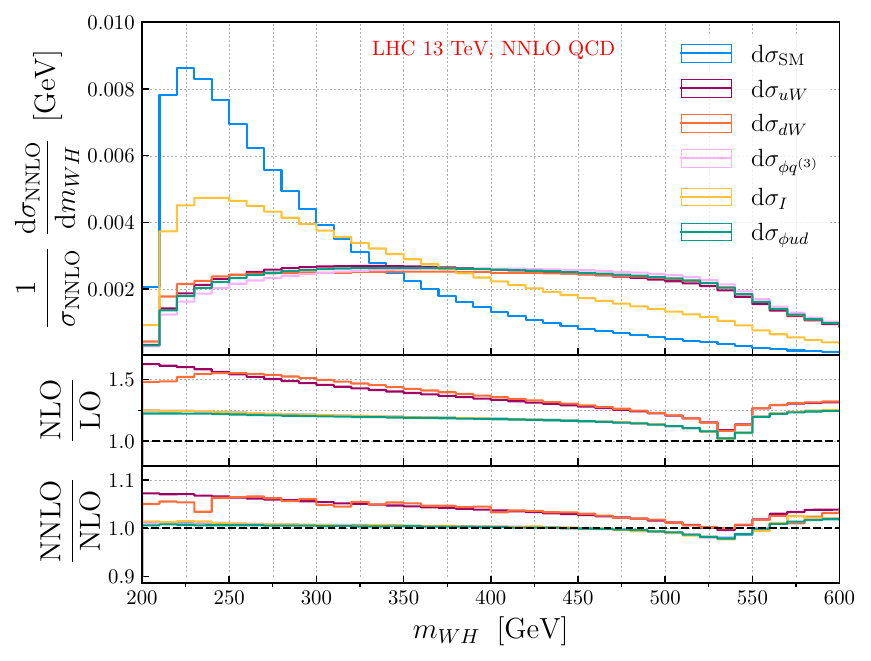}}}
    \hspace{0.05\textwidth}
    \caption{
    Normalized distribution of the invariant mass of the $W^+H$ system in $pp \to  W^+(\to \nu_e e^+) H$ at the 13 TeV LHC, for the six differential cross sections defined in Eq.~\eqref{eq:xsecs_coeffs}. The upper pane shows the results at NNLO in QCD. 
    The middle and lower panes show the NLO/LO and NNLO/NLO differential $k$-factors, respectively. Only the curves for the central factorization/renormalization scale are shown. See text for further details.}
    \label{fig:mhv}
\end{figure}

It is to be expected that cross sections driven by SMEFT operators contribute 
more at higher invariant masses because of their high mass dimension. However, in the current case there is another mechanism that enhances their contributions, namely the 
fact that the SM amplitude contains an intermediate $W^+$ boson propagator in the $s$-channel, whereas SMEFT 
contributions lead to a \emph{direct} transition from a $q \bar q$ initial state to the $\WH$ final state. Since the $W^+$ boson propagator decreases when 
the invariant mass of the $\WH$ system increases, the \sm\ contribution gets suppressed rapidly in comparison 
to the contribution of SMEFT operators.  This point is clearly seen 
in the normalized $\WH$ invariant mass distribution shown in Fig.~\ref{fig:mhv} where the very 
different shapes of the SM and SMEFT contributions are apparent. 

To understand this better, we can  estimate the dependence of the 
relevant amplitudes on  the invariant mass  of the $WH$
system.  Writing the weak coupling as $g \sim m_W/v$, we find 
\begin{equation}\label{eq:amp2_compare}
    |{\cal M}_{\rm SM}|^2 
    \sim \frac{m_W^4 m_{WH}^4}{(m_{WH}^2-m_W^2)^2 v^4} \,,
    \qquad
   |{\cal M}_{\rm SMEFT}|^2
   \sim \frac{m_{WH}^4}{\Lambda^4} \,.
\end{equation}
Hence, taking $m_{WH} \sim 300$~{\rm GeV}, we find that the SMEFT amplitude is comparable to the  SM one. 
We emphasize, however, that the energy scale $m_{WH} \sim 300~{\rm GeV}$ should be safe for the SMEFT expansion since it is well below the SMEFT energy scale $\Lambda = 1~{\rm TeV}$. 

Since the SM and the SMEFT cross sections receive contributions from rather different values of the $\WH$ invariant mass, their QCD corrections cannot be expected to be similar a priori.
In fact, as can be seen from Table~\ref{tab:xsecs2}, this is indeed not the case.  
The NLO QCD corrections 
increase the fiducial cross sections for the SM 
and a few of the SMEFT operators 
by $\order{20\%}$,  but 
lead to a stronger $\order{35\%}$ increase 
for the two CP-odd operators $\cuW$
and $\cdW$.
An additional cut $\pTWplus > 150~{\rm GeV}$ does not change the NLO corrections significantly, other than for the SM case, where they increase to $28\%$. 
The NLO QCD corrections are not captured by the scale variation, which amounts
to $1\%-5\%$ only for the six cross sections that we consider. This
underestimation of NLO corrections is a common feature of Drell-Yan like
processes, due in part to the opening of new channels at NLO and the absence
of an $\as(\muR)$ dependence at LO.

Turning now to the NNLO corrections, we observe an increase in the values $\siguW$ and $\sigdW$ by  $4\%$, largely independent of the presence of the $\pTWplus$ cut. For the other coupling scenarios and for the central renormalization and factorization scales, the NNLO corrections amount to less than a percent with the standard fiducial cuts, and between $1\%-2\%$  with the additional $\pTWplus$ cut.
These are captured by the scale uncertainties at NLO, and we note that the NLO uncertainties for $\siguW$ and $\sigdW$ are larger than for the other cross sections, reflecting their larger NNLO  corrections. At NNLO, the scale variations are around 5\% for these two quantities, but are at or below the percent level for the four other cross sections displayed 
in Table~\ref{tab:xsecs2}.

These differences  also get reflected in how QCD corrections affect the  kinematic distributions. Since we are chiefly interested in the shape differences
between distributions for different SMEFT operators, and the impact of QCD
corrections on them, we normalize all distributions to their NNLO cross
section, so that the area under each curve is unity. 
We also display results for the central scale only, in order to keep the plots as uncluttered as possible. 

The NLO and NNLO QCD corrections to the $\WH$ invariant mass distribution are shown in Fig.~\ref{fig:mhv}, which we mentioned above. 
Looking at the lower two panels of this figure, we can see that the NLO corrections for the quantities ${\rm d}\siguW$ and ${\rm d}\sigdW$ start off very large, enhancing the cross section by around 50\% for $m_{\WH} \approx 200$~GeV, and then decrease to about 10\%, before beginning to increase again for $m_{\WH} \gtrsim 540$~GeV. By contrast, the $k$-factors for the other differential cross sections are relatively flat, although they each reach a minimum at the same value $m_{\WH} \approx 540$~GeV and increase for larger values of the invariant mass. To understand the existence of this minimum, recall that our cut $p_{\rmT,W^+} <  250$~GeV suppresses  events with $m_{W^+H} \gtrsim 540$~GeV  at LO. However, such  events become more 
abundant at higher orders as the $W$ boson can recoil against radiated partons.
In general, the NNLO $k$-factor has a similar although less pronounced pattern compared to the NLO one.

\begin{figure}[pt]
\centering
    \subfloat[]  {{\includegraphics[width=0.5\linewidth]{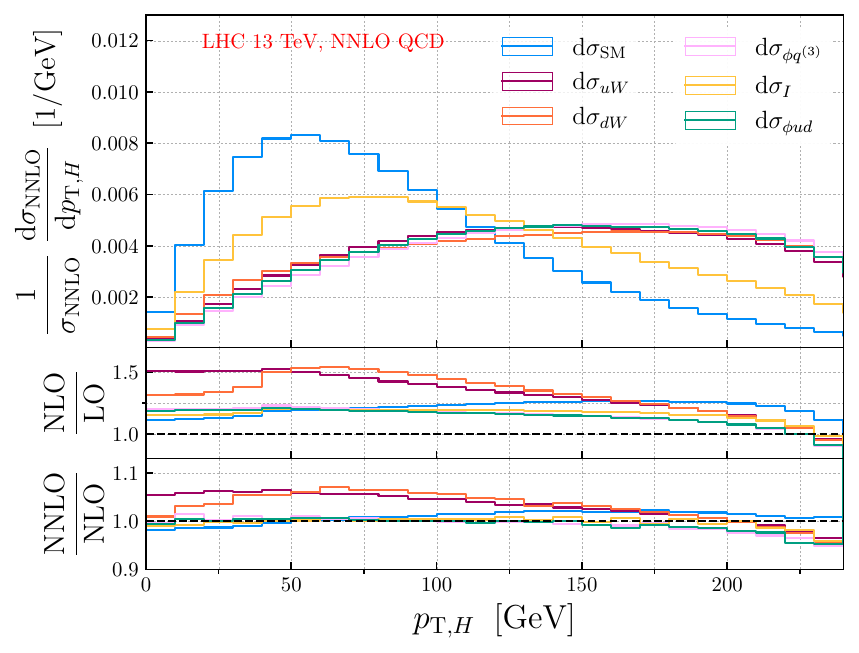}}}
    \subfloat[]{{\includegraphics[width=0.5\linewidth]{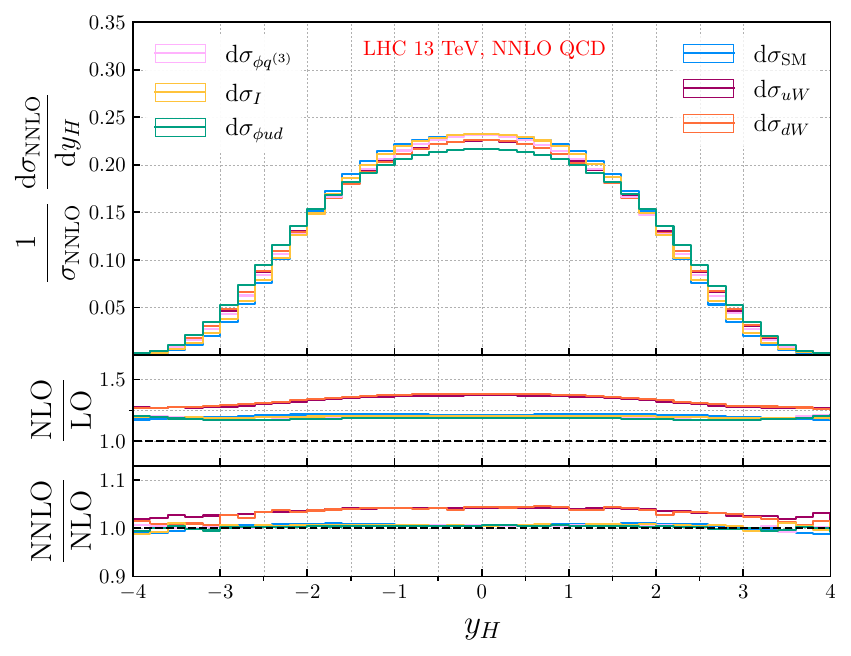}}}
    \caption{As for Fig.~\ref{fig:mhv} but displaying the normalized distributions of the transverse momentum (left) and rapidity (right) of the Higgs boson.}
    \label{fig:ptH_yH}
\end{figure} 
\begin{figure}[pt]
\centering
	\subfloat[]{{\includegraphics[width=0.5\linewidth]{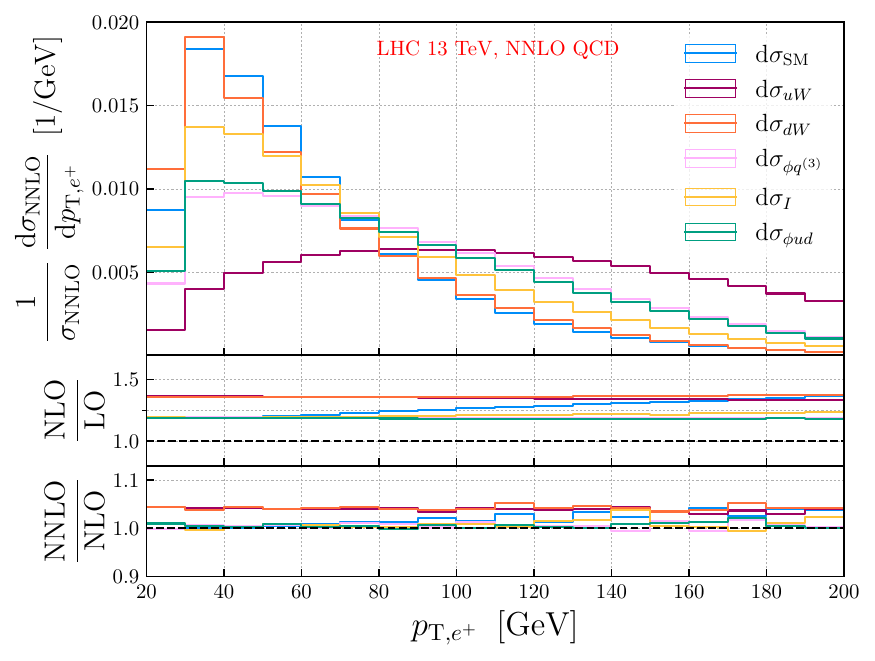}}}   
    \subfloat[]{{\includegraphics[width=0.5\linewidth]{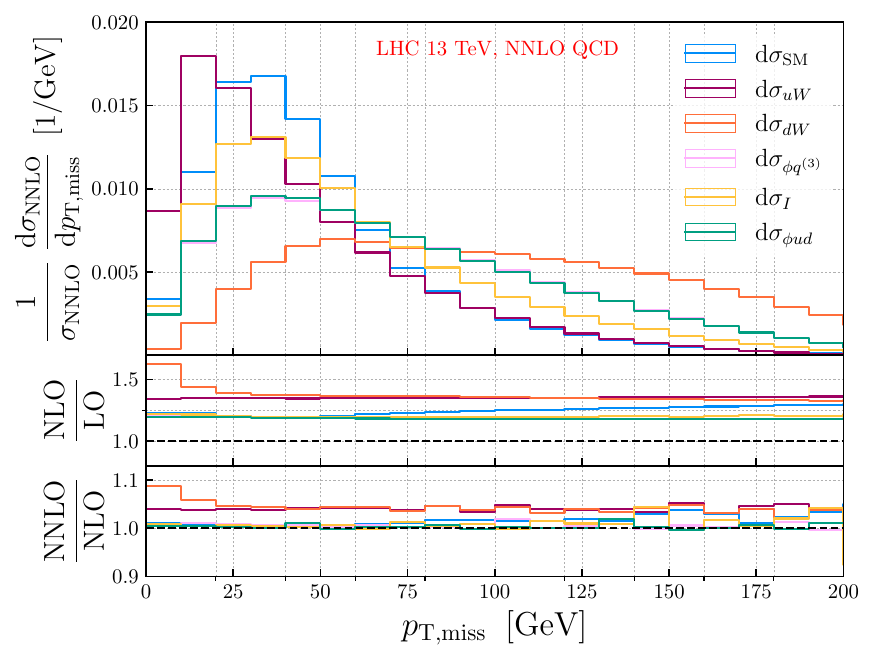}}}  
    \caption{As for Fig.~\ref{fig:mhv} but displaying the normalized
    distributions for the transverse momentum of the positron (left) and the
    missing transverse momentum (right). See main text for details.}
    \label{fig:ptl_ptmiss_yl}
\end{figure}

In Fig.~\ref{fig:ptH_yH}, we show the transverse momentum and the rapidity of the Higgs boson.  
We observe a significant shape difference in the $\pTH$ distribution between the SM and 
SMEFT results. However,  the different SMEFT operators  
lead to  very similar shapes of this distribution. On the contrary, the QCD corrections do depend on the SMEFT operator that mediates the $\WH$ production. At NLO, the corrections to $\rmd \siguW$ start off large and then decrease, becoming negative for $\pTH \gtrsim 230$~GeV. The corrections to $\rmd \sigdW$ increase at small $\pTH$, reaching a maximum value of about 50\% at $\pTH \approx 60$~GeV and then decrease. In contrast, the corrections to the SM have a very slow increase with $\pTH$ over most of the values considered. Again, these patterns persist at NNLO, although the effects are much milder there. The shape of the Higgs boson rapidity distribution is largely insensitive to 
the production mechanism, and we have checked that this is true for other rapidities, e.g.\ that of the positron. Furthermore, for all rapidity observables, the differential $k$-factors are largely constant and similar to  the $k$-factors of the fiducial cross sections shown in Table~\ref{tab:xsecs2}. 

We turn to Fig.~\ref{fig:ptl_ptmiss_yl}, where we show the transverse momentum of the positron and the missing transverse momentum. It is clear that these observables have different shapes for the different SMEFT operators,
except for $\rmd \sigHqthree$ and $\rmd \sigHud$. The $k$-factors for $\pTeplus$ are quite flat at NLO and NNLO, except for  $k_{\rm SM}$ at NLO which increases from around 1.2 at low $\pTeplus$ to almost 1.4 at $\pTeplus=200$~GeV.  The differential $k$-factors for $p_{\rmT,\rm miss}$ have a very similar behavior, at least in the region of phase space where the bulk of events lies. 

\begin{figure}[htbp]
\centering
	\subfloat[]{{\includegraphics[width=0.503\linewidth]{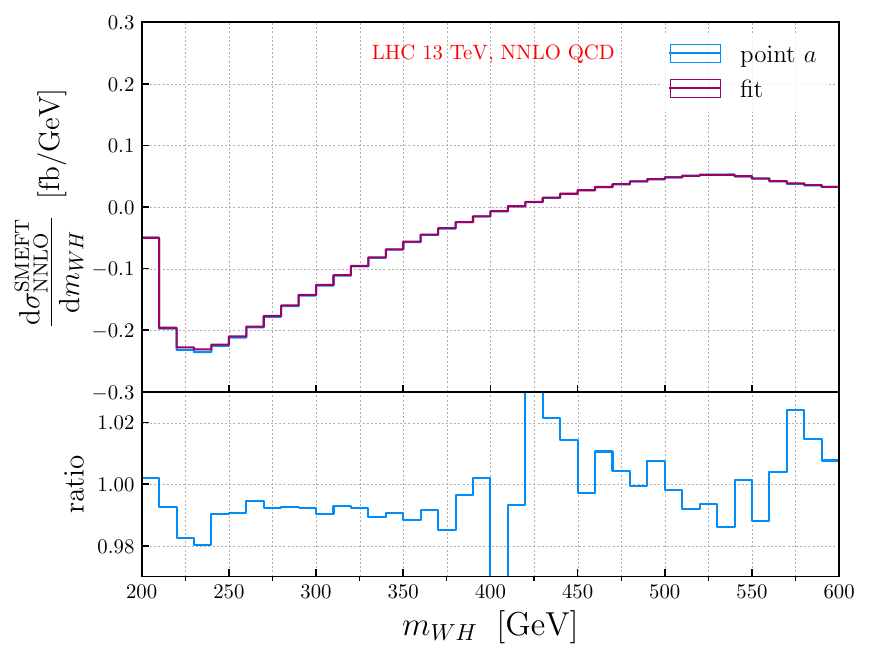}}}
    \subfloat[]{{\includegraphics[width=0.497\linewidth]{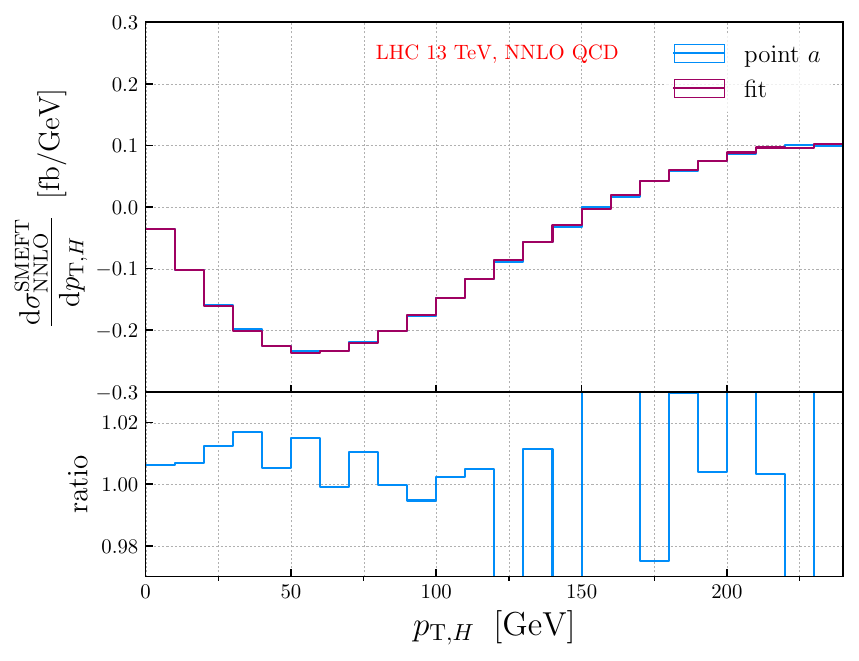}}}
    \caption{Exact calculation of the difference between the SMEFT and SM results for the invariant mass distribution of the $W^+H$ system (left) and the transverse momentum of the Higgs boson (right) compared with distributions obtained using a fit (cf.\ Eq.~\eqref{eq:xsecs_coeffs}). Distributions are obtained using the central factorization/renormalization scale, and the fiducial cuts described in the main text. }
    \label{fig:test_fit}
\end{figure}

Using  associated $WH$ production to search for deviations  from the SM within the SMEFT framework involves comparing data with theoretical predictions for many values of the 
Wilson coefficients. 
It may appear that performing such a comparison with NNLO accuracy for many combinations of SMEFT couplings is a daunting task, given the complexity  of NNLO computations,  but this 
is not necessarily the case, as we now explain. Indeed, 
since we can write 
the most general cross 
section for $\WH$ production as the sum of
six terms, c.f.\ in Eq.~(\ref{eq:xsecs_coeffs})),
it is clear that 
once individual
\emph{differential} cross sections are calculated, one can easily sum them  up including their respective Wilson coefficients and obtain any kinematic distribution for \emph{arbitrary} combinations of SMEFT couplings on the fly.  
To test the quality of  this procedure, 
we have chosen  a random set of Wilson coefficients 
\begin{equation}
    (a): \quad 
    \cuW = 0.321 \,, \quad 
    \cdW = 0.521 \,, \quad 
    \cHqthree = -0.718 \,, \quad
    \cHud = 0.451 \,,
\end{equation}
and evaluated the cross sections to NNLO accuracy by integrating matrix
elements squared with all Wilson coefficients included. We then compared this
result with the cross section that is reconstructed from the individual
``cross sections" using Eq.~\eqref{eq:xsecs_coeffs} and found agreement below
the permille level for the fiducial cross section using the cuts described
above. At a differential level, it is most important to study the
\emph{difference} $\rmd \Delta^{\rm SMEFT}_{\rm NNLO}$ between the \sm\ and
the \smeft\ result using this procedure. In Fig.~\ref{fig:test_fit}, we
display $\rmd \Delta^{\rm SMEFT}_{\rm NNLO}$ for the invariant mass
distribution of the $W^+ H$ system, $m_{\WH}$, and the transverse momentum of
the Higgs boson, $\pTH$. We see that Eq.~\eqref{eq:xsecs_coeffs} reproduces
these differences with around percent-level precision across a large range of
values of the observables (except where the \smeft\ and the \sm\ distributions
practically coincide), indicating that the numerical uncertainties are
well-controlled.
This suggests the following strategy: one computes the SM results to a given higher-order accuracy, and then one combines these with  SMEFT results generated for a wide range of couplings, at essentially zero computational cost. 
Since NNLO accuracy is most relevant when the impact of the SMEFT operators is small, this strategy may open up an opportunity to perform scans that include multiple Wilson coefficients efficiently.  

\section{Conclusion}
\label{sec:concl}
In this paper, we have studied the impact of dimension-six SMEFT operators on the associated Higgs boson production $pp \to \WH$ at the LHC. We have calculated  the NNLO   QCD corrections to this process for SMEFT-initiated contributions.  Although such computations are quite  straightforward by now,
they still require some effort. For example, 
 the  so-called CP-odd SMEFT operators  introduce  point-like interactions 
$q \bar q' W H$ which are  facilitated by a tensor ($\sigma^{\mu \nu} p_{W,\nu}$) 
current, for which  two-loop amplitudes are not 
known. Therefore, studying  contributions of such operators to $pp \to \WH$ production with NNLO QCD accuracy requires a dedicated computation of  new two- and one-loop amplitudes. 

Due to helicity selection rules, amplitudes facilitated by these    CP-odd SMEFT operators do not interfere with the Standard Model ones. As the result, only  their squares contribute to $pp \to \WH$ cross sections, thereby leading to a stronger suppression by the SMEFT energy scale.   Nevertheless, we find that contributions of these operators are quite  substantial, and  that QCD corrections to cross sections induced by these operators are significant and sometimes different from 
similar corrections in the SM. 

We also find that kinematic distributions show varying degrees of sensitivity to the different production mechanisms, with the Higgs boson rapidity distribution being the most insensitive.  What is also important is that the QCD corrections are not uniform for different SMEFT operators. This is especially true at NLO, whereas 
the remaining differences at NNLO are significantly less dramatic. 

It is clear that the large cross sections induced by the SMEFT operators, as well as the difference in the shapes of the $m_{W^+H}$ and $p_{\rmT,H}$ distributions, provide valuable information which should help to place stringent bounds on relevant Wilson coefficients.  
Although,
based on experience with SM calculations, one can argue that the scale variation at  NLO in cases with SM-like corrections reasonably captures the magnitude of NNLO QCD effects,
the same arguments could not  
be made for $\rmd \siguW$ and $\rmd \sigdW$ previously, since both the size of the  NLO QCD corrections as well as the  range of $WH$ invariant masses which provides the main contribution are different. Hence, to establish the perturbative stability of large NLO effects, the complete NNLO QCD computation was required. 
We have provided such an analysis in this paper and look forward to its  future usage for experimental studies  of  $W^+H$ production in the context of SMEFT.


\section*{Acknowledgments}
M.B.\ wishes to thank Federico Buccioni, Federica Devoto, Magnus Schaaf, Chiara Signorile-Signorile, Lukas Simon, and Hantian Zhang for support and fruitful discussions during different stages of this work. 
M.M.L.\ expresses gratitude to Marius Höfer for beneficial conversations and help with \texttt{GoSam}.
D.M.T.\ and R.R.\ are grateful to the Institute for Theoretical Particle Physics at KIT for hospitality extended to them during the course of the work on this paper. 
This work is supported by the Italian Ministry of
Universities and Research (MUR) through grant PRIN
2022BCXSW9 and the
\textit{Deutsche Forschungsgemeinschaft} (DFG, German Research Foundation) under grant no.\ 396021762 - TRR 257.
All Feynman diagrams in this paper are produced by
\texttt{FeynGame}~\cite{Harlander:2020cyh, Harlander:2024qbn, Bundgen:2025utt}.


\appendix

\section{Infrared behavior of the amplitudes}
\label{sec:app2}
The renormalized virtual NLO and NNLO amplitudes contain only IR divergences, which manifest themselves as poles in the dimensional regularization parameter $\epsilon$. 
These IR singularities are universal and  are described  by Catani's formula~\cite{Catani:1998bh}
\bnq
\begin{split}
\widetilde{\mathcal{M}}^{(1)}
&=
\mathbf{I}^{(1)} \widetilde{\mathcal{M}}^{(0)}
+\widetilde{\mathcal{M}}^{(1), \text{finite}}
\,,\\
\widetilde{\mathcal{M}}^{(2)}
&=
\mathbf{I}^{(2)} \widetilde{\mathcal{M}}^{(0)}
+\mathbf{I}^{(1)} \widetilde{\mathcal{M}}^{(1)}
+\widetilde{\mathcal{M}}^{(2), \text{finite}}
\,,
\end{split}
\label{eq:Catani_formula}
\enq
where $\widetilde{\mathcal{M}}$ indicates the renormalized $0 \to \overline{u} d \nu_e e^+ H$ amplitude, and the operators $\mathbf{I}^{(1,2)}$ read
\bnq
\begin{split}
    \mathbf{I}^{(1)}(p_1,p_2;\epsilon) = & 
    - C_{\epsilon} \; \CF \left(\frac{2}{\epsilon^2}+\frac{3}{\epsilon}\right)\eta_{12}
    \,,\\
    \mathbf{I}^{(2)}(p_1,p_2;\epsilon) = & -\frac{\left[\mathbf{I}^{(1)}(p_1,p_2;\epsilon)\right]^2}{2}
    -\frac{\beta_0}{\epsilon}\mathbf{I}^{(1)}(p_1,p_2;\epsilon) \\
    & + \mathrm{e}^{-\epsilon \gamma_{\mathrm{E}}}\frac{\Gamma(1-2\epsilon)}{\Gamma(1-\epsilon)}\left(\frac{\beta_0}{\epsilon}+\mathbf{K}\right)\mathbf{I}^{(1)}(p_1,p_2;2\epsilon) \\
    & + \mathbf{H}^{(2)}(p_1,p_2;\epsilon)
    \,,
\end{split}
\label{eqa2}
\enq
with
\begin{equation}
    C_{\epsilon} = \frac{\mathrm{e}^{\epsilon \gamma_{\mathrm{E}}}}{2\Gamma(1-\epsilon)}\,,
    \qquad
    \mathbf{K} = \left(\frac{67}{18}-\frac{\pi^2}{6}\right)\CA-\frac{10}{9}\nf \TR \,,
    \qquad
    \eta_{ij} = \left(-\frac{\mu^2}{s_{ij}}\right)^{\! \epsilon} \,,
\end{equation}
and~\cite{Becher:2009qa,Becher:2009cu}
\begin{equation}
\begin{split}
    \mathbf{H}^{(2)}(p_1,p_2;\epsilon) 
    =&\; \frac{C_{\epsilon} \CF}{\epsilon}\Bigg[
    \left(\frac{13 \zeta _3}{2}-\frac{23 \pi ^2}{48}+\frac{245}{216}\right) \CA
    \\
    & + \left(-6 \zeta _3+\frac{\pi ^2}{2}-\frac{3}{8}\right) \CF+\left(\frac{\pi ^2}{12}-\frac{25}{54}\right) \nf \TR\Bigg] \,.
\end{split}
\end{equation}

The IR structure of the renormalized real-virtual amplitude is described by Catani's formula as well. In this case, the decomposition reads
\bnq
\widetilde{\mathcal{M}}^{(1)}
=
\mathbf{I}^{(1)} \widetilde{\mathcal{M}}^{(0)}
+\widetilde{\mathcal{M}}^{(1), \text{finite}}
\,,
\label{eqa4}
\enq
where 
\bnq
\begin{split}
    \mathbf{I}^{(1)}(p_1,p_2,p_5;\epsilon) =&\; 
    C_{\epsilon} \bigg[ \left(\CA-2\CF\right)\left(\frac{1}{\epsilon^2}+\frac{3}{2\epsilon}\right) \eta_{12} \\
    & - \left(\frac{\CA}{\epsilon^2}+\frac{3\CA}{4\epsilon}+\frac{\beta_0}{2\epsilon}\right)\left(\eta_{15}+\eta_{25}\right)\bigg] \,.
\end{split}
\enq
We explicitly verify that the IR structure of the amplitudes computed in Section~\ref{sec:ampli} agrees with Eqs.~\eqref{eq:Catani_formula} and \eqref{eqa4}.


\bibliography{bibliography}

\begin{thebibliography}{63}%
\makeatletter
\providecommand \@ifxundefined [1]{%
 \@ifx{#1\undefined}
}%
\providecommand \@ifnum [1]{%
 \ifnum #1\expandafter \@firstoftwo
 \else \expandafter \@secondoftwo
 \fi
}%
\providecommand \@ifx [1]{%
 \ifx #1\expandafter \@firstoftwo
 \else \expandafter \@secondoftwo
 \fi
}%
\providecommand \natexlab [1]{#1}%
\providecommand \enquote  [1]{``#1''}%
\providecommand \bibnamefont  [1]{#1}%
\providecommand \bibfnamefont [1]{#1}%
\providecommand \citenamefont [1]{#1}%
\providecommand \href@noop [0]{\@secondoftwo}%
\providecommand \href [0]{\begingroup \@sanitize@url \@href}%
\providecommand \@href[1]{\@@startlink{#1}\@@href}%
\providecommand \@@href[1]{\endgroup#1\@@endlink}%
\providecommand \@sanitize@url [0]{\catcode `\\12\catcode `\$12\catcode
  `\&12\catcode `\#12\catcode `\^12\catcode `\_12\catcode `\%12\relax}%
\providecommand \@@startlink[1]{}%
\providecommand \@@endlink[0]{}%
\providecommand \url  [0]{\begingroup\@sanitize@url \@url }%
\providecommand \@url [1]{\endgroup\@href {#1}{\urlprefix }}%
\providecommand \urlprefix  [0]{URL }%
\providecommand \Eprint [0]{\href }%
\providecommand \doibase [0]{https://doi.org/}%
\providecommand \selectlanguage [0]{\@gobble}%
\providecommand \bibinfo  [0]{\@secondoftwo}%
\providecommand \bibfield  [0]{\@secondoftwo}%
\providecommand \translation [1]{[#1]}%
\providecommand \BibitemOpen [0]{}%
\providecommand \bibitemStop [0]{}%
\providecommand \bibitemNoStop [0]{.\EOS\space}%
\providecommand \EOS [0]{\spacefactor3000\relax}%
\providecommand \BibitemShut  [1]{\csname bibitem#1\endcsname}%
\let\auto@bib@innerbib\@empty
\bibitem [{\citenamefont {Workman}\ \emph {et~al.}(2022)\citenamefont {Workman}
  \emph {et~al.}}]{ParticleDataGroup:2022pth}%
  \BibitemOpen
  \bibfield  {author} {\bibinfo {author} {\bibfnamefont {R.~L.}\ \bibnamefont
  {Workman}} \emph {et~al.} (\bibinfo {collaboration} {Particle Data Group}),\
  }\bibfield  {title} {\bibinfo {title} {{Review of Particle Physics}},\ }\href
  {https://doi.org/10.1093/ptep/ptac097} {\bibfield  {journal} {\bibinfo
  {journal} {PTEP}\ }\textbf {\bibinfo {volume} {2022}},\ \bibinfo {pages}
  {083C01} (\bibinfo {year} {2022})}\BibitemShut {NoStop}%
\bibitem [{\citenamefont {Weinberg}(1979)}]{Weinberg:1979sa}%
  \BibitemOpen
  \bibfield  {author} {\bibinfo {author} {\bibfnamefont {S.}~\bibnamefont
  {Weinberg}},\ }\bibfield  {title} {\bibinfo {title} {{Baryon and lepton
  nonconserving processes}},\ }\href
  {https://doi.org/10.1103/PhysRevLett.43.1566} {\bibfield  {journal} {\bibinfo
   {journal} {Phys. Rev. Lett.}\ }\textbf {\bibinfo {volume} {43}},\ \bibinfo
  {pages} {1566} (\bibinfo {year} {1979})}\BibitemShut {NoStop}%
\bibitem [{\citenamefont {Buchm\"uller}\ and\ \citenamefont
  {Wyler}(1986)}]{Buchmuller:1985jz}%
  \BibitemOpen
  \bibfield  {author} {\bibinfo {author} {\bibfnamefont {W.}~\bibnamefont
  {Buchm\"uller}}\ and\ \bibinfo {author} {\bibfnamefont {D.}~\bibnamefont
  {Wyler}},\ }\bibfield  {title} {\bibinfo {title} {{Effective Lagrangian
  analysis of new interactions and flavor conservation}},\ }\href
  {https://doi.org/10.1016/0550-3213(86)90262-2} {\bibfield  {journal}
  {\bibinfo  {journal} {Nucl. Phys. B}\ }\textbf {\bibinfo {volume} {268}},\
  \bibinfo {pages} {621} (\bibinfo {year} {1986})}\BibitemShut {NoStop}%
\bibitem [{\citenamefont {Grzadkowski}\ \emph {et~al.}(2010)\citenamefont
  {Grzadkowski}, \citenamefont {Iskrzynski}, \citenamefont {Misiak},\ and\
  \citenamefont {Rosiek}}]{Grzadkowski:2010es}%
  \BibitemOpen
  \bibfield  {author} {\bibinfo {author} {\bibfnamefont {B.}~\bibnamefont
  {Grzadkowski}}, \bibinfo {author} {\bibfnamefont {M.}~\bibnamefont
  {Iskrzynski}}, \bibinfo {author} {\bibfnamefont {M.}~\bibnamefont {Misiak}},\
  and\ \bibinfo {author} {\bibfnamefont {J.}~\bibnamefont {Rosiek}},\
  }\bibfield  {title} {\bibinfo {title} {{Dimension-six terms in the Standard
  Model Lagrangian}},\ }\href {https://doi.org/10.1007/JHEP10(2010)085}
  {\bibfield  {journal} {\bibinfo  {journal} {JHEP}\ }\textbf {\bibinfo
  {volume} {10}},\ \bibinfo {pages} {085}},\ \Eprint
  {https://arxiv.org/abs/1008.4884} {arXiv:1008.4884 [hep-ph]} \BibitemShut
  {NoStop}%
\bibitem [{\citenamefont {Alonso}\ \emph {et~al.}(2014)\citenamefont {Alonso},
  \citenamefont {Jenkins}, \citenamefont {Manohar},\ and\ \citenamefont
  {Trott}}]{Alonso:2013hga}%
  \BibitemOpen
  \bibfield  {author} {\bibinfo {author} {\bibfnamefont {R.}~\bibnamefont
  {Alonso}}, \bibinfo {author} {\bibfnamefont {E.~E.}\ \bibnamefont {Jenkins}},
  \bibinfo {author} {\bibfnamefont {A.~V.}\ \bibnamefont {Manohar}},\ and\
  \bibinfo {author} {\bibfnamefont {M.}~\bibnamefont {Trott}},\ }\bibfield
  {title} {\bibinfo {title} {{Renormalization group evolution of the Standard
  Model dimension six operators III: gauge coupling dependence and
  phenomenology}},\ }\href {https://doi.org/10.1007/JHEP04(2014)159} {\bibfield
   {journal} {\bibinfo  {journal} {JHEP}\ }\textbf {\bibinfo {volume} {04}},\
  \bibinfo {pages} {159}},\ \Eprint {https://arxiv.org/abs/1312.2014}
  {arXiv:1312.2014 [hep-ph]} \BibitemShut {NoStop}%
\bibitem [{\citenamefont {Henning}\ \emph {et~al.}(2017)\citenamefont
  {Henning}, \citenamefont {Lu}, \citenamefont {Melia},\ and\ \citenamefont
  {Murayama}}]{Henning:2015alf}%
  \BibitemOpen
  \bibfield  {author} {\bibinfo {author} {\bibfnamefont {B.}~\bibnamefont
  {Henning}}, \bibinfo {author} {\bibfnamefont {X.}~\bibnamefont {Lu}},
  \bibinfo {author} {\bibfnamefont {T.}~\bibnamefont {Melia}},\ and\ \bibinfo
  {author} {\bibfnamefont {H.}~\bibnamefont {Murayama}},\ }\bibfield  {title}
  {\bibinfo {title} {{2, 84, 30, 993, 560, 15456, 11962, 261485, ...: Higher
  dimension operators in the SM EFT}},\ }\href
  {https://doi.org/10.1007/JHEP08(2017)016} {\bibfield  {journal} {\bibinfo
  {journal} {JHEP}\ }\textbf {\bibinfo {volume} {08}},\ \bibinfo {pages}
  {016}},\ \bibinfo {note} {[Erratum: JHEP 09, 019 (2019)]},\ \Eprint
  {https://arxiv.org/abs/1512.03433} {arXiv:1512.03433 [hep-ph]} \BibitemShut
  {NoStop}%
\bibitem [{\citenamefont {Harlander}\ \emph {et~al.}(2023)\citenamefont
  {Harlander}, \citenamefont {Kempkens},\ and\ \citenamefont
  {Schaaf}}]{Harlander:2023psl}%
  \BibitemOpen
  \bibfield  {author} {\bibinfo {author} {\bibfnamefont {R.~V.}\ \bibnamefont
  {Harlander}}, \bibinfo {author} {\bibfnamefont {T.}~\bibnamefont
  {Kempkens}},\ and\ \bibinfo {author} {\bibfnamefont {M.~C.}\ \bibnamefont
  {Schaaf}},\ }\bibfield  {title} {\bibinfo {title} {{Standard model effective
  field theory up to mass dimension 12}},\ }\href
  {https://doi.org/10.1103/PhysRevD.108.055020} {\bibfield  {journal} {\bibinfo
   {journal} {Phys. Rev. D}\ }\textbf {\bibinfo {volume} {108}},\ \bibinfo
  {pages} {055020} (\bibinfo {year} {2023})},\ \Eprint
  {https://arxiv.org/abs/2305.06832} {arXiv:2305.06832 [hep-ph]} \BibitemShut
  {NoStop}%
\bibitem [{\citenamefont {Butterworth}\ \emph {et~al.}(2008)\citenamefont
  {Butterworth}, \citenamefont {Davison}, \citenamefont {Rubin},\ and\
  \citenamefont {Salam}}]{Butterworth:2008iy}%
  \BibitemOpen
  \bibfield  {author} {\bibinfo {author} {\bibfnamefont {J.~M.}\ \bibnamefont
  {Butterworth}}, \bibinfo {author} {\bibfnamefont {A.~R.}\ \bibnamefont
  {Davison}}, \bibinfo {author} {\bibfnamefont {M.}~\bibnamefont {Rubin}},\
  and\ \bibinfo {author} {\bibfnamefont {G.~P.}\ \bibnamefont {Salam}},\
  }\bibfield  {title} {\bibinfo {title} {{Jet substructure as a new Higgs
  search channel at the LHC}},\ }\href
  {https://doi.org/10.1103/PhysRevLett.100.242001} {\bibfield  {journal}
  {\bibinfo  {journal} {Phys. Rev. Lett.}\ }\textbf {\bibinfo {volume} {100}},\
  \bibinfo {pages} {242001} (\bibinfo {year} {2008})},\ \Eprint
  {https://arxiv.org/abs/0802.2470} {arXiv:0802.2470 [hep-ph]} \BibitemShut
  {NoStop}%
\bibitem [{\citenamefont {Harlander}\ \emph {et~al.}(2018)\citenamefont
  {Harlander}, \citenamefont {Klappert}, \citenamefont {Pandini},\ and\
  \citenamefont {Papaefstathiou}}]{Harlander:2018yns}%
  \BibitemOpen
  \bibfield  {author} {\bibinfo {author} {\bibfnamefont {R.~V.}\ \bibnamefont
  {Harlander}}, \bibinfo {author} {\bibfnamefont {J.}~\bibnamefont {Klappert}},
  \bibinfo {author} {\bibfnamefont {C.}~\bibnamefont {Pandini}},\ and\ \bibinfo
  {author} {\bibfnamefont {A.}~\bibnamefont {Papaefstathiou}},\ }\bibfield
  {title} {\bibinfo {title} {{Exploiting the $WH/ZH$ symmetry in the search for
  New Physics}},\ }\href {https://doi.org/10.1140/epjc/s10052-018-6234-x}
  {\bibfield  {journal} {\bibinfo  {journal} {Eur. Phys. J. C}\ }\textbf
  {\bibinfo {volume} {78}},\ \bibinfo {pages} {760} (\bibinfo {year} {2018})},\
  \Eprint {https://arxiv.org/abs/1804.02299} {arXiv:1804.02299 [hep-ph]}
  \BibitemShut {NoStop}%
\bibitem [{\citenamefont {Ferrera}\ \emph {et~al.}(2011)\citenamefont
  {Ferrera}, \citenamefont {Grazzini},\ and\ \citenamefont
  {Tramontano}}]{Ferrera:2011bk}%
  \BibitemOpen
  \bibfield  {author} {\bibinfo {author} {\bibfnamefont {G.}~\bibnamefont
  {Ferrera}}, \bibinfo {author} {\bibfnamefont {M.}~\bibnamefont {Grazzini}},\
  and\ \bibinfo {author} {\bibfnamefont {F.}~\bibnamefont {Tramontano}},\
  }\bibfield  {title} {\bibinfo {title} {{Associated $WH$ production at hadron
  colliders: a fully exclusive QCD calculation at NNLO}},\ }\href
  {https://doi.org/10.1103/PhysRevLett.107.152003} {\bibfield  {journal}
  {\bibinfo  {journal} {Phys. Rev. Lett.}\ }\textbf {\bibinfo {volume} {107}},\
  \bibinfo {pages} {152003} (\bibinfo {year} {2011})},\ \Eprint
  {https://arxiv.org/abs/1107.1164} {arXiv:1107.1164 [hep-ph]} \BibitemShut
  {NoStop}%
\bibitem [{\citenamefont {Ferrera}\ \emph {et~al.}(2014)\citenamefont
  {Ferrera}, \citenamefont {Grazzini},\ and\ \citenamefont
  {Tramontano}}]{Ferrera:2013yga}%
  \BibitemOpen
  \bibfield  {author} {\bibinfo {author} {\bibfnamefont {G.}~\bibnamefont
  {Ferrera}}, \bibinfo {author} {\bibfnamefont {M.}~\bibnamefont {Grazzini}},\
  and\ \bibinfo {author} {\bibfnamefont {F.}~\bibnamefont {Tramontano}},\
  }\bibfield  {title} {\bibinfo {title} {{Higher-order QCD effects for
  associated $WH$ production and decay at the LHC}},\ }\href
  {https://doi.org/10.1007/JHEP04(2014)039} {\bibfield  {journal} {\bibinfo
  {journal} {JHEP}\ }\textbf {\bibinfo {volume} {04}},\ \bibinfo {pages}
  {039}},\ \Eprint {https://arxiv.org/abs/1312.1669} {arXiv:1312.1669 [hep-ph]}
  \BibitemShut {NoStop}%
\bibitem [{\citenamefont {Ferrera}\ \emph {et~al.}(2015)\citenamefont
  {Ferrera}, \citenamefont {Grazzini},\ and\ \citenamefont
  {Tramontano}}]{Ferrera:2014lca}%
  \BibitemOpen
  \bibfield  {author} {\bibinfo {author} {\bibfnamefont {G.}~\bibnamefont
  {Ferrera}}, \bibinfo {author} {\bibfnamefont {M.}~\bibnamefont {Grazzini}},\
  and\ \bibinfo {author} {\bibfnamefont {F.}~\bibnamefont {Tramontano}},\
  }\bibfield  {title} {\bibinfo {title} {{Associated $ZH$ production at hadron
  colliders: the fully differential NNLO QCD calculation}},\ }\href
  {https://doi.org/10.1016/j.physletb.2014.11.040} {\bibfield  {journal}
  {\bibinfo  {journal} {Phys. Lett. B}\ }\textbf {\bibinfo {volume} {740}},\
  \bibinfo {pages} {51} (\bibinfo {year} {2015})},\ \Eprint
  {https://arxiv.org/abs/1407.4747} {arXiv:1407.4747 [hep-ph]} \BibitemShut
  {NoStop}%
\bibitem [{\citenamefont {Campbell}\ \emph {et~al.}(2016)\citenamefont
  {Campbell}, \citenamefont {Ellis},\ and\ \citenamefont
  {Williams}}]{Campbell:2016jau}%
  \BibitemOpen
  \bibfield  {author} {\bibinfo {author} {\bibfnamefont {J.~M.}\ \bibnamefont
  {Campbell}}, \bibinfo {author} {\bibfnamefont {R.~K.}\ \bibnamefont
  {Ellis}},\ and\ \bibinfo {author} {\bibfnamefont {C.}~\bibnamefont
  {Williams}},\ }\bibfield  {title} {\bibinfo {title} {{Associated production
  of a Higgs boson at NNLO}},\ }\href {https://doi.org/10.1007/JHEP06(2016)179}
  {\bibfield  {journal} {\bibinfo  {journal} {JHEP}\ }\textbf {\bibinfo
  {volume} {06}},\ \bibinfo {pages} {179}},\ \Eprint
  {https://arxiv.org/abs/1601.00658} {arXiv:1601.00658 [hep-ph]} \BibitemShut
  {NoStop}%
\bibitem [{\citenamefont {Caola}\ \emph {et~al.}(2018)\citenamefont {Caola},
  \citenamefont {Luisoni}, \citenamefont {Melnikov},\ and\ \citenamefont
  {R\"ontsch}}]{Caola:2017xuq}%
  \BibitemOpen
  \bibfield  {author} {\bibinfo {author} {\bibfnamefont {F.}~\bibnamefont
  {Caola}}, \bibinfo {author} {\bibfnamefont {G.}~\bibnamefont {Luisoni}},
  \bibinfo {author} {\bibfnamefont {K.}~\bibnamefont {Melnikov}},\ and\
  \bibinfo {author} {\bibfnamefont {R.}~\bibnamefont {R\"ontsch}},\ }\bibfield
  {title} {\bibinfo {title} {{NNLO QCD corrections to associated $WH$
  production and $H \to b \bar b$ decay}},\ }\href
  {https://doi.org/10.1103/PhysRevD.97.074022} {\bibfield  {journal} {\bibinfo
  {journal} {Phys. Rev. D}\ }\textbf {\bibinfo {volume} {97}},\ \bibinfo
  {pages} {074022} (\bibinfo {year} {2018})},\ \Eprint
  {https://arxiv.org/abs/1712.06954} {arXiv:1712.06954 [hep-ph]} \BibitemShut
  {NoStop}%
\bibitem [{\citenamefont {Ferrera}\ \emph {et~al.}(2018)\citenamefont
  {Ferrera}, \citenamefont {Somogyi},\ and\ \citenamefont
  {Tramontano}}]{Ferrera:2017zex}%
  \BibitemOpen
  \bibfield  {author} {\bibinfo {author} {\bibfnamefont {G.}~\bibnamefont
  {Ferrera}}, \bibinfo {author} {\bibfnamefont {G.}~\bibnamefont {Somogyi}},\
  and\ \bibinfo {author} {\bibfnamefont {F.}~\bibnamefont {Tramontano}},\
  }\bibfield  {title} {\bibinfo {title} {{Associated production of a Higgs
  boson decaying into bottom quarks at the LHC in full NNLO QCD}},\ }\href
  {https://doi.org/10.1016/j.physletb.2018.03.021} {\bibfield  {journal}
  {\bibinfo  {journal} {Phys. Lett. B}\ }\textbf {\bibinfo {volume} {780}},\
  \bibinfo {pages} {346} (\bibinfo {year} {2018})},\ \Eprint
  {https://arxiv.org/abs/1705.10304} {arXiv:1705.10304 [hep-ph]} \BibitemShut
  {NoStop}%
\bibitem [{\citenamefont {Gauld}\ \emph {et~al.}(2019)\citenamefont {Gauld},
  \citenamefont {Gehrmann-De~Ridder}, \citenamefont {Glover}, \citenamefont
  {Huss},\ and\ \citenamefont {Majer}}]{Gauld:2019yng}%
  \BibitemOpen
  \bibfield  {author} {\bibinfo {author} {\bibfnamefont {R.}~\bibnamefont
  {Gauld}}, \bibinfo {author} {\bibfnamefont {A.}~\bibnamefont
  {Gehrmann-De~Ridder}}, \bibinfo {author} {\bibfnamefont {E.~W.~N.}\
  \bibnamefont {Glover}}, \bibinfo {author} {\bibfnamefont {A.}~\bibnamefont
  {Huss}},\ and\ \bibinfo {author} {\bibfnamefont {I.}~\bibnamefont {Majer}},\
  }\bibfield  {title} {\bibinfo {title} {{Associated production of a Higgs
  boson decaying into bottom quarks and a weak vector boson decaying
  leptonically at NNLO in QCD}},\ }\href
  {https://doi.org/10.1007/JHEP10(2019)002} {\bibfield  {journal} {\bibinfo
  {journal} {JHEP}\ }\textbf {\bibinfo {volume} {10}},\ \bibinfo {pages}
  {002}},\ \Eprint {https://arxiv.org/abs/1907.05836} {arXiv:1907.05836
  [hep-ph]} \BibitemShut {NoStop}%
\bibitem [{\citenamefont {Ciccolini}\ \emph {et~al.}(2003)\citenamefont
  {Ciccolini}, \citenamefont {Dittmaier},\ and\ \citenamefont
  {Kr\"amer}}]{Ciccolini:2003jy}%
  \BibitemOpen
  \bibfield  {author} {\bibinfo {author} {\bibfnamefont {M.~L.}\ \bibnamefont
  {Ciccolini}}, \bibinfo {author} {\bibfnamefont {S.}~\bibnamefont
  {Dittmaier}},\ and\ \bibinfo {author} {\bibfnamefont {M.}~\bibnamefont
  {Kr\"amer}},\ }\bibfield  {title} {\bibinfo {title} {{Electroweak radiative
  corrections to associated $WH$ and $ZH$ production at hadron colliders}},\
  }\href {https://doi.org/10.1103/PhysRevD.68.073003} {\bibfield  {journal}
  {\bibinfo  {journal} {Phys. Rev. D}\ }\textbf {\bibinfo {volume} {68}},\
  \bibinfo {pages} {073003} (\bibinfo {year} {2003})},\ \Eprint
  {https://arxiv.org/abs/hep-ph/0306234} {arXiv:hep-ph/0306234} \BibitemShut
  {NoStop}%
\bibitem [{\citenamefont {Denner}\ \emph {et~al.}(2012)\citenamefont {Denner},
  \citenamefont {Dittmaier}, \citenamefont {Kallweit},\ and\ \citenamefont
  {M\"uck}}]{Denner:2011id}%
  \BibitemOpen
  \bibfield  {author} {\bibinfo {author} {\bibfnamefont {A.}~\bibnamefont
  {Denner}}, \bibinfo {author} {\bibfnamefont {S.}~\bibnamefont {Dittmaier}},
  \bibinfo {author} {\bibfnamefont {S.}~\bibnamefont {Kallweit}},\ and\
  \bibinfo {author} {\bibfnamefont {A.}~\bibnamefont {M\"uck}},\ }\bibfield
  {title} {\bibinfo {title} {{Electroweak corrections to Higgs-strahlung off
  $W/Z$ bosons at the Tevatron and the LHC with HAWK}},\ }\href
  {https://doi.org/10.1007/JHEP03(2012)075} {\bibfield  {journal} {\bibinfo
  {journal} {JHEP}\ }\textbf {\bibinfo {volume} {03}},\ \bibinfo {pages}
  {075}},\ \Eprint {https://arxiv.org/abs/1112.5142} {arXiv:1112.5142 [hep-ph]}
  \BibitemShut {NoStop}%
\bibitem [{\citenamefont {Denner}\ \emph {et~al.}(2015)\citenamefont {Denner},
  \citenamefont {Dittmaier}, \citenamefont {Kallweit},\ and\ \citenamefont
  {M\"uck}}]{Denner:2014cla}%
  \BibitemOpen
  \bibfield  {author} {\bibinfo {author} {\bibfnamefont {A.}~\bibnamefont
  {Denner}}, \bibinfo {author} {\bibfnamefont {S.}~\bibnamefont {Dittmaier}},
  \bibinfo {author} {\bibfnamefont {S.}~\bibnamefont {Kallweit}},\ and\
  \bibinfo {author} {\bibfnamefont {A.}~\bibnamefont {M\"uck}},\ }\bibfield
  {title} {\bibinfo {title} {{HAWK 2.0: A Monte Carlo program for Higgs
  production in vector-boson fusion and Higgs strahlung at hadron colliders}},\
  }\href {https://doi.org/10.1016/j.cpc.2015.04.021} {\bibfield  {journal}
  {\bibinfo  {journal} {Comput. Phys. Commun.}\ }\textbf {\bibinfo {volume}
  {195}},\ \bibinfo {pages} {161} (\bibinfo {year} {2015})},\ \Eprint
  {https://arxiv.org/abs/1412.5390} {arXiv:1412.5390 [hep-ph]} \BibitemShut
  {NoStop}%
\bibitem [{\citenamefont {Baglio}\ \emph {et~al.}(2022)\citenamefont {Baglio},
  \citenamefont {Duhr}, \citenamefont {Mistlberger},\ and\ \citenamefont
  {Szafron}}]{Baglio:2022wzu}%
  \BibitemOpen
  \bibfield  {author} {\bibinfo {author} {\bibfnamefont {J.}~\bibnamefont
  {Baglio}}, \bibinfo {author} {\bibfnamefont {C.}~\bibnamefont {Duhr}},
  \bibinfo {author} {\bibfnamefont {B.}~\bibnamefont {Mistlberger}},\ and\
  \bibinfo {author} {\bibfnamefont {R.}~\bibnamefont {Szafron}},\ }\bibfield
  {title} {\bibinfo {title} {{Inclusive production cross sections at
  N$^{3}$LO}},\ }\href {https://doi.org/10.1007/JHEP12(2022)066} {\bibfield
  {journal} {\bibinfo  {journal} {JHEP}\ }\textbf {\bibinfo {volume} {12}},\
  \bibinfo {pages} {066}},\ \Eprint {https://arxiv.org/abs/2209.06138}
  {arXiv:2209.06138 [hep-ph]} \BibitemShut {NoStop}%
\bibitem [{\citenamefont {Alioli}\ \emph {et~al.}(2018)\citenamefont {Alioli},
  \citenamefont {Dekens}, \citenamefont {Girard},\ and\ \citenamefont
  {Mereghetti}}]{Alioli:2018ljm}%
  \BibitemOpen
  \bibfield  {author} {\bibinfo {author} {\bibfnamefont {S.}~\bibnamefont
  {Alioli}}, \bibinfo {author} {\bibfnamefont {W.}~\bibnamefont {Dekens}},
  \bibinfo {author} {\bibfnamefont {M.}~\bibnamefont {Girard}},\ and\ \bibinfo
  {author} {\bibfnamefont {E.}~\bibnamefont {Mereghetti}},\ }\bibfield  {title}
  {\bibinfo {title} {{NLO QCD corrections to SM-EFT dilepton and electroweak
  Higgs boson production, matched to parton shower in POWHEG}},\ }\href
  {https://doi.org/10.1007/JHEP08(2018)205} {\bibfield  {journal} {\bibinfo
  {journal} {JHEP}\ }\textbf {\bibinfo {volume} {08}},\ \bibinfo {pages}
  {205}},\ \Eprint {https://arxiv.org/abs/1804.07407} {arXiv:1804.07407
  [hep-ph]} \BibitemShut {NoStop}%
\bibitem [{\citenamefont {Bizo\'n}\ \emph {et~al.}(2022)\citenamefont
  {Bizo\'n}, \citenamefont {Caola}, \citenamefont {Melnikov},\ and\
  \citenamefont {R\"ontsch}}]{Bizon:2021rww}%
  \BibitemOpen
  \bibfield  {author} {\bibinfo {author} {\bibfnamefont {W.}~\bibnamefont
  {Bizo\'n}}, \bibinfo {author} {\bibfnamefont {F.}~\bibnamefont {Caola}},
  \bibinfo {author} {\bibfnamefont {K.}~\bibnamefont {Melnikov}},\ and\
  \bibinfo {author} {\bibfnamefont {R.}~\bibnamefont {R\"ontsch}},\ }\bibfield
  {title} {\bibinfo {title} {{Anomalous couplings in associated $VH$ production
  with Higgs boson decay to massive $b$ quarks at NNLO in QCD}},\ }\href
  {https://doi.org/10.1103/PhysRevD.105.014023} {\bibfield  {journal} {\bibinfo
   {journal} {Phys. Rev. D}\ }\textbf {\bibinfo {volume} {105}},\ \bibinfo
  {pages} {014023} (\bibinfo {year} {2022})},\ \Eprint
  {https://arxiv.org/abs/2106.06328} {arXiv:2106.06328 [hep-ph]} \BibitemShut
  {NoStop}%
\bibitem [{\citenamefont {Haisch}\ \emph {et~al.}(2022)\citenamefont {Haisch},
  \citenamefont {Scott}, \citenamefont {Wiesemann}, \citenamefont
  {Zanderighi},\ and\ \citenamefont {Zanoli}}]{Haisch:2022nwz}%
  \BibitemOpen
  \bibfield  {author} {\bibinfo {author} {\bibfnamefont {U.}~\bibnamefont
  {Haisch}}, \bibinfo {author} {\bibfnamefont {D.~J.}\ \bibnamefont {Scott}},
  \bibinfo {author} {\bibfnamefont {M.}~\bibnamefont {Wiesemann}}, \bibinfo
  {author} {\bibfnamefont {G.}~\bibnamefont {Zanderighi}},\ and\ \bibinfo
  {author} {\bibfnamefont {S.}~\bibnamefont {Zanoli}},\ }\bibfield  {title}
  {\bibinfo {title} {{NNLO event generation for $ pp\to Zh\to
  {\mathrm{\ell}}^{+}{\mathrm{\ell}}^{-}b\overline{b} $ production in the SM
  effective field theory}},\ }\href {https://doi.org/10.1007/JHEP07(2022)054}
  {\bibfield  {journal} {\bibinfo  {journal} {JHEP}\ }\textbf {\bibinfo
  {volume} {07}},\ \bibinfo {pages} {054}},\ \Eprint
  {https://arxiv.org/abs/2204.00663} {arXiv:2204.00663 [hep-ph]} \BibitemShut
  {NoStop}%
\bibitem [{\citenamefont {Franceschini}\ \emph {et~al.}(2018)\citenamefont
  {Franceschini}, \citenamefont {Panico}, \citenamefont {Pomarol},
  \citenamefont {Riva},\ and\ \citenamefont {Wulzer}}]{Franceschini:2017xkh}%
  \BibitemOpen
  \bibfield  {author} {\bibinfo {author} {\bibfnamefont {R.}~\bibnamefont
  {Franceschini}}, \bibinfo {author} {\bibfnamefont {G.}~\bibnamefont
  {Panico}}, \bibinfo {author} {\bibfnamefont {A.}~\bibnamefont {Pomarol}},
  \bibinfo {author} {\bibfnamefont {F.}~\bibnamefont {Riva}},\ and\ \bibinfo
  {author} {\bibfnamefont {A.}~\bibnamefont {Wulzer}},\ }\bibfield  {title}
  {\bibinfo {title} {{Electroweak precision tests in high-energy diboson
  processes}},\ }\href {https://doi.org/10.1007/JHEP02(2018)111} {\bibfield
  {journal} {\bibinfo  {journal} {JHEP}\ }\textbf {\bibinfo {volume} {02}},\
  \bibinfo {pages} {111}},\ \Eprint {https://arxiv.org/abs/1712.01310}
  {arXiv:1712.01310 [hep-ph]} \BibitemShut {NoStop}%
\bibitem [{\citenamefont {Gauld}\ \emph {et~al.}(2024)\citenamefont {Gauld},
  \citenamefont {Haisch},\ and\ \citenamefont {Schnell}}]{Gauld:2023gtb}%
  \BibitemOpen
  \bibfield  {author} {\bibinfo {author} {\bibfnamefont {R.}~\bibnamefont
  {Gauld}}, \bibinfo {author} {\bibfnamefont {U.}~\bibnamefont {Haisch}},\ and\
  \bibinfo {author} {\bibfnamefont {L.}~\bibnamefont {Schnell}},\ }\bibfield
  {title} {\bibinfo {title} {{SMEFT at NNLO+PS: $Vh$ production}},\ }\href
  {https://doi.org/10.1007/JHEP01(2024)192} {\bibfield  {journal} {\bibinfo
  {journal} {JHEP}\ }\textbf {\bibinfo {volume} {01}},\ \bibinfo {pages}
  {192}},\ \Eprint {https://arxiv.org/abs/2311.06107} {arXiv:2311.06107
  [hep-ph]} \BibitemShut {NoStop}%
\bibitem [{\citenamefont {Dedes}\ \emph {et~al.}(2017)\citenamefont {Dedes},
  \citenamefont {Materkowska}, \citenamefont {Paraskevas}, \citenamefont
  {Rosiek},\ and\ \citenamefont {Suxho}}]{Dedes:2017zog}%
  \BibitemOpen
  \bibfield  {author} {\bibinfo {author} {\bibfnamefont {A.}~\bibnamefont
  {Dedes}}, \bibinfo {author} {\bibfnamefont {W.}~\bibnamefont {Materkowska}},
  \bibinfo {author} {\bibfnamefont {M.}~\bibnamefont {Paraskevas}}, \bibinfo
  {author} {\bibfnamefont {J.}~\bibnamefont {Rosiek}},\ and\ \bibinfo {author}
  {\bibfnamefont {K.}~\bibnamefont {Suxho}},\ }\bibfield  {title} {\bibinfo
  {title} {{Feynman rules for the Standard Model Effective Field Theory in
  R$_\xi$-gauges}},\ }\href {https://doi.org/10.1007/JHEP06(2017)143}
  {\bibfield  {journal} {\bibinfo  {journal} {JHEP}\ }\textbf {\bibinfo
  {volume} {06}},\ \bibinfo {pages} {143}},\ \Eprint
  {https://arxiv.org/abs/1704.03888} {arXiv:1704.03888 [hep-ph]} \BibitemShut
  {NoStop}%
\bibitem [{\citenamefont {Brein}\ \emph {et~al.}(2012)\citenamefont {Brein},
  \citenamefont {Harlander}, \citenamefont {Wiesemann},\ and\ \citenamefont
  {Zirke}}]{Brein:2011vx}%
  \BibitemOpen
  \bibfield  {author} {\bibinfo {author} {\bibfnamefont {O.}~\bibnamefont
  {Brein}}, \bibinfo {author} {\bibfnamefont {R.}~\bibnamefont {Harlander}},
  \bibinfo {author} {\bibfnamefont {M.}~\bibnamefont {Wiesemann}},\ and\
  \bibinfo {author} {\bibfnamefont {T.}~\bibnamefont {Zirke}},\ }\bibfield
  {title} {\bibinfo {title} {{Top-quark mediated effects in hadronic
  Higgs-strahlung}},\ }\href {https://doi.org/10.1140/epjc/s10052-012-1868-6}
  {\bibfield  {journal} {\bibinfo  {journal} {Eur. Phys. J. C}\ }\textbf
  {\bibinfo {volume} {72}},\ \bibinfo {pages} {1868} (\bibinfo {year}
  {2012})},\ \Eprint {https://arxiv.org/abs/1111.0761} {arXiv:1111.0761
  [hep-ph]} \BibitemShut {NoStop}%
\bibitem [{\citenamefont {Kniehl}(1990)}]{Kniehl:1990iva}%
  \BibitemOpen
  \bibfield  {author} {\bibinfo {author} {\bibfnamefont {B.~A.}\ \bibnamefont
  {Kniehl}},\ }\bibfield  {title} {\bibinfo {title} {{Associated production of
  Higgs and $Z$ bosons from gluon fusion in hadron collisions}},\ }\href
  {https://doi.org/10.1103/PhysRevD.42.2253} {\bibfield  {journal} {\bibinfo
  {journal} {Phys. Rev. D}\ }\textbf {\bibinfo {volume} {42}},\ \bibinfo
  {pages} {2253} (\bibinfo {year} {1990})}\BibitemShut {NoStop}%
\bibitem [{\citenamefont {Brein}\ \emph {et~al.}(2004)\citenamefont {Brein},
  \citenamefont {Djouadi},\ and\ \citenamefont {Harlander}}]{Brein:2003wg}%
  \BibitemOpen
  \bibfield  {author} {\bibinfo {author} {\bibfnamefont {O.}~\bibnamefont
  {Brein}}, \bibinfo {author} {\bibfnamefont {A.}~\bibnamefont {Djouadi}},\
  and\ \bibinfo {author} {\bibfnamefont {R.}~\bibnamefont {Harlander}},\
  }\bibfield  {title} {\bibinfo {title} {{NNLO QCD corrections to the
  Higgs-strahlung processes at hadron colliders}},\ }\href
  {https://doi.org/10.1016/j.physletb.2003.10.112} {\bibfield  {journal}
  {\bibinfo  {journal} {Phys. Lett. B}\ }\textbf {\bibinfo {volume} {579}},\
  \bibinfo {pages} {149} (\bibinfo {year} {2004})},\ \Eprint
  {https://arxiv.org/abs/hep-ph/0307206} {arXiv:hep-ph/0307206} \BibitemShut
  {NoStop}%
\bibitem [{\citenamefont {Elvang}\ and\ \citenamefont
  {Huang}(2013)}]{Elvang:2013cua}%
  \BibitemOpen
  \bibfield  {author} {\bibinfo {author} {\bibfnamefont {H.}~\bibnamefont
  {Elvang}}\ and\ \bibinfo {author} {\bibfnamefont {Y.-t.}\ \bibnamefont
  {Huang}},\ }\bibfield  {title} {\bibinfo {title} {{Scattering amplitudes}},\
  }\Eprint {https://arxiv.org/abs/1308.1697} {arXiv:1308.1697 [hep-th]}
  (\bibinfo {year} {2013})\BibitemShut {NoStop}%
\bibitem [{\citenamefont {Nogueira}(1993)}]{Nogueira:1991ex}%
  \BibitemOpen
  \bibfield  {author} {\bibinfo {author} {\bibfnamefont {P.}~\bibnamefont
  {Nogueira}},\ }\bibfield  {title} {\bibinfo {title} {{Automatic Feynman graph
  generation}},\ }\href {https://doi.org/10.1006/jcph.1993.1074} {\bibfield
  {journal} {\bibinfo  {journal} {J. Comput. Phys.}\ }\textbf {\bibinfo
  {volume} {105}},\ \bibinfo {pages} {279} (\bibinfo {year}
  {1993})}\BibitemShut {NoStop}%
\bibitem [{\citenamefont {Ruijl}\ \emph {et~al.}(2017)\citenamefont {Ruijl},
  \citenamefont {Ueda},\ and\ \citenamefont {Vermaseren}}]{Ruijl:2017dtg}%
  \BibitemOpen
  \bibfield  {author} {\bibinfo {author} {\bibfnamefont {B.}~\bibnamefont
  {Ruijl}}, \bibinfo {author} {\bibfnamefont {T.}~\bibnamefont {Ueda}},\ and\
  \bibinfo {author} {\bibfnamefont {J.}~\bibnamefont {Vermaseren}},\ }\bibfield
   {title} {\bibinfo {title} {{FORM version 4.2}},\ }\Eprint
  {https://arxiv.org/abs/1707.06453} {arXiv:1707.06453 [hep-ph]}  (\bibinfo
  {year} {2017})\BibitemShut {NoStop}%
\bibitem [{\citenamefont {Inc.}()}]{Mathematica}%
  \BibitemOpen
  \bibfield  {author} {\bibinfo {author} {\bibfnamefont {W.~R.}\ \bibnamefont
  {Inc.}},\ }\href {https://www.wolfram.com/mathematica} {\bibinfo {title}
  {Mathematica, {V}ersion 13.0}},\ \bibinfo {note} {champaign, IL,
  2024}\BibitemShut {NoStop}%
\bibitem [{\citenamefont {Garland}\ \emph {et~al.}(2002)\citenamefont
  {Garland}, \citenamefont {Gehrmann}, \citenamefont {Glover}, \citenamefont
  {Koukoutsakis},\ and\ \citenamefont {Remiddi}}]{Garland:2002ak}%
  \BibitemOpen
  \bibfield  {author} {\bibinfo {author} {\bibfnamefont {L.~W.}\ \bibnamefont
  {Garland}}, \bibinfo {author} {\bibfnamefont {T.}~\bibnamefont {Gehrmann}},
  \bibinfo {author} {\bibfnamefont {E.~W.~N.}\ \bibnamefont {Glover}}, \bibinfo
  {author} {\bibfnamefont {A.}~\bibnamefont {Koukoutsakis}},\ and\ \bibinfo
  {author} {\bibfnamefont {E.}~\bibnamefont {Remiddi}},\ }\bibfield  {title}
  {\bibinfo {title} {{Two loop QCD helicity amplitudes for $e^+ e^- \to$ three
  jets}},\ }\href {https://doi.org/10.1016/S0550-3213(02)00627-2} {\bibfield
  {journal} {\bibinfo  {journal} {Nucl. Phys. B}\ }\textbf {\bibinfo {volume}
  {642}},\ \bibinfo {pages} {227} (\bibinfo {year} {2002})},\ \Eprint
  {https://arxiv.org/abs/hep-ph/0206067} {arXiv:hep-ph/0206067} \BibitemShut
  {NoStop}%
\bibitem [{\citenamefont {Gehrmann}\ \emph {et~al.}(2010)\citenamefont
  {Gehrmann}, \citenamefont {Glover}, \citenamefont {Huber}, \citenamefont
  {Ikizlerli},\ and\ \citenamefont {Studerus}}]{Gehrmann:2010ue}%
  \BibitemOpen
  \bibfield  {author} {\bibinfo {author} {\bibfnamefont {T.}~\bibnamefont
  {Gehrmann}}, \bibinfo {author} {\bibfnamefont {E.~W.~N.}\ \bibnamefont
  {Glover}}, \bibinfo {author} {\bibfnamefont {T.}~\bibnamefont {Huber}},
  \bibinfo {author} {\bibfnamefont {N.}~\bibnamefont {Ikizlerli}},\ and\
  \bibinfo {author} {\bibfnamefont {C.}~\bibnamefont {Studerus}},\ }\bibfield
  {title} {\bibinfo {title} {{Calculation of the quark and gluon form factors
  to three loops in QCD}},\ }\href {https://doi.org/10.1007/JHEP06(2010)094}
  {\bibfield  {journal} {\bibinfo  {journal} {JHEP}\ }\textbf {\bibinfo
  {volume} {06}},\ \bibinfo {pages} {094}},\ \Eprint
  {https://arxiv.org/abs/1004.3653} {arXiv:1004.3653 [hep-ph]} \BibitemShut
  {NoStop}%
\bibitem [{\citenamefont {Chetyrkin}\ and\ \citenamefont
  {Tkachov}(1981)}]{Chetyrkin:1981qh}%
  \BibitemOpen
  \bibfield  {author} {\bibinfo {author} {\bibfnamefont {K.~G.}\ \bibnamefont
  {Chetyrkin}}\ and\ \bibinfo {author} {\bibfnamefont {F.~V.}\ \bibnamefont
  {Tkachov}},\ }\bibfield  {title} {\bibinfo {title} {{Integration by Parts:
  The algorithm to calculate $\beta$ functions in 4 loops}},\ }\href
  {https://doi.org/10.1016/0550-3213(81)90199-1} {\bibfield  {journal}
  {\bibinfo  {journal} {Nucl. Phys. B}\ }\textbf {\bibinfo {volume} {192}},\
  \bibinfo {pages} {159} (\bibinfo {year} {1981})}\BibitemShut {NoStop}%
\bibitem [{\citenamefont {Laporta}(2000)}]{Laporta:2000dsw}%
  \BibitemOpen
  \bibfield  {author} {\bibinfo {author} {\bibfnamefont {S.}~\bibnamefont
  {Laporta}},\ }\bibfield  {title} {\bibinfo {title} {{High-precision
  calculation of multiloop Feynman integrals by difference equations}},\ }\href
  {https://doi.org/10.1142/S0217751X00002159} {\bibfield  {journal} {\bibinfo
  {journal} {Int. J. Mod. Phys. A}\ }\textbf {\bibinfo {volume} {15}},\
  \bibinfo {pages} {5087} (\bibinfo {year} {2000})},\ \Eprint
  {https://arxiv.org/abs/hep-ph/0102033} {arXiv:hep-ph/0102033} \BibitemShut
  {NoStop}%
\bibitem [{\citenamefont {von Manteuffel}\ and\ \citenamefont
  {Studerus}(2012)}]{vonManteuffel:2012np}%
  \BibitemOpen
  \bibfield  {author} {\bibinfo {author} {\bibfnamefont {A.}~\bibnamefont {von
  Manteuffel}}\ and\ \bibinfo {author} {\bibfnamefont {C.}~\bibnamefont
  {Studerus}},\ }\bibfield  {title} {\bibinfo {title} {{Reduze
  2\textemdash{}Distributed Feynman integral reduction}},\ }\href@noop {} {\
  (\bibinfo {year} {2012})},\ \Eprint {https://arxiv.org/abs/1201.4330}
  {arXiv:1201.4330 [hep-ph]} \BibitemShut {NoStop}%
\bibitem [{\citenamefont {Maierh\"ofer}\ \emph {et~al.}(2018)\citenamefont
  {Maierh\"ofer}, \citenamefont {Usovitsch},\ and\ \citenamefont
  {Uwer}}]{Maierhofer:2017gsa}%
  \BibitemOpen
  \bibfield  {author} {\bibinfo {author} {\bibfnamefont {P.}~\bibnamefont
  {Maierh\"ofer}}, \bibinfo {author} {\bibfnamefont {J.}~\bibnamefont
  {Usovitsch}},\ and\ \bibinfo {author} {\bibfnamefont {P.}~\bibnamefont
  {Uwer}},\ }\bibfield  {title} {\bibinfo {title} {{Kira\textemdash{}A Feynman
  integral reduction program}},\ }\href
  {https://doi.org/10.1016/j.cpc.2018.04.012} {\bibfield  {journal} {\bibinfo
  {journal} {Comput. Phys. Commun.}\ }\textbf {\bibinfo {volume} {230}},\
  \bibinfo {pages} {99} (\bibinfo {year} {2018})},\ \Eprint
  {https://arxiv.org/abs/1705.05610} {arXiv:1705.05610 [hep-ph]} \BibitemShut
  {NoStop}%
\bibitem [{\citenamefont {Klappert}\ \emph
  {et~al.}(2021{\natexlab{a}})\citenamefont {Klappert}, \citenamefont {Lange},
  \citenamefont {Maierh\"ofer},\ and\ \citenamefont
  {Usovitsch}}]{Klappert:2020nbg}%
  \BibitemOpen
  \bibfield  {author} {\bibinfo {author} {\bibfnamefont {J.}~\bibnamefont
  {Klappert}}, \bibinfo {author} {\bibfnamefont {F.}~\bibnamefont {Lange}},
  \bibinfo {author} {\bibfnamefont {P.}~\bibnamefont {Maierh\"ofer}},\ and\
  \bibinfo {author} {\bibfnamefont {J.}~\bibnamefont {Usovitsch}},\ }\bibfield
  {title} {\bibinfo {title} {{Integral reduction with Kira 2.0 and finite field
  methods}},\ }\href {https://doi.org/10.1016/j.cpc.2021.108024} {\bibfield
  {journal} {\bibinfo  {journal} {Comput. Phys. Commun.}\ }\textbf {\bibinfo
  {volume} {266}},\ \bibinfo {pages} {108024} (\bibinfo {year}
  {2021}{\natexlab{a}})},\ \Eprint {https://arxiv.org/abs/2008.06494}
  {arXiv:2008.06494 [hep-ph]} \BibitemShut {NoStop}%
\bibitem [{\citenamefont {Lewis}()}]{Lewis}%
  \BibitemOpen
  \bibfield  {author} {\bibinfo {author} {\bibfnamefont {R.~H.}\ \bibnamefont
  {Lewis}},\ }\href {https://home.bway.net/lewis} {\bibinfo {title} {Fermat: A
  computer algebra system for polynomial and matrix multiplication}},\ \bibinfo
  {note} {computer Software}\BibitemShut {NoStop}%
\bibitem [{\citenamefont {Klappert}\ and\ \citenamefont
  {Lange}(2020)}]{Klappert:2019emp}%
  \BibitemOpen
  \bibfield  {author} {\bibinfo {author} {\bibfnamefont {J.}~\bibnamefont
  {Klappert}}\ and\ \bibinfo {author} {\bibfnamefont {F.}~\bibnamefont
  {Lange}},\ }\bibfield  {title} {\bibinfo {title} {{Reconstructing rational
  functions with FireFly}},\ }\href {https://doi.org/10.1016/j.cpc.2019.106951}
  {\bibfield  {journal} {\bibinfo  {journal} {Comput. Phys. Commun.}\ }\textbf
  {\bibinfo {volume} {247}},\ \bibinfo {pages} {106951} (\bibinfo {year}
  {2020})},\ \Eprint {https://arxiv.org/abs/1904.00009} {arXiv:1904.00009
  [cs.SC]} \BibitemShut {NoStop}%
\bibitem [{\citenamefont {Klappert}\ \emph
  {et~al.}(2021{\natexlab{b}})\citenamefont {Klappert}, \citenamefont {Klein},\
  and\ \citenamefont {Lange}}]{Klappert:2020aqs}%
  \BibitemOpen
  \bibfield  {author} {\bibinfo {author} {\bibfnamefont {J.}~\bibnamefont
  {Klappert}}, \bibinfo {author} {\bibfnamefont {S.~Y.}\ \bibnamefont
  {Klein}},\ and\ \bibinfo {author} {\bibfnamefont {F.}~\bibnamefont {Lange}},\
  }\bibfield  {title} {\bibinfo {title} {{Interpolation of dense and sparse
  rational functions and other improvements in FireFly}},\ }\href
  {https://doi.org/10.1016/j.cpc.2021.107968} {\bibfield  {journal} {\bibinfo
  {journal} {Comput. Phys. Commun.}\ }\textbf {\bibinfo {volume} {264}},\
  \bibinfo {pages} {107968} (\bibinfo {year} {2021}{\natexlab{b}})},\ \Eprint
  {https://arxiv.org/abs/2004.01463} {arXiv:2004.01463 [cs.MS]} \BibitemShut
  {NoStop}%
\bibitem [{\citenamefont {Gehrmann}\ and\ \citenamefont
  {Remiddi}(2000)}]{Gehrmann:1999as}%
  \BibitemOpen
  \bibfield  {author} {\bibinfo {author} {\bibfnamefont {T.}~\bibnamefont
  {Gehrmann}}\ and\ \bibinfo {author} {\bibfnamefont {E.}~\bibnamefont
  {Remiddi}},\ }\bibfield  {title} {\bibinfo {title} {{Differential equations
  for two loop four point functions}},\ }\href
  {https://doi.org/10.1016/S0550-3213(00)00223-6} {\bibfield  {journal}
  {\bibinfo  {journal} {Nucl. Phys. B}\ }\textbf {\bibinfo {volume} {580}},\
  \bibinfo {pages} {485} (\bibinfo {year} {2000})},\ \Eprint
  {https://arxiv.org/abs/hep-ph/9912329} {arXiv:hep-ph/9912329} \BibitemShut
  {NoStop}%
\bibitem [{\citenamefont {Catani}(1998)}]{Catani:1998bh}%
  \BibitemOpen
  \bibfield  {author} {\bibinfo {author} {\bibfnamefont {S.}~\bibnamefont
  {Catani}},\ }\bibfield  {title} {\bibinfo {title} {{The singular behavior of
  QCD amplitudes at two loop order}},\ }\href
  {https://doi.org/10.1016/S0370-2693(98)00332-3} {\bibfield  {journal}
  {\bibinfo  {journal} {Phys. Lett. B}\ }\textbf {\bibinfo {volume} {427}},\
  \bibinfo {pages} {161} (\bibinfo {year} {1998})},\ \Eprint
  {https://arxiv.org/abs/hep-ph/9802439} {arXiv:hep-ph/9802439} \BibitemShut
  {NoStop}%
\bibitem [{\citenamefont {Passarino}\ and\ \citenamefont
  {Veltman}(1979)}]{Passarino:1978jh}%
  \BibitemOpen
  \bibfield  {author} {\bibinfo {author} {\bibfnamefont {G.}~\bibnamefont
  {Passarino}}\ and\ \bibinfo {author} {\bibfnamefont {M.~J.~G.}\ \bibnamefont
  {Veltman}},\ }\bibfield  {title} {\bibinfo {title} {{One loop corrections for
  $e^+ e^-$ annihilation into $\mu^+ \mu^-$ in the Weinberg model}},\ }\href
  {https://doi.org/10.1016/0550-3213(79)90234-7} {\bibfield  {journal}
  {\bibinfo  {journal} {Nucl. Phys. B}\ }\textbf {\bibinfo {volume} {160}},\
  \bibinfo {pages} {151} (\bibinfo {year} {1979})}\BibitemShut {NoStop}%
\bibitem [{\citenamefont {Buccioni}\ \emph {et~al.}(2019)\citenamefont
  {Buccioni}, \citenamefont {Lang}, \citenamefont {Lindert}, \citenamefont
  {Maierh\"ofer}, \citenamefont {Pozzorini}, \citenamefont {Zhang},\ and\
  \citenamefont {Zoller}}]{Buccioni:2019sur}%
  \BibitemOpen
  \bibfield  {author} {\bibinfo {author} {\bibfnamefont {F.}~\bibnamefont
  {Buccioni}}, \bibinfo {author} {\bibfnamefont {J.-N.}\ \bibnamefont {Lang}},
  \bibinfo {author} {\bibfnamefont {J.~M.}\ \bibnamefont {Lindert}}, \bibinfo
  {author} {\bibfnamefont {P.}~\bibnamefont {Maierh\"ofer}}, \bibinfo {author}
  {\bibfnamefont {S.}~\bibnamefont {Pozzorini}}, \bibinfo {author}
  {\bibfnamefont {H.}~\bibnamefont {Zhang}},\ and\ \bibinfo {author}
  {\bibfnamefont {M.~F.}\ \bibnamefont {Zoller}},\ }\bibfield  {title}
  {\bibinfo {title} {{OpenLoops 2}},\ }\href
  {https://doi.org/10.1140/epjc/s10052-019-7306-2} {\bibfield  {journal}
  {\bibinfo  {journal} {Eur. Phys. J. C}\ }\textbf {\bibinfo {volume} {79}},\
  \bibinfo {pages} {866} (\bibinfo {year} {2019})},\ \Eprint
  {https://arxiv.org/abs/1907.13071} {arXiv:1907.13071 [hep-ph]} \BibitemShut
  {NoStop}%
\bibitem [{\citenamefont {Alwall}\ \emph {et~al.}(2014)\citenamefont {Alwall},
  \citenamefont {Frederix}, \citenamefont {Frixione}, \citenamefont {Hirschi},
  \citenamefont {Maltoni}, \citenamefont {Mattelaer}, \citenamefont {Shao},
  \citenamefont {Stelzer}, \citenamefont {Torrielli},\ and\ \citenamefont
  {Zaro}}]{Alwall:2014hca}%
  \BibitemOpen
  \bibfield  {author} {\bibinfo {author} {\bibfnamefont {J.}~\bibnamefont
  {Alwall}}, \bibinfo {author} {\bibfnamefont {R.}~\bibnamefont {Frederix}},
  \bibinfo {author} {\bibfnamefont {S.}~\bibnamefont {Frixione}}, \bibinfo
  {author} {\bibfnamefont {V.}~\bibnamefont {Hirschi}}, \bibinfo {author}
  {\bibfnamefont {F.}~\bibnamefont {Maltoni}}, \bibinfo {author} {\bibfnamefont
  {O.}~\bibnamefont {Mattelaer}}, \bibinfo {author} {\bibfnamefont {H.~S.}\
  \bibnamefont {Shao}}, \bibinfo {author} {\bibfnamefont {T.}~\bibnamefont
  {Stelzer}}, \bibinfo {author} {\bibfnamefont {P.}~\bibnamefont {Torrielli}},\
  and\ \bibinfo {author} {\bibfnamefont {M.}~\bibnamefont {Zaro}},\ }\bibfield
  {title} {\bibinfo {title} {{The automated computation of tree-level and
  next-to-leading order differential cross sections, and their matching to
  parton shower simulations}},\ }\href
  {https://doi.org/10.1007/JHEP07(2014)079} {\bibfield  {journal} {\bibinfo
  {journal} {JHEP}\ }\textbf {\bibinfo {volume} {07}},\ \bibinfo {pages}
  {079}},\ \Eprint {https://arxiv.org/abs/1405.0301} {arXiv:1405.0301 [hep-ph]}
  \BibitemShut {NoStop}%
\bibitem [{\citenamefont {Cullen}\ \emph {et~al.}(2012)\citenamefont {Cullen},
  \citenamefont {Greiner}, \citenamefont {Heinrich}, \citenamefont {Luisoni},
  \citenamefont {Mastrolia}, \citenamefont {Ossola}, \citenamefont {Reiter},\
  and\ \citenamefont {Tramontano}}]{Cullen:2011ac}%
  \BibitemOpen
  \bibfield  {author} {\bibinfo {author} {\bibfnamefont {G.}~\bibnamefont
  {Cullen}}, \bibinfo {author} {\bibfnamefont {N.}~\bibnamefont {Greiner}},
  \bibinfo {author} {\bibfnamefont {G.}~\bibnamefont {Heinrich}}, \bibinfo
  {author} {\bibfnamefont {G.}~\bibnamefont {Luisoni}}, \bibinfo {author}
  {\bibfnamefont {P.}~\bibnamefont {Mastrolia}}, \bibinfo {author}
  {\bibfnamefont {G.}~\bibnamefont {Ossola}}, \bibinfo {author} {\bibfnamefont
  {T.}~\bibnamefont {Reiter}},\ and\ \bibinfo {author} {\bibfnamefont
  {F.}~\bibnamefont {Tramontano}} (\bibinfo {collaboration} {GoSam}),\
  }\bibfield  {title} {\bibinfo {title} {{Automated one-loop calculations with
  GoSam}},\ }\href {https://doi.org/10.1140/epjc/s10052-012-1889-1} {\bibfield
  {journal} {\bibinfo  {journal} {Eur. Phys. J. C}\ }\textbf {\bibinfo {volume}
  {72}},\ \bibinfo {pages} {1889} (\bibinfo {year} {2012})},\ \Eprint
  {https://arxiv.org/abs/1111.2034} {arXiv:1111.2034 [hep-ph]} \BibitemShut
  {NoStop}%
\bibitem [{\citenamefont {Cullen}\ \emph {et~al.}(2014)\citenamefont {Cullen}
  \emph {et~al.}}]{GoSam:2014iqq}%
  \BibitemOpen
  \bibfield  {author} {\bibinfo {author} {\bibfnamefont {G.}~\bibnamefont
  {Cullen}} \emph {et~al.} (\bibinfo {collaboration} {GoSam}),\ }\bibfield
  {title} {\bibinfo {title} {{G$\scriptsize{O}$S$\scriptsize{AM}$-2.0: a tool
  for automated one-loop calculations within the Standard Model and beyond}},\
  }\href {https://doi.org/10.1140/epjc/s10052-014-3001-5} {\bibfield  {journal}
  {\bibinfo  {journal} {Eur. Phys. J. C}\ }\textbf {\bibinfo {volume} {74}},\
  \bibinfo {pages} {3001} (\bibinfo {year} {2014})},\ \Eprint
  {https://arxiv.org/abs/1404.7096} {arXiv:1404.7096 [hep-ph]} \BibitemShut
  {NoStop}%
\bibitem [{\citenamefont {Peraro}(2019)}]{Peraro:2019svx}%
  \BibitemOpen
  \bibfield  {author} {\bibinfo {author} {\bibfnamefont {T.}~\bibnamefont
  {Peraro}},\ }\bibfield  {title} {\bibinfo {title} {{FiniteFlow: multivariate
  functional reconstruction using finite fields and dataflow graphs}},\ }\href
  {https://doi.org/10.1007/JHEP07(2019)031} {\bibfield  {journal} {\bibinfo
  {journal} {JHEP}\ }\textbf {\bibinfo {volume} {07}},\ \bibinfo {pages}
  {031}},\ \Eprint {https://arxiv.org/abs/1905.08019} {arXiv:1905.08019
  [hep-ph]} \BibitemShut {NoStop}%
\bibitem [{\citenamefont {Gehrmann}\ and\ \citenamefont
  {Remiddi}(2001{\natexlab{a}})}]{Gehrmann:2000zt}%
  \BibitemOpen
  \bibfield  {author} {\bibinfo {author} {\bibfnamefont {T.}~\bibnamefont
  {Gehrmann}}\ and\ \bibinfo {author} {\bibfnamefont {E.}~\bibnamefont
  {Remiddi}},\ }\bibfield  {title} {\bibinfo {title} {{Two loop master
  integrals for $\gamma^\ast\to$ 3 jets: The planar topologies}},\ }\href
  {https://doi.org/10.1016/S0550-3213(01)00057-8} {\bibfield  {journal}
  {\bibinfo  {journal} {Nucl. Phys. B}\ }\textbf {\bibinfo {volume} {601}},\
  \bibinfo {pages} {248} (\bibinfo {year} {2001}{\natexlab{a}})},\ \Eprint
  {https://arxiv.org/abs/hep-ph/0008287} {arXiv:hep-ph/0008287} \BibitemShut
  {NoStop}%
\bibitem [{\citenamefont {Gehrmann}\ and\ \citenamefont
  {Remiddi}(2001{\natexlab{b}})}]{Gehrmann:2001ck}%
  \BibitemOpen
  \bibfield  {author} {\bibinfo {author} {\bibfnamefont {T.}~\bibnamefont
  {Gehrmann}}\ and\ \bibinfo {author} {\bibfnamefont {E.}~\bibnamefont
  {Remiddi}},\ }\bibfield  {title} {\bibinfo {title} {{Two loop master
  integrals for $\gamma^\ast \to$ 3 jets: The nonplanar topologies}},\ }\href
  {https://doi.org/10.1016/S0550-3213(01)00074-8} {\bibfield  {journal}
  {\bibinfo  {journal} {Nucl. Phys. B}\ }\textbf {\bibinfo {volume} {601}},\
  \bibinfo {pages} {287} (\bibinfo {year} {2001}{\natexlab{b}})},\ \Eprint
  {https://arxiv.org/abs/hep-ph/0101124} {arXiv:hep-ph/0101124} \BibitemShut
  {NoStop}%
\bibitem [{\citenamefont {Brein}\ \emph {et~al.}(2013)\citenamefont {Brein},
  \citenamefont {Harlander},\ and\ \citenamefont {Zirke}}]{Brein:2012ne}%
  \BibitemOpen
  \bibfield  {author} {\bibinfo {author} {\bibfnamefont {O.}~\bibnamefont
  {Brein}}, \bibinfo {author} {\bibfnamefont {R.~V.}\ \bibnamefont
  {Harlander}},\ and\ \bibinfo {author} {\bibfnamefont {T.~J.~E.}\ \bibnamefont
  {Zirke}},\ }\bibfield  {title} {\bibinfo {title} {{vh@nnlo\textemdash{}Higgs
  strahlung at hadron colliders}},\ }\href
  {https://doi.org/10.1016/j.cpc.2012.11.002} {\bibfield  {journal} {\bibinfo
  {journal} {Comput. Phys. Commun.}\ }\textbf {\bibinfo {volume} {184}},\
  \bibinfo {pages} {998} (\bibinfo {year} {2013})},\ \Eprint
  {https://arxiv.org/abs/1210.5347} {arXiv:1210.5347 [hep-ph]} \BibitemShut
  {NoStop}%
\bibitem [{\citenamefont {Boughezal}\ \emph {et~al.}(2017)\citenamefont
  {Boughezal}, \citenamefont {Campbell}, \citenamefont {Ellis}, \citenamefont
  {Focke}, \citenamefont {Giele}, \citenamefont {Liu}, \citenamefont
  {Petriello},\ and\ \citenamefont {Williams}}]{Boughezal:2016wmq}%
  \BibitemOpen
  \bibfield  {author} {\bibinfo {author} {\bibfnamefont {R.}~\bibnamefont
  {Boughezal}}, \bibinfo {author} {\bibfnamefont {J.~M.}\ \bibnamefont
  {Campbell}}, \bibinfo {author} {\bibfnamefont {R.~K.}\ \bibnamefont {Ellis}},
  \bibinfo {author} {\bibfnamefont {C.}~\bibnamefont {Focke}}, \bibinfo
  {author} {\bibfnamefont {W.}~\bibnamefont {Giele}}, \bibinfo {author}
  {\bibfnamefont {X.}~\bibnamefont {Liu}}, \bibinfo {author} {\bibfnamefont
  {F.}~\bibnamefont {Petriello}},\ and\ \bibinfo {author} {\bibfnamefont
  {C.}~\bibnamefont {Williams}},\ }\bibfield  {title} {\bibinfo {title} {{Color
  singlet production at NNLO in MCFM}},\ }\href
  {https://doi.org/10.1140/epjc/s10052-016-4558-y} {\bibfield  {journal}
  {\bibinfo  {journal} {Eur. Phys. J. C}\ }\textbf {\bibinfo {volume} {77}},\
  \bibinfo {pages} {7} (\bibinfo {year} {2017})},\ \Eprint
  {https://arxiv.org/abs/1605.08011} {arXiv:1605.08011 [hep-ph]} \BibitemShut
  {NoStop}%
\bibitem [{\citenamefont {Caola}\ \emph {et~al.}(2017)\citenamefont {Caola},
  \citenamefont {Melnikov},\ and\ \citenamefont {R\"ontsch}}]{Caola:2017dug}%
  \BibitemOpen
  \bibfield  {author} {\bibinfo {author} {\bibfnamefont {F.}~\bibnamefont
  {Caola}}, \bibinfo {author} {\bibfnamefont {K.}~\bibnamefont {Melnikov}},\
  and\ \bibinfo {author} {\bibfnamefont {R.}~\bibnamefont {R\"ontsch}},\
  }\bibfield  {title} {\bibinfo {title} {{Nested soft-collinear subtractions in
  NNLO QCD computations}},\ }\href
  {https://doi.org/10.1140/epjc/s10052-017-4774-0} {\bibfield  {journal}
  {\bibinfo  {journal} {Eur. Phys. J. C}\ }\textbf {\bibinfo {volume} {77}},\
  \bibinfo {pages} {248} (\bibinfo {year} {2017})},\ \Eprint
  {https://arxiv.org/abs/1702.01352} {arXiv:1702.01352 [hep-ph]} \BibitemShut
  {NoStop}%
\bibitem [{\citenamefont {Caola}\ \emph {et~al.}(2019)\citenamefont {Caola},
  \citenamefont {Melnikov},\ and\ \citenamefont {R\"ontsch}}]{Caola:2019nzf}%
  \BibitemOpen
  \bibfield  {author} {\bibinfo {author} {\bibfnamefont {F.}~\bibnamefont
  {Caola}}, \bibinfo {author} {\bibfnamefont {K.}~\bibnamefont {Melnikov}},\
  and\ \bibinfo {author} {\bibfnamefont {R.}~\bibnamefont {R\"ontsch}},\
  }\bibfield  {title} {\bibinfo {title} {{Analytic results for color-singlet
  production at NNLO QCD with the nested soft-collinear subtraction scheme}},\
  }\href {https://doi.org/10.1140/epjc/s10052-019-6880-7} {\bibfield  {journal}
  {\bibinfo  {journal} {Eur. Phys. J. C}\ }\textbf {\bibinfo {volume} {79}},\
  \bibinfo {pages} {386} (\bibinfo {year} {2019})},\ \Eprint
  {https://arxiv.org/abs/1902.02081} {arXiv:1902.02081 [hep-ph]} \BibitemShut
  {NoStop}%
\bibitem [{\citenamefont {Behring}\ \emph {et~al.}(2020)\citenamefont
  {Behring}, \citenamefont {Bizo\'n}, \citenamefont {Caola}, \citenamefont
  {Melnikov},\ and\ \citenamefont {R\"ontsch}}]{Behring:2020uzq}%
  \BibitemOpen
  \bibfield  {author} {\bibinfo {author} {\bibfnamefont {A.}~\bibnamefont
  {Behring}}, \bibinfo {author} {\bibfnamefont {W.}~\bibnamefont {Bizo\'n}},
  \bibinfo {author} {\bibfnamefont {F.}~\bibnamefont {Caola}}, \bibinfo
  {author} {\bibfnamefont {K.}~\bibnamefont {Melnikov}},\ and\ \bibinfo
  {author} {\bibfnamefont {R.}~\bibnamefont {R\"ontsch}},\ }\bibfield  {title}
  {\bibinfo {title} {{Bottom quark mass effects in associated $WH$ production
  with the $H \to b\bar{b}$ decay through NNLO QCD}},\ }\href
  {https://doi.org/10.1103/PhysRevD.101.114012} {\bibfield  {journal} {\bibinfo
   {journal} {Phys. Rev. D}\ }\textbf {\bibinfo {volume} {101}},\ \bibinfo
  {pages} {114012} (\bibinfo {year} {2020})},\ \Eprint
  {https://arxiv.org/abs/2003.08321} {arXiv:2003.08321 [hep-ph]} \BibitemShut
  {NoStop}%
\bibitem [{\citenamefont {Harlander}\ \emph {et~al.}(2020)\citenamefont
  {Harlander}, \citenamefont {Klein},\ and\ \citenamefont
  {Lipp}}]{Harlander:2020cyh}%
  \BibitemOpen
  \bibfield  {author} {\bibinfo {author} {\bibfnamefont {R.~V.}\ \bibnamefont
  {Harlander}}, \bibinfo {author} {\bibfnamefont {S.~Y.}\ \bibnamefont
  {Klein}},\ and\ \bibinfo {author} {\bibfnamefont {M.}~\bibnamefont {Lipp}},\
  }\bibfield  {title} {\bibinfo {title} {{FeynGame}},\ }\href
  {https://doi.org/10.1016/j.cpc.2020.107465} {\bibfield  {journal} {\bibinfo
  {journal} {Comput. Phys. Commun.}\ }\textbf {\bibinfo {volume} {256}},\
  \bibinfo {pages} {107465} (\bibinfo {year} {2020})},\ \Eprint
  {https://arxiv.org/abs/2003.00896} {arXiv:2003.00896 [physics.ed-ph]}
  \BibitemShut {NoStop}%
\bibitem [{\citenamefont {Harlander}\ \emph {et~al.}(2024)\citenamefont
  {Harlander}, \citenamefont {Klein},\ and\ \citenamefont
  {Schaaf}}]{Harlander:2024qbn}%
  \BibitemOpen
  \bibfield  {author} {\bibinfo {author} {\bibfnamefont {R.}~\bibnamefont
  {Harlander}}, \bibinfo {author} {\bibfnamefont {S.~Y.}\ \bibnamefont
  {Klein}},\ and\ \bibinfo {author} {\bibfnamefont {M.~C.}\ \bibnamefont
  {Schaaf}},\ }\bibfield  {title} {\bibinfo {title}
  {{FeynGame-2.1\textemdash{}Feynman diagrams made easy}},\ }\href
  {https://doi.org/10.22323/1.449.0657} {\bibfield  {journal} {\bibinfo
  {journal} {PoS}\ }\textbf {\bibinfo {volume} {EPS-HEP2023}},\ \bibinfo
  {pages} {657} (\bibinfo {year} {2024})},\ \Eprint
  {https://arxiv.org/abs/2401.12778} {arXiv:2401.12778 [hep-ph]} \BibitemShut
  {NoStop}%
\bibitem [{\citenamefont {B\"undgen}\ \emph {et~al.}(2025)\citenamefont
  {B\"undgen}, \citenamefont {Harlander}, \citenamefont {Klein},\ and\
  \citenamefont {Schaaf}}]{Bundgen:2025utt}%
  \BibitemOpen
  \bibfield  {author} {\bibinfo {author} {\bibfnamefont {L.}~\bibnamefont
  {B\"undgen}}, \bibinfo {author} {\bibfnamefont {R.~V.}\ \bibnamefont
  {Harlander}}, \bibinfo {author} {\bibfnamefont {S.~Y.}\ \bibnamefont
  {Klein}},\ and\ \bibinfo {author} {\bibfnamefont {M.~C.}\ \bibnamefont
  {Schaaf}},\ }\bibfield  {title} {\bibinfo {title} {{FeynGame 3.0}},\ }\Eprint
  {https://arxiv.org/abs/2501.04651} {arXiv:2501.04651 [hep-ph]}  (\bibinfo
  {year} {2025})\BibitemShut {NoStop}%
\bibitem [{\citenamefont {Becher}\ and\ \citenamefont
  {Neubert}(2009{\natexlab{a}})}]{Becher:2009qa}%
  \BibitemOpen
  \bibfield  {author} {\bibinfo {author} {\bibfnamefont {T.}~\bibnamefont
  {Becher}}\ and\ \bibinfo {author} {\bibfnamefont {M.}~\bibnamefont
  {Neubert}},\ }\bibfield  {title} {\bibinfo {title} {{On the Structure of
  Infrared Singularities of Gauge-Theory Amplitudes}},\ }\href
  {https://doi.org/10.1088/1126-6708/2009/06/081} {\bibfield  {journal}
  {\bibinfo  {journal} {JHEP}\ }\textbf {\bibinfo {volume} {06}},\ \bibinfo
  {pages} {081}},\ \bibinfo {note} {[Erratum: JHEP 11, 024 (2013)]},\ \Eprint
  {https://arxiv.org/abs/0903.1126} {arXiv:0903.1126 [hep-ph]} \BibitemShut
  {NoStop}%
\bibitem [{\citenamefont {Becher}\ and\ \citenamefont
  {Neubert}(2009{\natexlab{b}})}]{Becher:2009cu}%
  \BibitemOpen
  \bibfield  {author} {\bibinfo {author} {\bibfnamefont {T.}~\bibnamefont
  {Becher}}\ and\ \bibinfo {author} {\bibfnamefont {M.}~\bibnamefont
  {Neubert}},\ }\bibfield  {title} {\bibinfo {title} {{Infrared singularities
  of scattering amplitudes in perturbative QCD}},\ }\href
  {https://doi.org/10.1103/PhysRevLett.102.162001} {\bibfield  {journal}
  {\bibinfo  {journal} {Phys. Rev. Lett.}\ }\textbf {\bibinfo {volume} {102}},\
  \bibinfo {pages} {162001} (\bibinfo {year} {2009}{\natexlab{b}})},\ \bibinfo
  {note} {[Erratum: Phys.Rev.Lett. 111, 199905 (2013)]},\ \Eprint
  {https://arxiv.org/abs/0901.0722} {arXiv:0901.0722 [hep-ph]} \BibitemShut
  {NoStop}%
\end{thebibliography}%

\end{document}